\documentclass[usenatbib]{mn2e}

\usepackage{graphicx}
\usepackage{amsmath}
\usepackage{lscape}
\usepackage{afterpage}

\title[The most extreme ULXs: evidence for IMBHs?]{The most extreme ultraluminous X-ray sources: evidence for intermediate-mass black holes?}

\author[A.\,D. Sutton, T.\,P. Roberts, D.\,J. Walton, J.\,C. Gladstone, A.\,E. Scott]{Andrew D. Sutton$^1$\thanks{Email: andrew.sutton@durham.ac.uk}, Timothy P. Roberts$^1$, Dominic J. Walton$^2$,\and Jeanette C. Gladstone$^3$, Amy E. Scott$^4$\\
 \\
$^1$Department of Physics, University of Durham, South Road, Durham, DH1 3LE, UK\\
$^2$Institute of Astronomy, University of Cambridge, Madingley Road, Cambridge, CB3 0HA, UK\\
$^3$Department of Physics, University of Alberta, Edmonton, Alberta, T6G 2G7, Canada\\
$^4$X-ray and Observational Astronomy Group, University of Leicester, University Road, Leicester, LE1 7RH, UK}

\pagerange{\pageref{firstpage}--\pageref{lastpage}} \pubyear{2011}

\newcommand{\xmmn}{{\it XMM-Newton~\/}}

\newcommand{\chan}{{\it Chandra~\/}}
\newcommand{\hst}{{\it HST~\/}}

\def\Msun{\hbox{$\rm M_{\odot}$}}
\def\ergcms{{\rm ~erg~cm^{-2}~s^{-1}}}

\def\ergsec{{\rm ~erg~s^{-1}}}

\def\H0{{\rm ~km~s^{-1}~Mpc^{-1}}}

\def\eg{{e.g.~\/}}

\def\ie{{i.e.,~\/}}
\def\cf{{cf.~\/}}
\def\la{\mathrel{\hbox{\rlap{\hbox{\lower4pt\hbox{$\sim$}}}{\raise2pt\hbox{$<$}}
}}}
\def\ga{\mathrel{\hbox{\rlap{\hbox{\lower4pt\hbox{$\sim$}}}{\raise2pt\hbox{$>$}}
}}}
\def\d25{$D_{25}$}

\def\srcseven{Src. 7}
\def\srcnine{Src. 9}
\def\srctwo{Src. 2}
\def\srcthree{Src. 3}
\def\srcfour{Src. 4}
\def\srcfive{Src. 5}
\def\srcone{Src. 1}
\def\srcsix{Src. 6}
\def\srceight{Src. 8}
\def\srcten{Src. 10}

\begin{document}

\maketitle

\label{firstpage}

\begin{abstract}
We present the results from an X-ray and optical study of a new sample of eight extreme luminosity ultraluminous X-ray source (ULX) candidates, which were selected as the brightest ULXs (with $L_{\rm X} > 5 \times 10^{40}~{\rm erg~s^{-1}}$) located within 100 Mpc identified in a cross correlation of the 2XMM-DR1 and RC3 catalogues.  These objects are so luminous that they are difficult to describe with current models of super-Eddington accretion onto all but the 
most massive stellar remnants; hence they are amongst the most plausible candidates to host larger, intermediate-mass black holes (IMBHs).  Two objects are luminous enough in at least one observation to be classed as hyperluminous X-ray source (HLX) candidates, including one persistent HLX in an S0 galaxy that (at $3 \times 10^{41} \ergsec$) is the second most luminous HLX yet detected.  The remaining seven sources are located in spiral galaxies, and several appear to be closely associated with regions of star formation as is common for many less luminous ULXs.  However, the X-ray characteristics of these extreme ULXs appear to diverge from the less luminous objects. They are typically harder, possessing absorbed power-law continuum spectra with $\Gamma \sim 1.7$, and are potentially more variable on short timescales, with data consistent with $\sim$ 10--20 per cent rms variability on timescales of 0.2--2 ks (albeit at low to moderate significance in many datasets).  These properties appear consistent with the sub-Eddington hard state, which given the observed luminosities of these objects suggests the presence of IMBHs with masses in the range $10^3 - 10^4~\Msun$.  As such, this strengthens the case for these brightest ULXs as good candidates for the eventual conclusive detection of the highly elusive IMBHs in the present-day Universe.  However, we caution that a combination of the highest plausible super-Eddington accretion rates and the largest permitted stellar black hole remnants cannot be ruled out without future, improved observations.

\end{abstract}


\section{Introduction}






Although the existence of a population of black holes with masses between those in Galactic black hole X-ray binaries (BHXRBs; $ \sim 10~\Msun$) and in active galactic nuclei (AGNs; $> 10^5~\Msun$) has long been postulated, these {\it intermediate-mass black holes\/} (IMBHs) have remained observationally elusive, with dynamical evidence for such objects in large globular clusters still the subject of some dispute (e.g. \citealt*{gebhardt_etal_2005}; \citealt{anderson_van_der_marel_2010}). Yet such objects appear necessary to explain the initial seeding of the supermassive black holes (SMBHs) now present in most galaxy nuclei \citep{volonteri_2010}, whose mass was subsequently built up either through the accretion of gas (\eg \citealt{yu_and_tremaine_2002}), or through hierarchical black hole mergers (\eg \citealt{schneider_etal_2002}). It is likely that the processes forming these seeds -- for example, the collapse of Population III stars, or dynamical interactions in dense stellar environments -- would have left relic IMBHs in the halos of present-day galaxies (e.g. \citealt{madau_and_rees_2001}; \citealt{ebisuzaki_etal_2001}; \citealt{islam_etal_2004}). Indeed, the latter process may still be ongoing in some young, dense stellar clusters (\citealt{portegies_zwart_etal_2004}), adding to the primordial IMBH population. Furthermore, the tidal stripping of merging satellite galaxies containing nuclear black holes may also add to this population in large galaxy haloes (e.g. \citealt{Bellovary_etal_2010}). It is therefore puzzling that the existence of such IMBHs remains to be confirmed.


The detection of a population of unusually luminous X-ray sources -- in excess of $10^{39} \ergsec$ -- located outside the nuclei of nearby galaxies initially led to suggestions that these objects could harbour the long sought-after IMBHs (e.g. \citealt{colbert_and_mushotzky_1999}). Early \chan \& \xmmn spectral results for these ultraluminous X-ray sources (ULXs; see \citealt{feng_and_soria_2011} for a recent review) appeared to support this diagnosis with the detection of apparent cool disc signatures, indicative of IMBHs (e.g. \citealt{kaaret_etal_2003}; \citealt{miller_etal_2003}; \citealt*{miller_etal_2004}). However, subsequent analyses of high quality \xmmn data have instead revealed spectral and temporal characteristics in many ULXs that appear inconsistent with the sub-Eddington accretion regime expected for IMBHs at typical ULX luminosities. Instead, these analyses support models where ULXs are powered by super-Eddington accretion onto stellar remnant black holes ($\la 100~\Msun$, \citealt{roberts_2007}; \citealt*{gladstone_etal_2009}; \citealt{middleton_etal_2011a}; \citealt*{middleton_etal_2011b}).

Further observational support for ULXs as stellar-mass objects comes from the close spatial and temporal association of ULXs with ongoing star formation, as most notably seen in the Cartwheel galaxy (\citealt{gao_etal_2003}; \citealt{king_2004}; \citealt{mapelli_etal_2008}), suggesting they are a type of high-mass X-ray binary (HMXB) which naturally provides them with the large mass reservoir required to sustain ULX luminosities \citep*{rappaport_etal_2005}.  It also comes from the power-law form of the X-ray luminosity function (XLF) for HMXBs in nearby star forming galaxies, that is smooth across the transition from standard HMXBs to ULXs, but then breaks at $\sim 2 \times 10^{40} \ergsec$ (\citealt*{grimm_etal_2003}; \citealt{swartz_etal_2004}; \citealt{swartz_etal_2011}).  This break is important, as its luminosity is only $\sim 10$ per cent of the Eddington luminosity for a $\sim 10^3~\Msun$ IMBH; a similar turn-off at such a low Eddington fraction is not seen in other accretion-powered populations, making it difficult to reconcile with a population of IMBHs, but is again supportive of most ULXs as a population of super-Eddington BHXRBs.

However, there remain some plausible IMBH candidates in the ULX population. Recent work by \cite{swartz_etal_2011} has demonstrated that an extrapolation of the local XLF out to 100 Mpc underpredicts the observed numbers of high luminosity ULXs, particularly those that possess luminosities greater than $10^{41} \ergsec$, suggesting that such objects originate in a separate physical population to the HMXBs that likely underlie most ULXs.  Indeed, while a combination of massive stellar black holes, possibly formed in low metallicity regions (up to $80~\Msun$; \citealt{zampieri_and_roberts_2009}, \citealt{belczynski_etal_2010}) and close to maximal radiation from super-Eddington accretion ($\sim 20 ~L_{\rm Edd}$ from a face-on disc; \citealt{ohsuga_and_mineshige_2011}) could explain ULXs up to $\sim 10^{41} \ergsec$, but objects exceeding this luminosity appear to require a different physical scenario. It is these {\it hyperluminous X-ray sources\/} (HLXs; $L_{\rm X} > 10^{41}~{\rm erg~s^{-1}}$) that may present us with the best evidence for IMBHs to date.

Very few {\it bona fide\/} HLXs are known: the best candidates include M82 X-1 (e.g. \citealt{matsumoto_etal_2001}), ESO 243-49 HLX-1 \citep{farrell_etal_2009}, Cartwheel N10 \citep{pizzolato_etal_2010} and CXO J122518.6+144545 \citep{jonker_etal_2010}. Most observations have focussed on the nearest HLX, M82 X-1 at $d \sim 4$ Mpc, and the most luminous, ESO 243-49 HLX-1 at $L_{\rm X,peak} \sim 10^{42} \ergsec$. M82 X-1 is notable for several reasons, including its co-location with a massive, young stellar cluster \citep{portegies_zwart_etal_2004}, the detection of a 50 -- 100 mHz QPO in its power spectrum \citep{strohmayer_and_mushotzky_2003}, a reported 62-day periodicity in its X-ray light curve \citep{kaaret_and_feng_2007} and plausible transitions between a hard state and a thermal dominant state \citep{feng_and_kaaret_2010}, all of which contribute to the picture of this object as an excellent IMBH candidate. ESO 243-49 HLX-1 is most remarkable for its high peak luminosity and regular outbursts, which appear to possess a fast rise/exponential decay profile and to repeat on timescales of $\sim 1$ year \citep{lasota_etal_2011}. This object also appears to mimic BHXRB behaviour in its hardness-intensity diagram as it transits through its outbursts, indicative of sub-Eddington state changes \citep{servillat_etal_2011}. 


It is evident that the most luminous ULXs are rare, and little is known about them as a class; yet we have summarised good arguments as to why these are potentially important targets in the ongoing search for IMBHs.  One way to learn more of this population is to extend the number of known objects using archival searches.  Here, we report a study of ten candidate extreme ULXs selected from the \citeauthor{walton_etal_2011b} (2011b) catalogue of ULXs detected in the 2XMM-DR1 survey, that display observed luminosities well in excess of the break in the XLF, including two new HLXs\footnote{This work builds and improves on the early results from this sample, published in \cite{sutton_etal_2011}.}.  

The paper is laid out as follows.  In section 2 we describe the selection of the new sample, and the reduction of the recent optical and X-ray data associated with them.  Section 3 details the imaging, spectral and timing results from all available \xmmn and \chan datasets for the sources, and describes an optical follow 
up study.  The characteristics of these objects are then compared to those of the less luminous bulk of the ULX population in section 4, before the implications of the findings are discussed with respect to the nature of the objects in section 5, and the paper is concluded in section 6.


\section{Sample selection \& data reduction}

\subsection{Sample selection}

\citeauthor{walton_etal_2011b} (2011b)  present a catalogue of 470 candidate ULXs in 238 galaxies, detected in archival 
\xmmn observations, produced from a cross-correlation of the RC3 catalogue of galaxies \citep{deVaucouleurs_etal_1991} and 
the 2XMM-DR1 catalogue \citep{watson_etal_2009}.  This was reduced to a small sample of the most luminous, relatively nearby ULXs by filtering on 
luminosity ($L_{\rm X} > 5 \times 10^{40}~{\rm erg~s^{-1}}$) and distance ($d < 100$ Mpc).  A sample of 12 sources were found to match these criteria.  Additional quality checks of the \xmmn data for the selected sources were then performed, and 2 of the 12 sources were excluded from the final sample.  We excluded a ULX candidate detected near to the cD galaxy NGC 4889, in the galaxy cluster Abell 1656, where we could not separate sufficient source counts from the strong diffuse emission in which the candidate ULX was embedded to permit detailed study.  The other excluded source was M82 X-1, as it has been extensively studied by other authors (\eg \citealt{matsumoto_etal_2001}; \citealt*{kaaret_etal_2006}; \citealt{feng_and_kaaret_2010}), 
and the \xmmn data is somewhat complicated by the presence of a strong ISM component around the position of the ULX, and bright and variable nearby objects\footnote{Of the other notable HLXs mentioned in the introduction, ESO 243-49 HLX-1 was not included in the sample as its host is not in the RC3 catalogue, and both Cartwheel N10 and CXO J122518.6+144545 lie beyond the 100 Mpc limit of this sample.}.  

The 10 remaining objects are listed in Table~\ref{srcs}, along with some characteristics of their presumed host galaxies.  The X-ray sources are assumed to be at the distance of the host galaxy, and for the majority of sources cosmology-corrected 
distances are used (${\rm H_0} = 71~{\rm km~s^{-1}~Mpc^{-1}}$, $\Omega_{\rm M} = 0.27$, $\Omega_{\Lambda} = 0.73$, 
\citealt{komatsu_etal_2009}), calculated using redshifts from \citet{deVaucouleurs_etal_1991}.  An exception to this is \srceight, whose host galaxy is sufficiently nearby that its peculiar motion could dominate its measured recession velocity, so we instead use the distance of 14.9 Mpc from \citet{tully_1988}.
Table~\ref{srcs} also provides a luminosity for each object, calculated using this distance and the 2XMM 0.2--12 keV 
flux.  It is immediately evident that the extreme luminosity sample presented here includes 3 candidate HLXs - \srcone, \srcfour~and 
\srcseven~(additionally, \srcthree~had previously been identified as a HLX by \citealt{davis_and_mushotzky_2004}, see section 
\ref{resolved}). 

\begin{table*}
\caption{The extremely luminous candidate ULX sample.}
\centering
\begin{tabular}{cccccccc}
\hline
Source ID & 2XMM source & Host galaxy & Galaxy type$^a$ & Distance$^b$ & Separation$^c$ & ${N_{\rm H}}^d$ & ${L_{\rm X, max}}^e$
\\
\hline
\srcone & 2XMM J011942.7+032421 & NGC 470 & SA(rs)b & 32.7 & 33 & 3.09 & $15.3 \pm 0.8$
\\
\srctwo & 2XMM J024025.6$-$082428 & NGC 1042 & SAB(rs)cd & 18.9 & 96& 2.61& $5.1 \pm 0.2$
\\
\srcthree & 2XMM J072647.9+854550 & NGC 2276 & SAB(rs)c & 33.3 & 45 & 5.52 & $6.4 \pm 0.3$
\\
\srcfour & 2XMM J120405.8+201345 & NGC 4065 & E & 87.9 & 18 & 2.40 & $13.1 \pm 1.7$
\\
\srcfive & 2XMM J121856.1+142419 & NGC 4254 & SA(s)c & 33.2 & 103 & 2.70 & $6.0 \pm 0.3$
\\
\srcsix & 2XMM J125939.8+275718 & NGC 4874 & cD0 & 99.8 & 57 & 0.89 & $5.8 \pm 1.3$
\\
\srcseven & 2XMM J134404.1$-$271410 & IC 4320 & S0? & 95.1 & 18 & 5.04 & $28.2 \pm 2.4$
\\
\srceight & 2XMM J151558.6+561810 & NGC 5907$^f$ & SA(s)c: edge-on & 14.9 & 102 & 1.38 & $5.6 \pm 0.1$
\\
\srcnine & 2XMM J163614.0+661410 & MCG 11-20-19 & Sa & 96.2 & 16 & 3.14 & $ 6.5 \pm 1.2$
\\
\srcten & 2XMM J230457.6+122028 & NGC 7479 & SB(s)c & 32.8 & 68 & 5.08 & $7.1 \pm 0.3$
\\
\hline
\end{tabular}
\begin{minipage}{\linewidth}

Notes: 
$^a$ Galaxy morphology from \citet{deVaucouleurs_etal_1991}.
$^b$ Host galaxy distances in Mpc.  These are cosmology-corrected distances using
redshifts from \citet{deVaucouleurs_etal_1991}, with the exception of NGC 5907 where
the distance from \citet{tully_1988} was used.  
$^c$ Separation of the candidate ULX position from the nucleus of its host galaxy, in units of arcseconds.
$^d$ Galactic column densities interpolated from \citet{dickey_and_lockman_1990}, in units of $10^{20}~{\rm cm^{-2}}$.  
$^e$ Maximum detected 0.2--12 keV luminosity of the ULX candidate, based on the 2XMM broad band fluxes and the quoted distance, in units of $10^{40} \ergsec$.
$^f$ The part of NGC 5907 west of its dust lane, that hosts the ULX candidate, is also known as NGC 5906.

\end{minipage}
\label{srcs}
\end{table*}

In addition to the data sets identified in the 2XMM catalogue, the \xmmn science archive\footnote{\tt http://xmm.esac.esa.int/xsa/} and \chan data archive\footnote{\tt http://cxc.harvard.edu/cda/} were searched for further observations covering the objects.  All the archived \xmmn EPIC and \chan ACIS 
observations potentially covering the positions of the 10 candidate ULXs were extracted from the archive, and examined for further detections of the 10 objects.  Details of all relevant observations are shown in
Table~\ref{obs}, with the exceptions of the \xmmn observations 0300530401 and 0301651201 (plus several additional short observations of \srcsix, see caption for details).  These were pointed in the proximity of \srcsix~and \srcten~respectively, but were not used as they contained no useful data due to the presence of heavy background flaring throughout the duration of the EPIC exposures.  In all cases, the ULX in question was detected.  All but one of the sources (\srcseven) are present in multiple archived observations, including 
six with archival \chan data.  We were awarded additional \chan time during cycle 12 to observe the other four 
sources (proposal ID 12620389, PI Sutton; results for \srctwo, \srcfour~and \srcseven~have been obtained and are 
included here, but an observation of \srceight~awaits scheduling at the time of writing).

\begin{table*}
\caption{Observation details}
\centering
\begin{tabular}{ccccccc}
\hline
Source ID & Observatory & ObsID$^a$ & Date & $t_{\rm exp}$$^b$ & Total counts$^c$ & Off-axis angle$^d$\\
& & & (yyyy/mm/dd) & (ks) & (0.3--10.0 keV) & (arcmin)\\
\hline
\srcone         & \xmmn & 0200780101 & 2004/01/24 & 4.0/3.8/3.6 & $850 \pm 30$ & 5.25\\
                & \chan & 10562 & 2008/11/12 & 28.1 & $610 \pm 20$ & 0.29\\
                & \xmmn & 0601670101 & 2009/06/27 & 40.1/53.9/47.1 & $2550 \pm 50$ & 1.10\\
\srctwo         & \xmmn & 0093630101 & 2001/08/15 & 11.6/15.0/~-~ & $920 \pm 30$ & 14.41\\
                & \xmmn & 0306230101 & 2006/01/12 & ~-~/53.1/52.1 & $2860 \pm 60$ & 12.18\\
                & \xmmn & 0553300401 & 2009/08/12 & 42.2/48.4/~-~ & $4310 \pm 70$ & 14.38\\
                & \chan & 12988 & 2010/10/09 & 29.0 & $2770 \pm 50$ & 0.32\\
\srcthree       & \xmmn & 0022340201 & 2001/03/16 & ~-~/52.3/52.3 & $2070 \pm 50$ & 6.08\\
\srcthree a$^e$ & \chan & 4968 & 2004/06/23 & 45.6 & $320 \pm 20$ & 1.13\\
\srcthree b$^e$ & \chan & 4968 & 2004/06/23 & 45.6 & $190 \pm 10$ & 1.10\\
\srcthree c$^e$ & \chan & 4968 & 2004/06/23 & 45.6 & $120 \pm 10$ & 1.06\\
\srcfour        & \xmmn & 0112270601 & 2003/01/02 & ~-~/6.5/6.5 & $75 \pm 9$ & 2.10\\
                & \xmmn & 0112271001 & 2003/06/09 & ~-~/6.9/6.9 & $140 \pm 10$ & 0.48\\
                & \xmmn & 0112271101 & 2003/06/30 & 3.1/3.6/3.6 & $210 \pm 10$ & 0.60\\
                & \chan & 12990 & 2011/07/13 & 5.0 & $57 \pm 8$ & 0.30 \\
\srcfive        & \xmmn & 0147610101 & 2003/06/29 & 13.1/17.3/19.4 & $1770 \pm 40$ & 2.75\\
                & \chan & 7863 & 2007/11/21 & 5.1 & $270 \pm 20$ & 1.84\\
\srcsix$^f$     & \chan & 556 & 1999/11/04 & 9.6 & $60 \pm 10$ & 2.28\\
                & \xmmn & 0124711401 & 2000/05/29 & 11.0/16.8/16.8 & $350 \pm 80$ & 0.42\\
                & \xmmn & 0153750101 & 2001/12/04 & 17.0/21.0/21.0 & $580 \pm 90$ & 2.69\\
                & \xmmn & 0300530701 & 2005/06/06 & 16.0/23.6/24.3 & $430 \pm 90$ & 2.14\\
                & \xmmn & 0300530601 & 2005/06/07 & 14.3/~-~/22.7 & $410 \pm 80$ & 2.21\\
                & \xmmn & 0300530501 & 2005/06/08 & 15.5/24.8/24.9 & $480 \pm 90$ & 2.31\\
                & \xmmn & 0300530301 & 2005/06/11 & ~-~/30.4/30.4 & $230 \pm 70$ & 1.60\\
                & \xmmn & 0300530101 & 2005/06/18 & 15.7/18.0/18.0 & $480 \pm 90$ & 1.42\\
                & \chan & 9714 & 2008/03/20 & 29.7 & $40 \pm 10$ & 1.86\\
                & \chan & 10672 & 2009/03/15 & 28.5 & $110 \pm 10$ & 1.93\\
\srcseven       & \xmmn & 0092970801 & 2003/01/04 & 3.5/5.3/5.2 & $390 \pm 20$ & 1.10\\
                & \chan & 12989 & 2011/03/21 & 49.4 & $1730 \pm 40$ & 0.31\\
\srceight       & \xmmn & 0145190201 & 2003/02/20 & 10.6/24.5/25.2 & $7640 \pm 90$ & 2.40\\
                & \xmmn & 0145190101 & 2003/02/28 & 9.4/18.8/19.1 & $5460 \pm 80$ & 2.41\\
\srcnine        & \chan & 553 & 1999/10/19 & 5.9 & $14 \pm 4$ & 4.83\\
                & \chan & 1454 & 1999/10/19 & 11.4 & $29 \pm 5$ & 4.87\\
                & \chan & 1666 & 2001/08/30 & 48.6 & $120 \pm 10$ & 2.29\\
                & \xmmn & 0112980101 & 2002/09/28 & 13.1/14.7/14.7 & $130 \pm 20$ & 4.11\\
                & \xmmn & 0112980401 & 2002/09/30 & 9.0/10.4/10.4 & $70 \pm 10$ & 4.15\\
                & \xmmn & 0112980501 & 2002/10/04 & 7.5/8.4/8.4 & $60 \pm 10$ & 4.12\\
                & \chan & 7698 & 2007/06/13 & 5.1 & $8 \pm 3$ & 4.07\\
\srcten         & \xmmn & 0025541001 & 2001/06/19 & 6.0/9.1/9.2 & $1270 \pm 40$ & 0.55\\
                & \chan & 11230 & 2009/08/11 & 24.8 & $160 \pm 10$ & 2.30\\
                & \chan & 10120 & 2009/10/24 & 10.0 & $56 \pm 8$ & 1.21\\
\hline
\end{tabular}
\begin{minipage}{\linewidth}

Notes: 
$^a$ Observation identifiers.
$^b$ The sum of the good time intervals for each detector.  For \xmmn observations it is shown in the order PN/MOS1/MOS2, a dash indicates that the source was not detected.
$^c$ Total background-subtracted source counts accumulated in the observation (sum of all 3 EPIC detectors for \xmmn observations).
$^d$ Angular separation between the on-axis position of the observation and the 2XMM
source position.
$^e$ The \chan observation of \srcthree~was able to resolve the candidate HLX 
into multiple sources (Figure \ref{NGC2276}).
$^f$ NGC 4874 is the central galaxy in the Coma cluster, and the Coma cluster has been extensively observed by both \chan and \xmmn (in many cases for calibration purposes). There are a number of similar observations from both observatories that are not 
included in this analysis, due to the short exposure time ($\la 10$ ks), but are 
available in the archives.

\end{minipage}
\label{obs}
\end{table*}

\subsection{{\bfseries{\em XMM-Newton~\/}}data reduction}

The \xmmn data were reduced and data products were extracted using the standard tools available in the \xmmn {\sc sas}\footnote{\tt http://xmm.esac.esa.int/sas/}
(version 9.0.0).  Severe background flaring resulted in multiple 
exposures being obtained in a number of the observations\footnote{This affected both MOS and PN
detectors during observation 0145190201 of \srceight, and MOS 
detectors alone were affected during observations 0200780101, 0147610101 and 
0145190101 of \srcone, \srcfive~and \srceight~respectively.}.  However, in all cases only one of the exposures contained a significant length of good time.  In each case we identified the sole exposure containing useful science data per observation and focused our analysis upon it.  
In two 
observations of \srcsix~the solar proton flaring was severe enough to render the data from whole instruments unusable, in particular the PN detector in observation 0300530301 and the MOS1 detector in 0300530601.  
However, background flaring was present to some significant level in almost all the \xmmn observations, with the only exceptions being observations 0300530301 and 0112980101
 of the ULX candidates \srcsix~and \srcnine~respectively.  Good Time Interval (GTI) files were therefore created to allow high background epochs to be filtered out of the data, based on high energy (10--15 keV) full field light curves.  The limiting count rate used to define the GTIs varied between observations, dependent on the best balance between minimising the source data
loss whilst maximising the excision of background flares, with typical thresholds of  $1.0$--$1.5~\rm{cts}~\rm{s}^{-1}$ for PN fields
and $\sim 0.6~\rm{cts}~\rm{s}^{-1}$ for MOS fields.

Source spectra and light curves were then extracted from circular
apertures centred on the ULX candidate. Source extraction 
region radii were typically between $\sim 15$--$40$ arcseconds 
(which corresponds to an on axis fractional encircled energy of $\sim$ 70--90 per cent at 1.5 keV), in order 
to maximise source counts whilst avoiding contamination from nearby sources.
In three cases the source was positioned close to a chip gap, which would 
have severely restricted the radius of a circular aperture.  In each of these cases
elliptical source regions (with semi-major axes equivalent to the circular aperture radii) were used to extract the source data.  In both the MOS1 data from \srcthree~in the 0022340201 dataset, and the PN data for \srcfive~in observation 0147610101, the ellipse was aligned with the nearby chip gap; however in the PN data for \srctwo~in 0093630101, the source was sufficiently far off-axis that the ellipse was aligned with the direction of maximum point spread function (PSF) extent.
Larger circular apertures were used to provide source background 
data in all cases.  These were positioned on the same chip, approximately the same distance from the readout nodes if possible, and were sized to include a minimum of 
$\sim 50$ per cent more area on the detector than source apertures.
In cases where nearby sources placed limitations on the 
size of a background region, multiple circular apertures were used.

The final data products used in the analysis (\ie spectra and light curves) were then extracted from the source apertures, using the GTI file and standard event patterns for the detectors ({\tt PATTERN} $\le 4$ for the PN, {\tt PATTERN} $\le 12$ for the MOS).  Spectral extractions also used the conservative {\tt FLAG} $= 0$ filter, although we alleviated this somewhat for the light curve extractions by using the less restrictive {\tt \#xmmea\_ep} or {\tt \#xmmea\_em} filters for the PN or MOS detectors respectively instead.  The appropriate response matrices for the spectral analysis were automatically generated during the spectral extraction in the {\sc sas} task {\sc xmmselect}.

\subsection{{\bfseries {\em Chandra~\/}}data reduction}

\chan data were reduced using {\sc ciao}\footnote{\tt http://cxc.harvard.edu/ciao/} version 4.1.1, and calibration files from
the \chan {\sc caldb} version 4.1.0.  Circular apertures with radii of 4 arcseconds 
(corresponding to an on axis fractional encircled energy of $\ga 99$ per cent at 1 keV) 
were used to extract the source data, with the only exception being in the case of \srcthree, as this ULX candidate was 
resolved by \chan into multiple objects separated by less than
the typical extraction region radius (see section~\ref{resolved}).  Instead, apertures of $\sim 1$ arcsecond radius 
($\sim$ 90 per cent on axis fractional encircled energy at 1 keV)
were used to avoid 
overlapping the source regions. Annular 
apertures, centred on the source, but with inner radii outside of the 
source region were used to characterise the background for each source.
In the case of \srcthree, a single background
annulus surrounding the resolved sources
was used to avoid cross-contamination of the background regions by the neighbouring point sources.  The data products were then extracted from the 
archived 
level 2 event files, to include all standard good event detections with energies between
0.3--10 keV, within the source data extraction apertures.  The \chan spectra were extracted using the {\sc ciao} script {\sc specextract}, which also produces the appropriate response matrices for spectral analysis, and the light curves were extracted using {\sc dmextract}.

\subsection{Optical data}

This work also utilises optical data to 
further probe these extreme objects.  As an initial step we retrieved Digitised Sky Survey version 2 (hereafter DSS2) data from the ESO archive\footnote{\tt http://archive.eso.org/dss/dss}, within a 5 arcminute square centred on the host galaxy of each ULX candidate.  DSS2-blue images were used preferentially when available, but due to its limited coverage (only $45$ per cent of the sky), the use of the alternative all-sky DSS2-red images was necessary for  \srctwo~and \srcseven.

An examination of the Hubble Legacy Archive (HLA)\footnote{\tt http://hla.stsci.edu/} revealed 
that high resolution space-based optical imaging from \hst was available for four of our objects.  Enhanced {\it HLA} data products covering the source region were 
available for \srcthree, \srcsix~and \srceight, and two data sets processed using the standard \hst 
single image calibration pipeline were available
for \srcten.  All this data was downloaded from the archive for further study; we tabulate the basic observation details in Table~\ref{hst_bands}.  As an initial step the astrometry of the images was improved using the {\sc iraf} tools {\sc ccfind}, {\sc ccmap} and {\sc ccsetwcs} and objects common to the \hst images and either the 2MASS (\srceight) or USNO (\srcthree, \srcsix~and \srcten) catalogues.  Where individual counterparts to the X-ray sources were detected, aperture photometry was performed using {\sc gaia} and the appropriate Vega magnitude zero points from either \citet{sirianni_etal_2005}
for ACS/WFC observations, 
or the WFPC2 Data Handbook\footnote{\tt http://www.stsci.edu/hst/wfpc2/documents/handbook/\\WFPC2\_DHB.html}.  Corrections to a standard 0.5 arcsecond aperture were calculated using a 
mean magnitude correction from ten bright point sources in each field, and these were extrapolated to an infinite radius aperture, again using values from \citet{sirianni_etal_2005} and the WFPC2 Data Handbook.

\begin{table}
\caption{Details of archival \hst observations, and Galactic extinction in the direction of the targets}
\begin{center}
\begin{tabular}{lcccc}
\hline
Source & $E(B-V)$ & Prop. ID & Instrument & Filter\\
\hline
\srcthree & 0.1 & 6355 & WFPC2 & F300W \\
	& & 8597 & WFPC2 & F606W \\
	& & 10877 & WFPC2 & F555W\\
	& & & & F814W\\
	\\
\srcsix & 0.009 & 10861 & ACS/WFC & F475W\\
	& & & & F814W\\
	\\
\srceight & 0.011 & 6092 & WFPC2 & F450W\\
	& & & & F814W\\
	\\
\srcten & 0.112 & 11575 & ACS/WFC & F555W\\
	& & & & F814W\\
\hline
\end{tabular}
\end{center}
\label{hst_bands}
\end{table}

In addition to the study of archival imaging data for the sources, a low dispersion Gemini GMOS spectrum was obtained to further constrain the nature of an optical counterpart to \srcsix. The 
data were reduced using standard tools in the {\sc gmos} sub-package from the {\sc gemini} package in {\sc iraf}.  
The specific tools used were: {\sc gsflat} to extract the flat field; {\sc gsreduce} to reduce both the target and calibration arc exposures; {\sc gswavelength} to establish the wavelength calibration, then {\sc gstransform} to apply it to the exposures; {\sc gsskysub} to subtract the sky background; {\sc gscrrej} to reject cosmic ray events; {\sc gsextract} to extract one dimensional spectra of the source, which were then combined using the tool {\sc gemarith}.
Gaussian line profiles were fitted where appropriate to the reduced optical spectrum of the counterpart using the {\sc curvefit} function in {\sc IDL} version 6.2.

\section{Data analysis \& results}

\subsection{Spatial analyses}

\subsubsection{High resolution X-ray imaging}
\label{resolved}

\begin{figure}
\centering
\includegraphics[width=5.cm, angle=0]{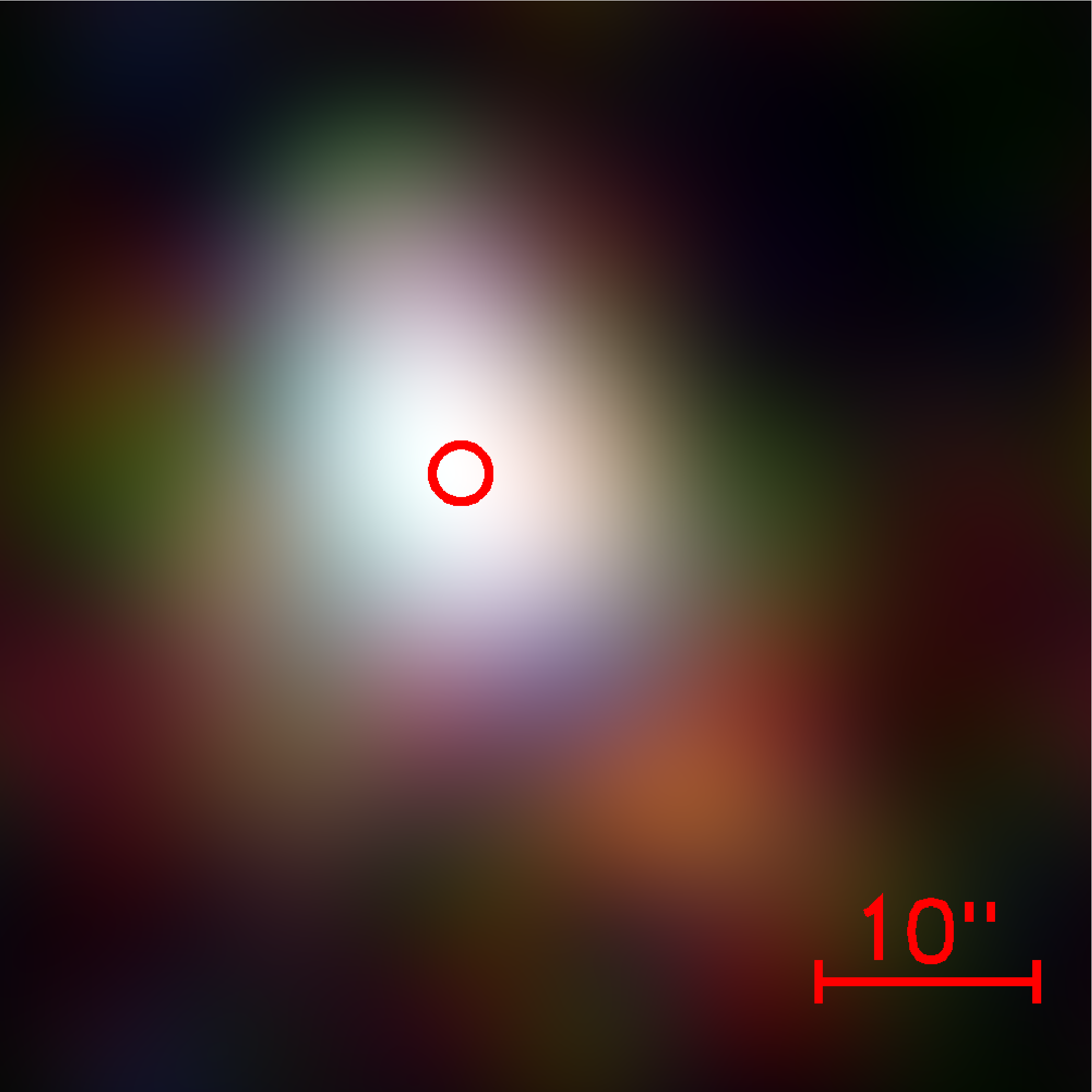}\vspace*{0.4cm}
\includegraphics[width=5.cm, angle=0]{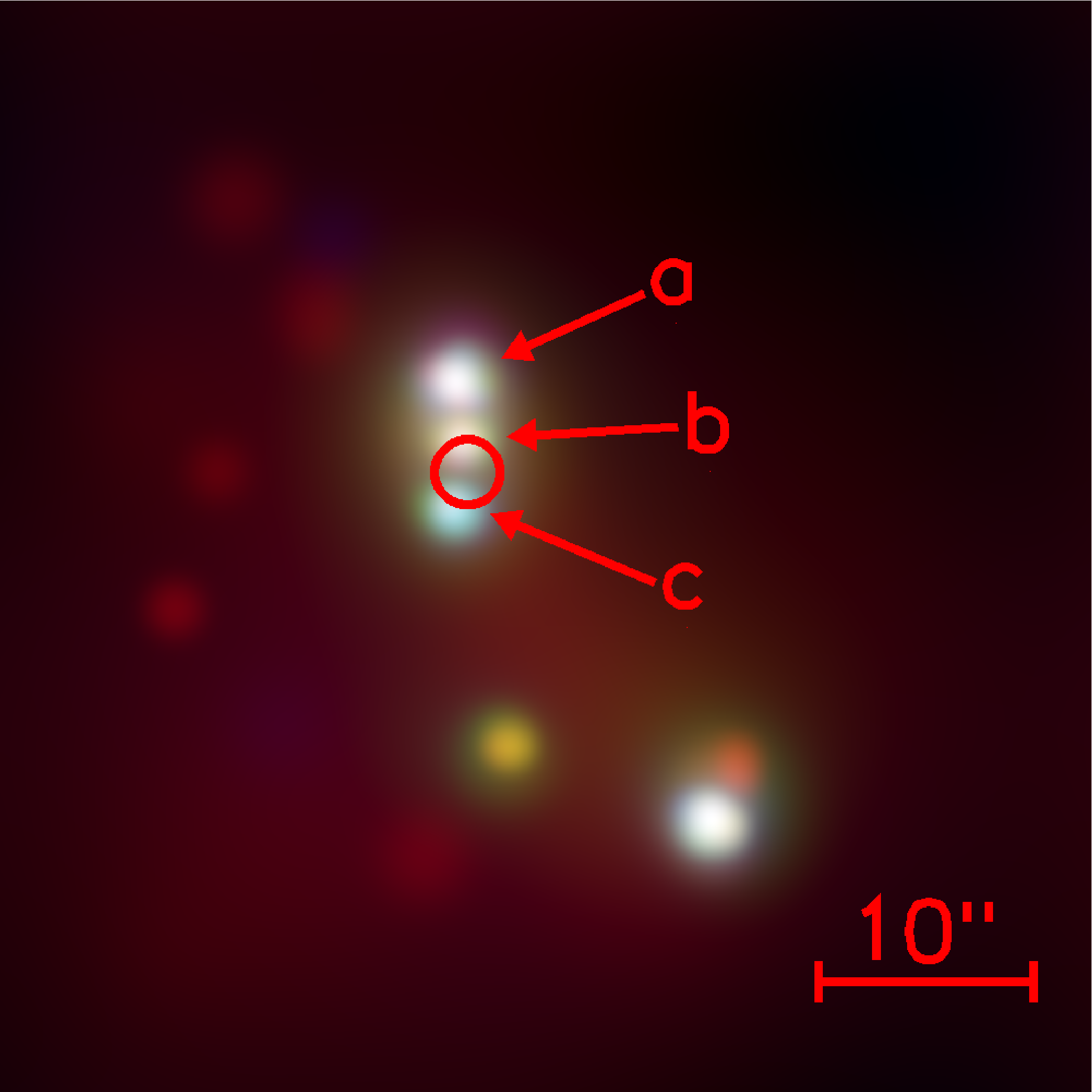}
\caption{({\it top}) True colour \xmmn image of \srcthree, where the rgb colours correspond 
to the 0.3--1.5, 1.5--2.5 \&~2.5--10 keV energy bands respectively. The region circled is the 
effective 90 per cent uncertainty in the \xmmn position of the source.
({\it bottom}) Equivalent \chan image.  The \xmmn detection is resolved into three separate objects, two of which border on the
90 per cent \xmmn error region.  They are referred to as labelled here throughout the rest of the paper.  These and all subsequent images are aligned such that north is up.}
\label{NGC2276}
\end{figure}

As an initial step, we investigated the spatial characteristics of the X-ray source detections.  One particular issue for the \xmmn detections of these objects, especially given their relatively large distances, is whether we are seeing a single point-like object, or whether multiple objects and/or spatially extended components are present within the \xmmn beam.  \chan observations offer an excellent means of settling this issue, with the much smaller \chan beam able to resolve objects that are confused by {\it XMM-Newton\/}, even in regions of high diffuse emission (\eg in the centre of the starburst galaxies NGC 3256 \& NGC 3310, \citealt{jenkins_etal_2004}).  We therefore began by examining the deepest \chan detection for each object in our sample, in order to search for resolved structure.  

Only one \xmmn detection was resolved to reveal finer structure by {\it Chandra\/}, namely \srcthree.  This object had previously been identified as a HLX candidate in the spiral galaxy
NGC 2276 by \citet{davis_and_mushotzky_2004} using the \xmmn data (Figure \ref{NGC2276}, {\it top}). 
However, their assumed distance for NGC 2276 of 45.7 Mpc is somewhat discrepant compared to more recent measurements and overestimates the distance to NGC 2276 (\cf Table \ref{srcs}).  This in turn inflates the measured luminosity.  The assumptions in this paper, and the redshift from \citet{deVaucouleurs_etal_1991}, lead to a distance of 33.3 Mpc and hence a lower (although still extreme) 2XMM luminosity estimate of $\sim 6 \times 10^{40} \ergsec$.   In the later \chan observation of this source, the ULX
candidate was shown to be resolved into three point sources 
(Figure \ref{NGC2276}, {\it bottom}; see also \citealt{wolter_etal_2011}), although only the southern two lie coincident with the measured \xmmn source centroid position.  Remarkably, each of the point sources is
sufficiently luminous to be classified as a ULX, if at the distance
of NGC 2276, with $L_{\rm X} \ga 5 \times 10^{39} \ergsec$.  To find 3 such luminous ULXs within a small region of a galaxy -- the projected separation of the 3 objects is $< 150$ pc -- appears very remarkable, although this should be tempered by the realisation that the western regions of NGC 2276 have heightened star formation, with a galaxy-wide star formation rate estimated at $5 - 15~\Msun~{\rm yr^{-1}}$ (\citealt{wolter_etal_2011} and references therein).  This is plausibly the result of gravitational interaction with NGC 2300 and/or ram pressure interaction as NGC 2276 transits through  the bright intra-group medium of the NGC 2300 group (\eg \citealt{gruendl_etal_1993}).

We then used an additional diagnostic test for extension on the other, apparently point-like objects.  The {\sc ciao srcextent} script quantifies the size of each PSF, which it does by effectively calculating an average radius for the PSF using a Mexican Hat optimisation algorithm.  We show the results of this in Table~\ref{pointsource}.  These results are consistent with expectations for the \chan PSF for both on-axis and slightly off-axis objects, where an extent of $\sim 0.5$ arcseconds is equivalent to the 75 per cent encircled energy radius for 1.5 keV photons (see Fig. 4.6 of the \chan Proposer's Observatory Guide; also the \chan High Resolution Mirror Array calibration documentation\footnote{\tt http://cxc.harvard.edu/proposer/POG, http://cxc.harvard.edu/cal/Hrma/Index.html}).  We have verified the off-axis extents are consistent with point-like objects by comparison to the population of point-like ULXs in the galaxy NGC 4490 (see \citealt{roberts_etal_2002}; \citealt{fridriksson_etal_2008}; \citealt{gladstone_and_roberts_2009}; \citealt{yoshida_etal_2010}), 
eight objects that lie up to 3 arcminutes off-axis in three separate \chan ACIS-S observations.  Our extents are consistent with the {\sc srcextent}-measured extents for these ULXs over the full range of off-axis angles seen.  The only outlier is \srcnine, but even this is consistent at the 90 per cent confidence level.  We conclude that there is no evidence for significant extended emission for any object, meaning that each ULX candidate remains consistent with a single, point-like object at the high X-ray spatial resolution of {\it Chandra\/}.

\begin{table}
\caption{\chan source characteristics}
\begin{center}
\begin{tabular}{lcccc}
\hline
Source & Obs ID & RA, dec	& $\theta ^a$ & Extent\\
 & & (J2000) & (arcmin) & (arcsec)\\
\hline
\srcone & 10562 & $01 19 42.8 +03 24 22$ & 0.29 & $0.48 \pm 0.03$\\
\srctwo & 12988 & $02 40 25.6 -08 24 30$ & 0.32 & $0.49^{+0.02}_{-0.01}$\\
\srcthree a & 4968 & $07 26 48.1 +85 45 54$ & 1.13 & $0.49 \pm 0.04$\\
\srcthree b & 4968 & $07 26 47.9 +85 45 52$ & 1.10 & -$^b$ \\
\srcthree c	 & 4968 & $07 26 48.3 +85 45 49$ & 1.06 & $0.45^{+0.07}_{-0.06}$\\
\srcfour	& 12990 & $120405.8+201345$ & 0.30 & $0.50 \pm 0.1$ \\
\srcfive & 7863 & $12 18 56.1 +14 24 19$ & 1.84 & $0.55 \pm 0.05$\\
\srcsix	&10672 & $12 59 39.8 +27 57 16$ & 1.93 & $0.62^{+0.1}_{-0.09}$\\
\srcseven & 12989 & $13 44 04.2 -27 14 11$ & 0.31 & $0.50 \pm 0.02$\\
\srceight$^c$	& - & - & - & - \\
\srcnine & 1666 & $16 36 14.1 +66 14 10$ & 2.29 & $0.7 \pm 0.1$\\
\srcten & 11230 & $23 04 57.7 +12 20 29$ & 2.30 & $0.54 \pm 0.07$\\
\hline
\end{tabular}
\end{center}
\begin{minipage}{\linewidth}
Notes: Source characteristics are measured from the \chan observation with the greatest number of counts accumulated from the ULX candidate in question.
$^a$ Off-axis angle of source.
$^b$ {\sc srcextent} was unable to constrain the size of this point-like object due to the proximity of the brighter \srcthree a.  However, inspection shows its PSF appears very similar to its two verified point-like neighbours.
$^c$ \srceight~has not previously been observed with {\it Chandra}. However, a \chan observation of the source is scheduled.
\end{minipage}
\label{pointsource}
\end{table}

\subsubsection{Source locations and optical counterparts}

\begin{figure*}
\centering
\includegraphics[width=15.cm, angle=0]{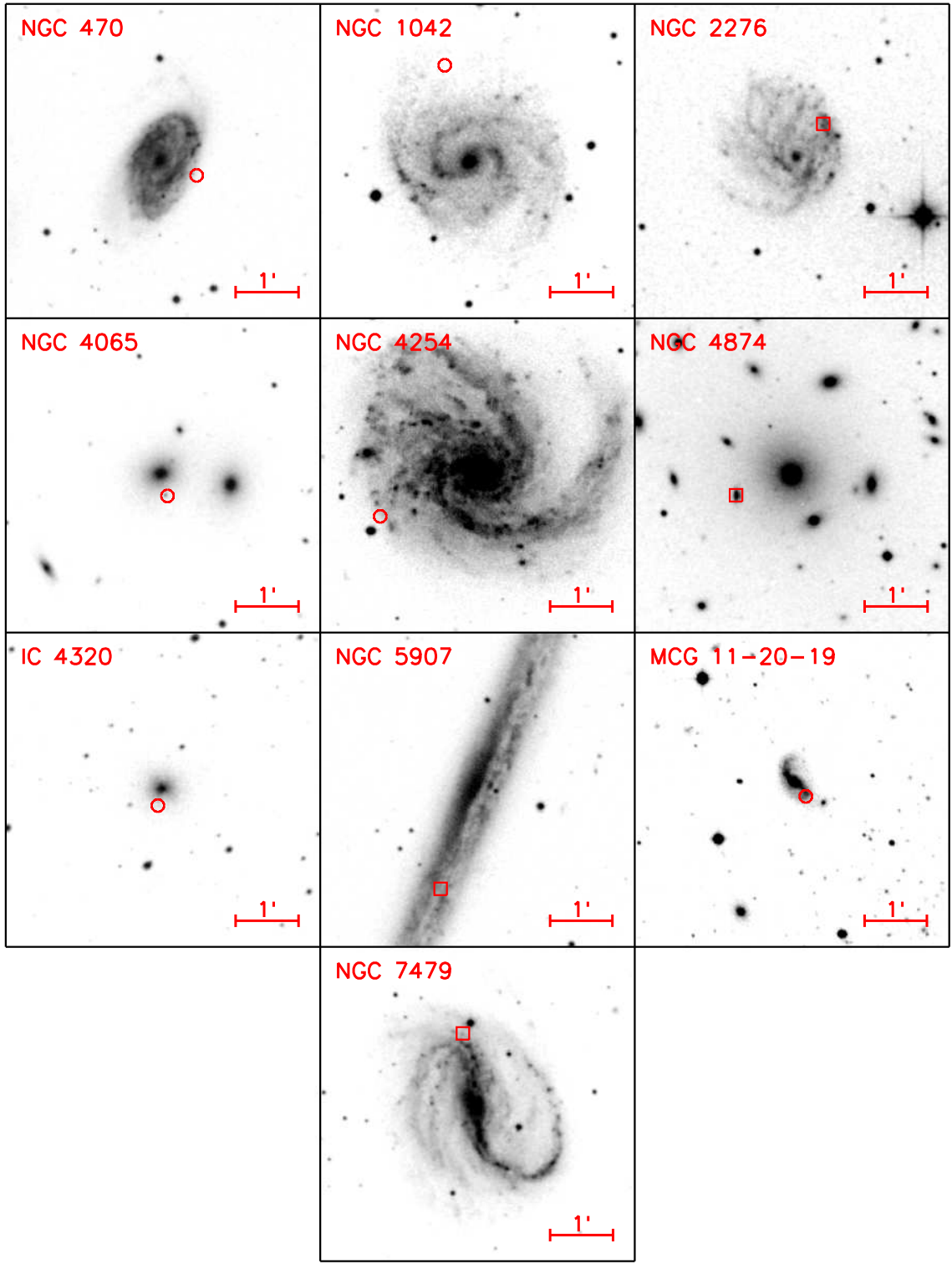}\hspace*{5mm}
\caption{DSS2 images of the habitats of the 10 objects in the selected sample.  The ordering of panels matches the numbering of the sources, starting at top left with \srcone~and reading across each row.  Data from the DSS2-blue survey is shown where available, but in the two cases this was not possible (\srctwo~\& \srcseven). The DSS2-red filter data is shown instead. Each image is 5 arcminutes across, aligned such that North is up, and centred on the nucleus of the host galaxy.  The greyscale is arbitrary in each panel, and displayed such that the internal structure of each galaxy can be seen.  The position of each object is marked by a circle, except where \hst data is available where instead a box is shown matching the size of the image in Fig.~\ref{HSTimgs}.  Note that the actual position errors are considerably smaller than the size of the circle shown.}
\label{DSSimgs}
\end{figure*}

A second benefit of obtaining \chan data for the objects in our sample is the accuracy of its astrometry, with sub-arcsecond positioning expected even without applying astrometric corrections.  We were therefore able to derive accurate positions (shown in Table~\ref{pointsource}) and so examine multi-wavelength data in order to search for counterparts, and identify the environments hosting these objects.  Initially we investigated the locations of the objects in our sample by overlaying their positions on DSS2 images, which we show here as Fig.~\ref{DSSimgs}.  Seven objects are associated with spiral galaxies; in the six cases where the galaxy is close to face-on, the ULX candidate appears to be located on or close to the spiral arms, in most cases in the outer regions of the disc.  In the one edge-on case (NGC 5907, the galaxy containing \srceight) the source is located in the plane of the disc, with a projected location coincident with a strong dust lane.  We also searched for catalogued objects near to the positions of our objects using the NED and SIMBAD databases\footnote{{\tt http://ned.ipac.caltech.edu}, {\tt http://simbad.u-strasbg.fr}}.  In the case of the spiral-hosted objects, three were in close proximity to catalogued H{\small II} regions: \srcone~is $\sim 2$ arcseconds from H{\small II} region 46 of \citet{hodge_and_kennicutt_1983}; \srcthree~is within $\sim 6$ arcseconds of regions 61 \& 65 of \citet{hodge_and_kennicutt_1983}; and \srcten~is within $\sim 2$ arcseconds of regions 197 \& 898 of \citet{rozas_etal_1999}.  
These associations imply a strong possibility that these objects are associated with ongoing star formation, as appears the case for many less luminous ULXs (\eg \citealt*{swartz_etal_2009}).

The three objects associated with elliptical galaxies present a more diverse subsample.  As Fig.~\ref{DSSimgs} shows, the object near to NGC 4065 (\srcfour) appears to have a faint, point-like counterpart.  Subsequent to the selection of the sample this object was revealed to be a background QSO by SDSS spectroscopy (object SDSS J120405.83+201345.0).  We therefore do not consider it further in the main body of this paper, although we briefly report its properties in Appendix A.  The second object associated with an elliptical, \srcsix, is actually found to be in the close proximity of a satellite galaxy (SDSS J125939.65+275714.0) of the central dominant galaxy of the Coma cluster, although it is displaced from the satellite galaxy nucleus by $\sim 4$ arcseconds.  The final object, \srcseven, is the most X-ray luminous object in the sample, although also amongst the most distant.  Its host is a peculiar S0 object that appears to be crossed by a dust lane \citep{lauberts_1982}.  
However, there is no obvious counterpart to this object in the DSS image.

\begin{figure*}
\centering
\includegraphics[width=6.cm, angle=0]{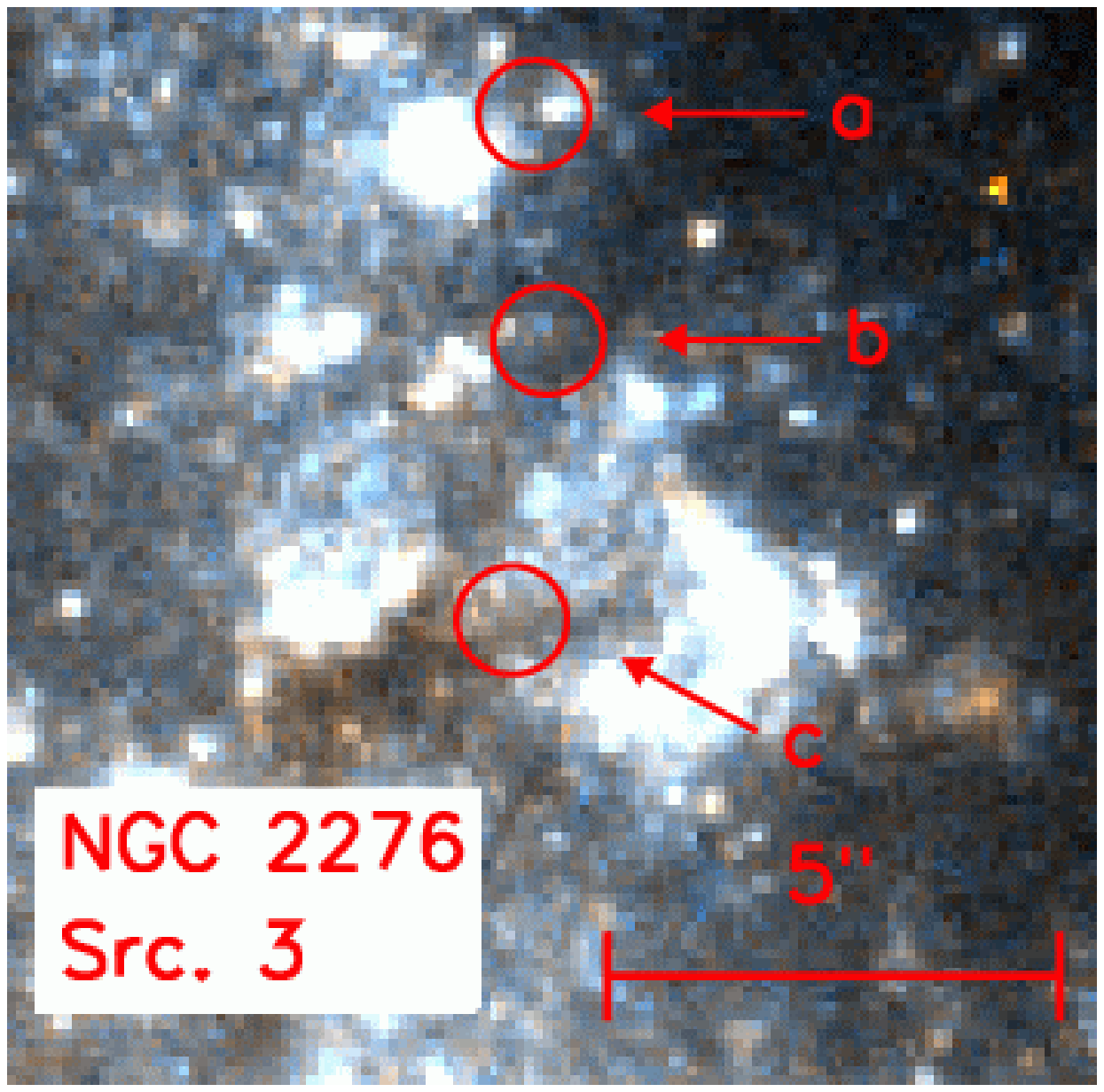}\hspace*{5mm}
\includegraphics[width=6.cm, angle=0]{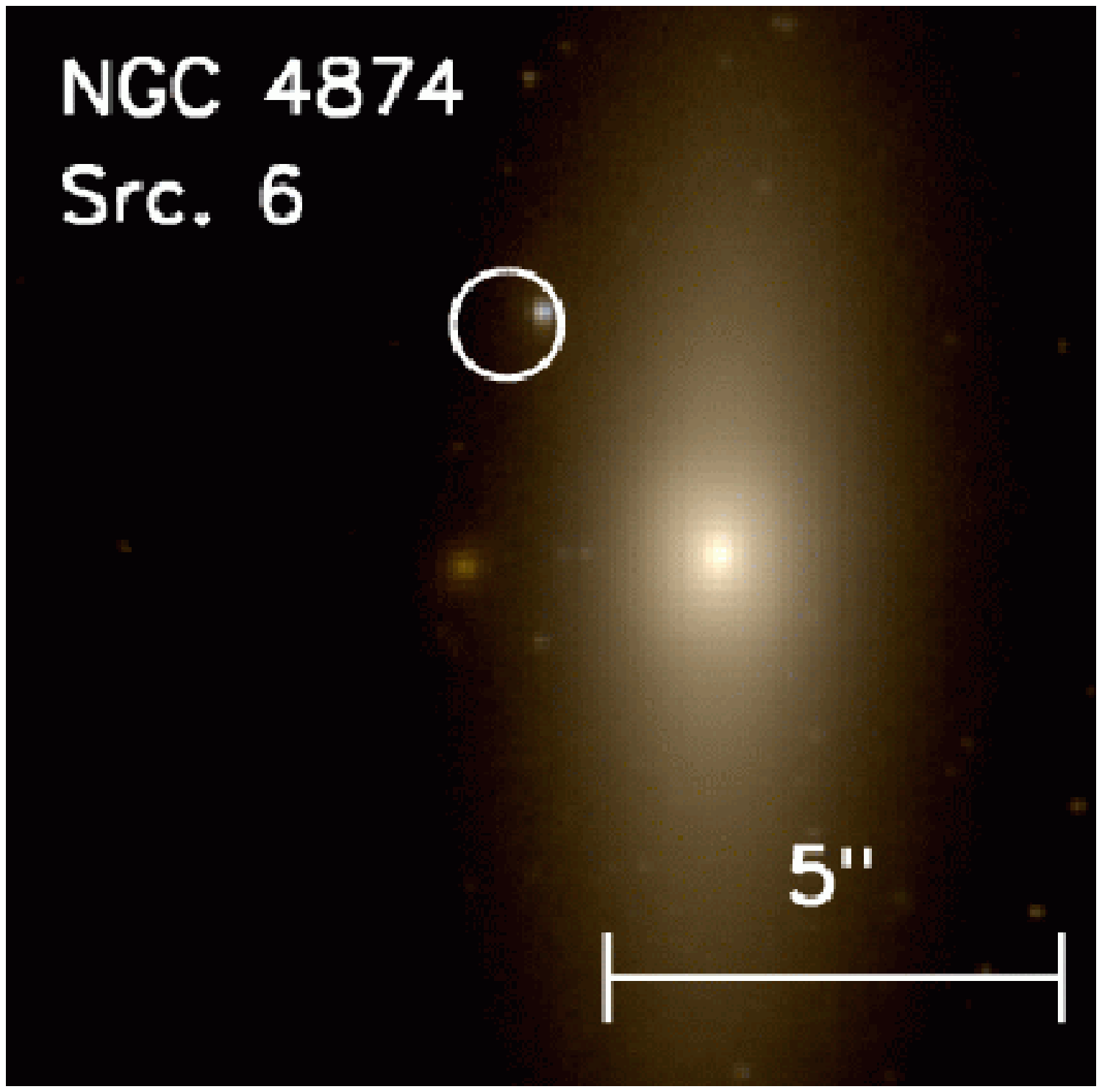}\\\vspace*{5mm}
\includegraphics[width=6.cm, angle=0]{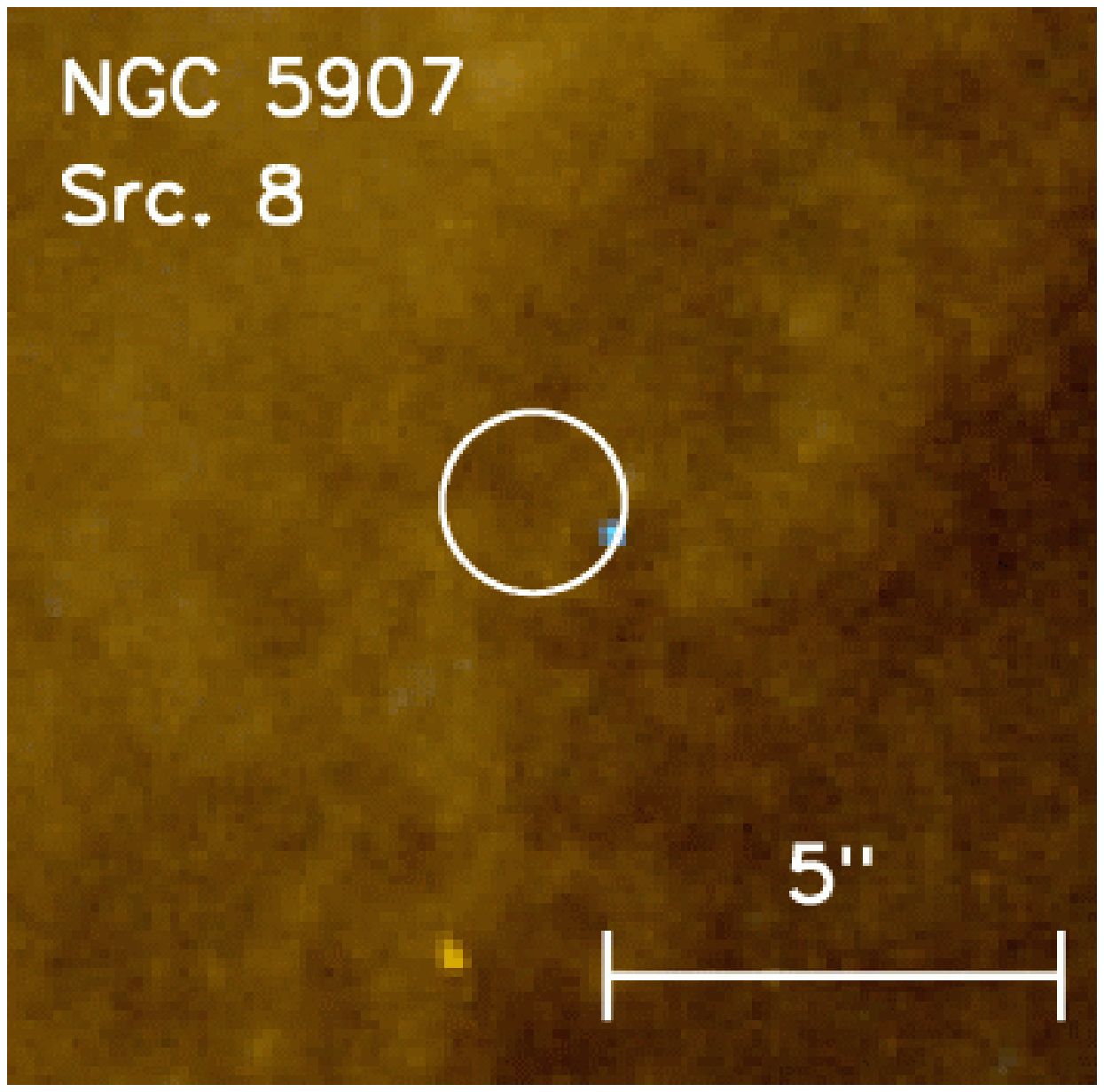}\hspace*{5mm}
\includegraphics[width=6.cm, angle=0]{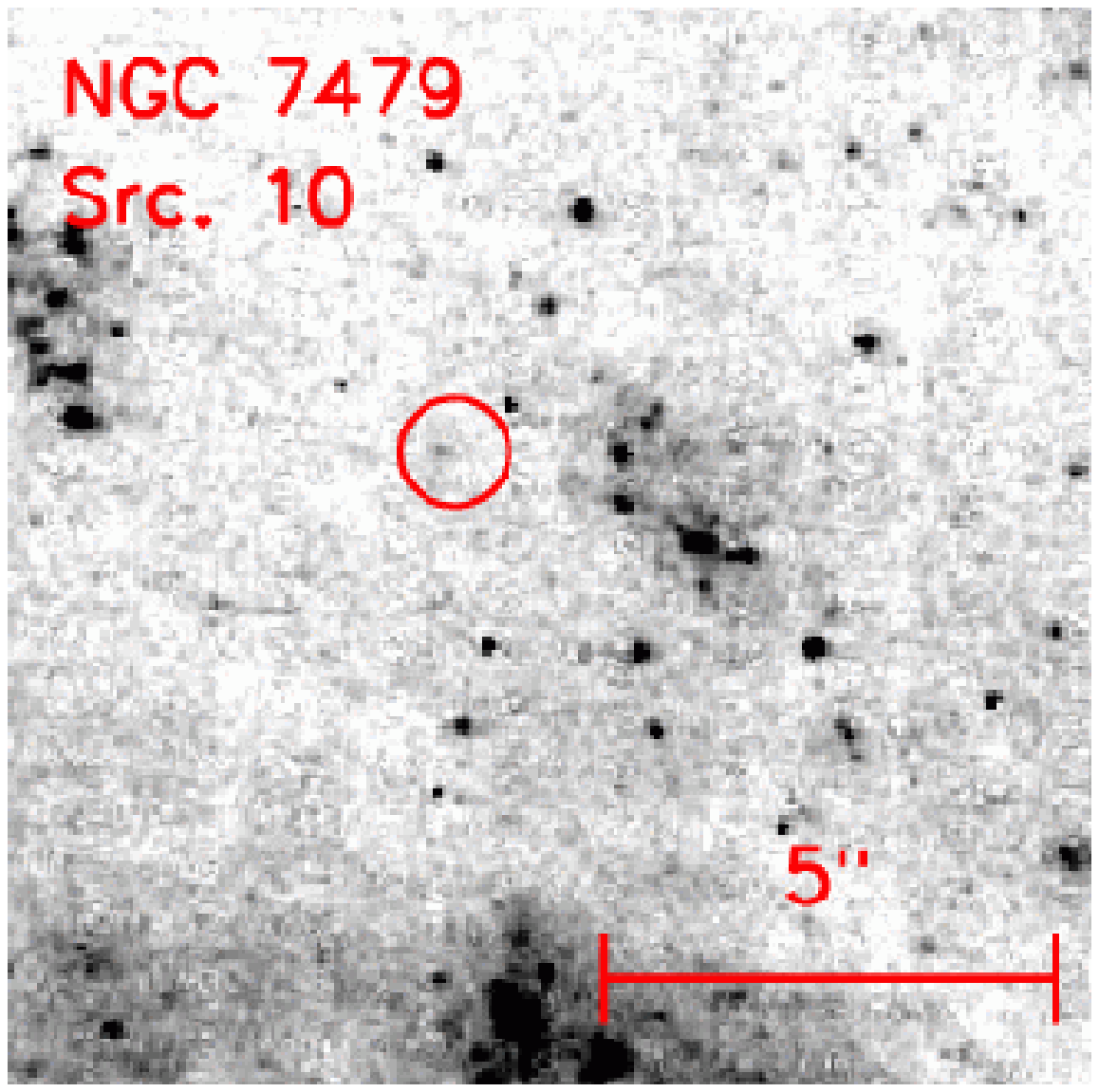}
\caption{\hst images of the regions containing four of the candidate ULXs.  {\it Top left} -- \srcthree a, b \& c.  This \hst data is displayed as a 3-colour image, with the F814W filter providing the red data, F555W the blue, and green taken as an average of the two.  The data is displayed on a linear scale to aid the visualisation of the counterparts.  {\it Top right} -- \srcsix.  The image is 3-colour, with the F814W data in red, F475W in blue, and an average of the two filters in green.  The image is displayed logarithmically to permit the separation of the counterpart from the galaxy to be seen.  {\it Bottom left} -- \srceight.  A 3-colour image using F814W as red, F450W as blue, and an average of the two as green.  This is scaled linearly for display purposes.  {\it Bottom right} -- \srcten.  A single, linearly greyscaled image in the F555W band is presented to emphasize the faint counterpart.}
\label{HSTimgs}
\end{figure*}

\begin{table}
\caption{\hst photometry for candidate counterparts}
\centering
\begin{tabular}{lccc}
\hline
Source ID	& Filter	& $m_f$ $^a$ & $M_f$ $^b$\\
\hline
\srcthree a & F300W & $22 \pm 4$ & $-11.1$\\
		& F555W & $23.1 \pm 0.9$ & $-9.8$\\
		& F606W & $22.9 \pm 0.8$ & $-10.0$ \\
		& F814W & $21.8 \pm 0.8$ & $-11.0$\\
	& \\
\srcthree b & F300W & - & - \\
		& F555W & $25 \pm 2$ & $-7.9$\\
		& F606W & $25 \pm 3$ & $-7.9$ \\
		& F814W & $23 \pm 5$ & $-9.8$\\
	& \\
\srcthree c & F300W & - & - \\
		& F555W & - & - \\
		& F606W & $24 \pm 3$ & $-8.9$ \\
		& F814W & $24 \pm 5$ & $-8.8$\\
	& \\
\srcsix	& F475W	& $22.0 \pm 0.2$ & $-13.0$ \\
		& F814W	& $21.2 \pm 0.2$ & $-13.8$ \\
	& \\		
\srceight	& F450W 	& $21.5 \pm 0.4$ & $-9.4$ \\
		& F814W	& - & - \\
	& \\
\srcten	& F555W	& $26 \pm 2$ & $-6.9$ \\
		& F814W	& $26 \pm 5$ & $-6.8$ \\
\hline
\end{tabular}
\begin{minipage}{\linewidth}
Notes: 
$^a$ Apparent magnitudes, for the filter$f$, in the Vega photometric system. 
$^b$ Absolute magnitudes, corrected for foreground extinction as per Table~\ref{hst_bands}.
\end{minipage}
\label{HSTphot}
\end{table}

A minority of the objects in the sample are also covered by \hst imaging observations, as listed in Table~\ref{hst_bands}.  We show these four datasets in Fig~\ref{HSTimgs}, where we display the immediate region covering the candidate ULX position at the full \hst spatial resolution.  We overlay an X-ray position error circle on top of the image to indicate the likely position of each ULX candidate.  This is calculated as the combination of position errors from the astrometric uncertainty of the \chan data, the X-ray source centroid, and the astrometric registration of the \hst data, as per \citet*{roberts_etal_2008}.  
Possible point-like candidate counterparts are apparent within the X-ray position error circle for all of the objects.  Aperture photometry was performed on the brightest identified counterpart in each case (which in most cases is the only counterpart, although there are possibly fainter objects within the crowded regions in NGC 2276), as described in Section 2.4, and we list the derived magnitudes of the counterparts in Table~\ref{HSTphot}.

The brightest individual objects within the error circles for the ULXs in NGC 2276 (\srcthree a--c)  display absolute magnitudes in the ranges
$M_{F606W} \sim -8 \rightarrow -10$ and $M_{F814W} \sim -9 \rightarrow -11$, albeit with large uncertainties.  Archival \hst F300W and F555W observations covering the field of \srcthree a--c are also available, but the magnitude of the brightest optical source within the error circle of \srcthree c is not constrained by the F555W observation, and similarly for both \srcthree b and c in the F300W observation.  Such magnitudes are more luminous than typical individual supergiant stars, although examples of luminous blue variable stars with similar magnitudes are known (e.g. Cyg OB2-304 with $M_V \sim -10.6$; \citealt{massey_etal_2001}).  However, these magnitudes are also reasonable for young massive star clusters, such as one might find in an intense star forming region, and even somewhat unremarkable -- for example more than 50 young clusters in the Antennae have absolute magnitudes of $M_V < -11.3$ \citep{whitmore_etal_1999}.  
\srcsix~is a very interesting case, as it is located in 
an early type galaxy. However, the possibility of a globular cluster host seems rather unlikely, as its absolute magnitude is larger than would be expected for such an object.
A very interesting alternative is that it could be located in an Ultra-Compact Dwarf (UCD) galaxy, a number of which are also known to be located within the central regions of the Coma cluster \citep{madrid_etal_2010}.  
Given that UCDs are thought to fill an evolutionary step between the largest globular clusters and dwarf spheroidal galaxies, such an environment might be an ideal place to look for IMBHs.  However, the colour of the object is anomalously blue (F475W$-$F814W $\approx 0.8$) compared to other UCDs in the same field, which typically have F475W$-$F814W colours of $\sim 1.5$ to 2, and so is perhaps more indicative of a background QSO.  The next object, \srceight, is a very blue object, not being detected in the F814W filter, and hence must be located at the near side of the dust lane in its host galaxy (NGC 5907).  Its absolute magnitude of $\sim -9.5$ in the F450W filter is too large for a single stellar object, but is potentially indicative of a young, blue stellar cluster (\eg \citealt{whitmore_etal_1999}).  Finally, the counterpart to \srcten~appears consistent with a single giant star, although again its magnitudes are poorly constrained by the data and so no further constraints from its colour information are forthcoming.


\subsubsection{Optical spectroscopy of a counterpart}

Given the interesting nature of the counterparts to these objects, further follow-up observations to identify their nature are critical.  To this end, we have already obtained a follow-up low-dispersion optical spectrum of the possible UCD galaxy counterpart to \srcsix~using the GMOS long slit spectrometer on the Gemini-North telescope.  Unfortunately, in this case the spectrum revealed emission lines consistent with a relatively high redshift QSO ($z \approx 3.25$), thus explaining its blue colours.  We therefore identify this object as the second contaminant in our sample, and do not consider it further in the analysis (although, we again describe its properties in Appendix A).

\subsubsection{Background contamination estimate}

Out of the ten objects in our initial sample we have therefore identified two background contaminants.  But how many should we expect?  To answer this question we have revisited the background contamination calculations presented for the parent 2XMM-DR1/RC3 sample in \citeauthor{walton_etal_2011b} (2011b).  
We repeat the process presented in that paper, with appropriate changes for consistency with this sample, \ie we set the luminosity limit to be $5 \times 10^{40} \ergsec$, and only consider galaxies within 100 Mpc.  From this we predict a false detection rate of $\sim$ three objects with apparent luminosities above $5 \times 10^{40} \ergsec$ across our entire sample, albeit with large uncertainties stemming from the small number statistics.  As with \citeauthor{walton_etal_2011b} (2011b), the background contamination is expected to be higher for elliptical galaxies, where we predict $\sim$ two QSO contaminants, compared to $\sim$ one for spiral galaxies.  This compares favourably to what we have detected; the two identified QSOs are indeed located close to elliptical galaxies, matching the prediction.  On the other hand, the low predicted background rate for spiral galaxies (one out of seven detections) provides further confidence that the majority of these objects are real luminous ULXs, associated with their parent galaxies.

\subsection{X-ray spectral analysis} \label{spec_analysis}

\begin{table*}
\caption{X-ray spectral modelling}
\begin{center}
\begin{tabular}{lcccccccc}
\hline
& & \multicolumn{3}{c}{Absorbed power-law} & \multicolumn{3}{c}{Absorbed MCD}
\\
Source ID & Obs ID & ${N_H} ^a$ & $\Gamma ^b$ & $\chi ^2 / \rm{dof}^c$ & ${N_H} ^a$ & $kT_{\rm{in}}$ $^d$ & $\chi ^2 / \rm{dof}^c$ & $f_{\rm X, PL}$ $^e$
\\
\hline
\srcone & 0200780101 & $0.26 ^{+0.1} _{-0.08}$ & $2.2 \pm 0.2$ & 54.8/43 & $<0.09$ & $1.0 \pm 0.1$ & {\bf 34.5/43} & $8.0^{+0.4}_{-0.8}$\\
        & 10562 & $0.14 ^{+0.07} _{-0.08}$ & $1.8 \pm 0.2$ & {\bf 27.9/25} & $<0.03$ & $1.2 ^{+0.2} _{-0.1}$ & [39.9/25] & $1.95^{+0.08}_{-0.2}$\\
        & 0601670101 & $0.10 \pm 0.02$ & $1.80 \pm 0.08 $ & {\bf 132.3/131} & $< 0.004$ & $1.22^{+0.1}_{-0.08}$ & [227.8/131] & $1.49^{+0.05}_{-0.05}$\\
\srctwo & 0093630101 & $0.20 ^{+0.06} _{-0.05}$ & $2.2 \pm 0.2$ & {\bf 43.3/41} & $<0.04$ & $0.9 \pm 0.1$ & [74.4/41] & $6.6^{+0.5}_{-0.5}$\\
        & 0306230101 & $0.18 ^{+0.04} _{-0.03}$ & $1.61 \pm 0.09$ & {\bf 126.6/133} & $0.03 \pm 0.02$ & $1.6 \pm 0.1$ & 154.7/133 & $8.4^{+0.2}_{-0.5}$\\
        & 0553300401 & $0.16 \pm 0.03$ & $1.59^{+0.08}_{-0.07}$ & {\bf 219.6/214} & $0.02^{+0.02}_{-0.01}$ & $1.6 \pm 0.1$ & [281.6/214] & $10.8^{+0.3}_{-0.4}$\\
        & 12988 & $0.26 \pm 0.04$ & $1.9^{+0.1}_{-0.09}$ & {\bf 122.4/132} & $0.04 \pm 0.03$ & $1.3^{+0.1}_{-0.09}$ & 152.6/132 & $8.0^{+0.2}_{-0.5}$\\
\srcthree & 0022340201 & $0.10 \pm 0.04$ & $1.42^{+0.09}_{-0.08}$ & 107.1/92 & $< 0.01$ & $1.9 ^{+0.2} _{-0.1}$ & {\bf 104.8/92} & $4.6^{+0.2}_{-0.2}$\\
\srcthree a & 4968 & $0.5^{+0.4}_{-0.2}$ & $1.9 \pm 0.3$ & 16.4/12 & $<0.4$ & $1.2^{+0.5}_{-0.2}$ & {\bf 15.1/12} & $0.76^{+0.06}_{-0.2}$\\
\srcthree b & 4968 & $0.4 \pm 0.2$ & $2.1^{+0.3}_{-0.4}$ & {\bf 10.9\%$^f$} & $<0.2$ & $1.3^{+0.4}_{-0.2}$ & 13.9\%$^f$ & $0.35^{+0.05}_{-0.1}$\\
\srcthree c & 4968 & $1.5 ^{+1} _{-0.7}$ & $1.7\pm0.6$ & [100\%$^f$] & $0.9 ^{+0.7} _{-0.4}$ & $2.3 ^{+2} _{-0.8}$ & [100\%$^f$] & $0.39^{+0.06}_{-0.3}$\\
\srcfive & 0147610101 & $0.26^{+0.06}_{-0.05}$ & $1.5 \pm 0.1$ & {\bf 74.1/86} & $0.08 ^{+0.03} _{-0.04}$ & $1.8 \pm 0.2$ & 94.1/86 & $4.4^{+0.2}_{-0.2}$\\
        & 7863 & $<0.5$ & $1.4 ^{+0.5} _{-0.4}$ & {\bf 6.8/10} & $<0.2$ & $1.7 ^{+0.9} _{-0.4}$ & 8.7/10 & $6.8^{+0.8}_{-2}$\\
\srcseven & 0092970801 & $<0.2$ & $1.7 \pm 0.3$ & {\bf 15.0/16} & $<0.05$ & $1.2 \pm 0.2$ & 16.7/16 & $2.8^{+0.2}_{-0.4}$\\
         & 12989 & $0.08 ^{+0.05} _{-0.04}$ & $1.63 ^{+0.09} _{-0.1}$ & {\bf 78.8/69} & $<0.007$ & $1.34 ^{+0.1} _{-0.09}$ & [103.0/69] & $3.2^{+0.2}_{-0.3}$\\
\srceight & 0145190201 & $0.97 ^{+0.07} _{-0.06}$ & $1.88 ^{+0.07} _{-0.06}$ & {\bf 409.5/366} & $0.54 \pm 0.04$ & $1.66 ^{+0.08} _{-0.07}$ & [420.0/366] & $15.8^{+0.3}_{-0.4}$\\
        & 0145190101 & $0.84 ^{+0.07} _{-0.06}$ & $1.68 \pm 0.07$ & 267.0/251 & $0.49 \pm 0.04$ & $1.9 \pm 0.1$ & {\bf 258.5/251} & $14.7^{+0.4}_{-0.4}$\\
\srcnine & 1666 & $<0.2$ & $1.6 ^{+0.4} _{-0.3}$ & {\bf 1.1\%$^f$} & $<0.06$ & $1.4 ^{+0.5} _{-0.3}$ & [98.4\%$^f$] & $0.34^{+0.06}_{-0.09}$\\
        & 0112980101 & $<0.2$ & $1.8 ^{+0.4} _{-0.3}$ & {\bf 83.3\%$^f$} & $<0.03$ & $1.5 ^{+0.4} _{-0.3}$ & [97.8\%$^f$] & $0.61^{+0.06}_{-0.09}$\\
\srcten & 0025541001 & $0.27 ^{+0.08} _{-0.07}$ & $1.9 ^{+0.2} _{-0.1}$ & {\bf 61.3/62} & $<0.07$ & $1.4 ^{+0.2} _{-0.1}$ & 66.1/62 & $4.7^{+0.2}_{-0.2}$\\
        & 11230 & $0.6 \pm 0.2$ & $4.2^{+0.9}_{-0.8}$ & [96.7\%$^f$] & $<0.3$ & $0.40^{+0.1}_{-0.08}$ & {\bf 57.3\%$^f$} & $0.28^{+0.04}_{-0.2}$\\
\hline

\end{tabular}
\end{center}
\begin{minipage}{\linewidth}
Notes: Errors shown for spectral parameters are 90 per cent confidence ranges.
$^a$ Absorption column density external to our Galaxy ($\times 10^{22}~\rm{cm}^{-2}$).
$^b$ Power-law photon index.
$^c$ Statistical goodness of the fit, in terms of the $\chi^2$ statistic and number of degrees of freedom.  Numbers in bold indicate which model provided the best fit to the data, whilst numbers in square brackets indicate a fit that is rejected at 2$\sigma$ significance.
$^d$ Inner-disc temperature (keV).
$^e$ Observed 0.3--10 keV flux, in units of $10^{-13} \ergcms$, based on the power-law fit.
$^f$ Total counts from this observation were insufficient for $\chi ^2$ fitting, 
instead Cash statistics were used \citep{cash_1979}. Goodness of fit is shown
instead of $\chi^2/\rm{dof}$.
\end{minipage}
\label{powerlaw}
\end{table*}

\begin{figure*}
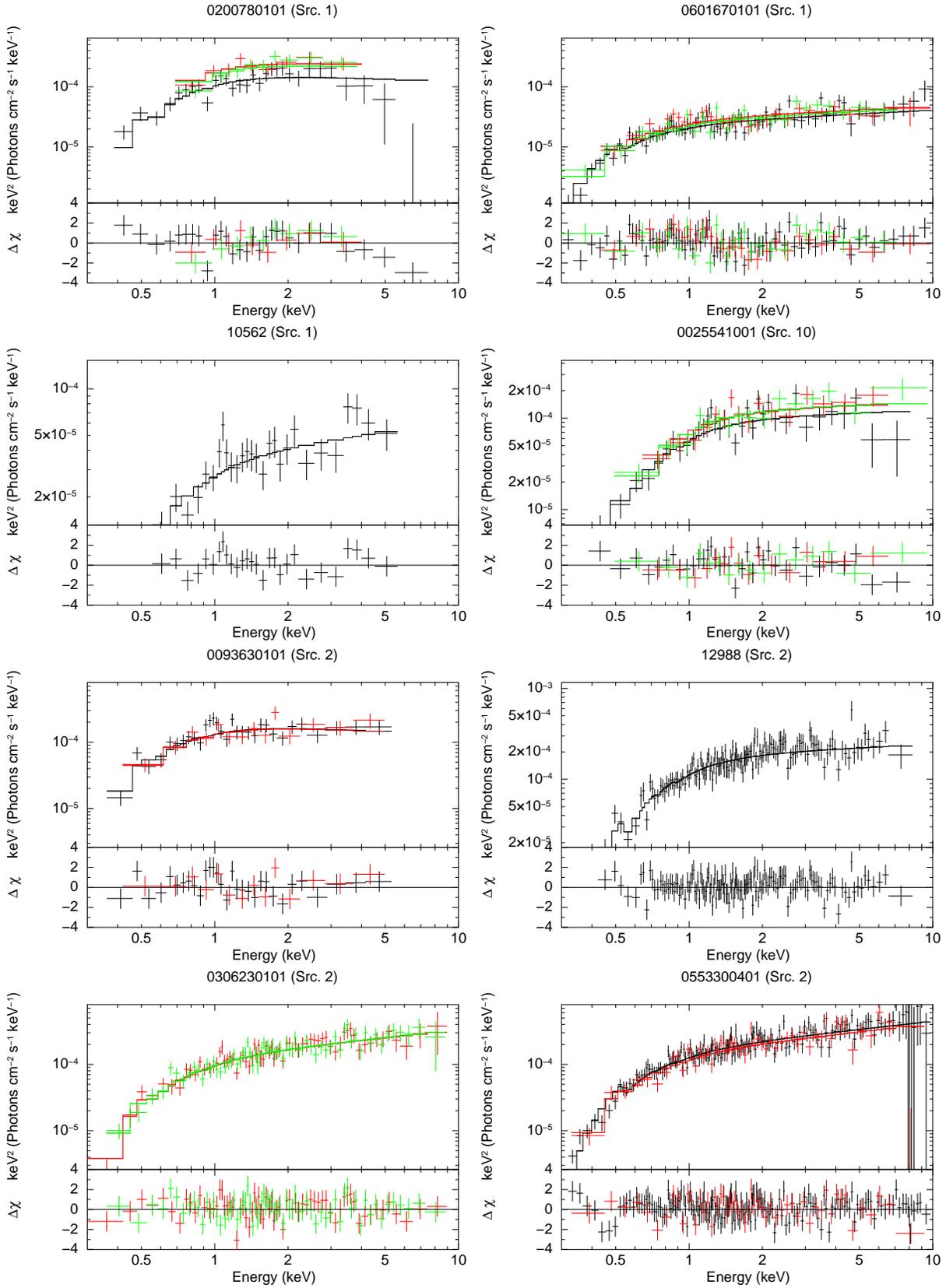

\includegraphics[height=8.cm, angle=-90]{./images/NGC470_powerlaw_1.ps}
\includegraphics[height=8.cm, angle=-90]{./images/NGC470_powerlaw_3.ps}
\includegraphics[height=8.cm, angle=-90]{./images/NGC470_powerlaw_2.ps}
\includegraphics[height=8.cm, angle=-90]{./images/NGC7479_powerlaw.ps}
\includegraphics[height=8.cm, angle=-90]{./images/NGC1042_powerlaw1.ps}
\includegraphics[height=8.cm, angle=-90]{./images/NGC1042_powerlaw4.ps}
\includegraphics[height=8.cm, angle=-90]{./images/NGC1042_powerlaw2.ps}
\includegraphics[height=8.cm, angle=-90]{./images/NGC1042_powerlaw3.ps}
\caption{\xmmn EPIC PN (black), MOS1 (red), MOS2 (green) or \chan spectra
for observations with greater than 250 source counts, all of which are unfolded from the detector response. 
Multiple observations of sources are shown side-by-side where possible.
Where there are two or more observations of a source with the same 
detector, plots are shown on the same scale for direct comparison.
Over-plotted lines show the best fitting absorbed 
power-law spectra (Table \ref{powerlaw}), 
with one absorption component set to the Galactic value and another 
free to model intrinsic absorption in the ULX candidates. A range of
photon indexes between 1.4-2.2 were obtained.}
\label{spectra}
\end{figure*}
\begin{figure*}
\includegraphics[height=8.cm, angle=-90]{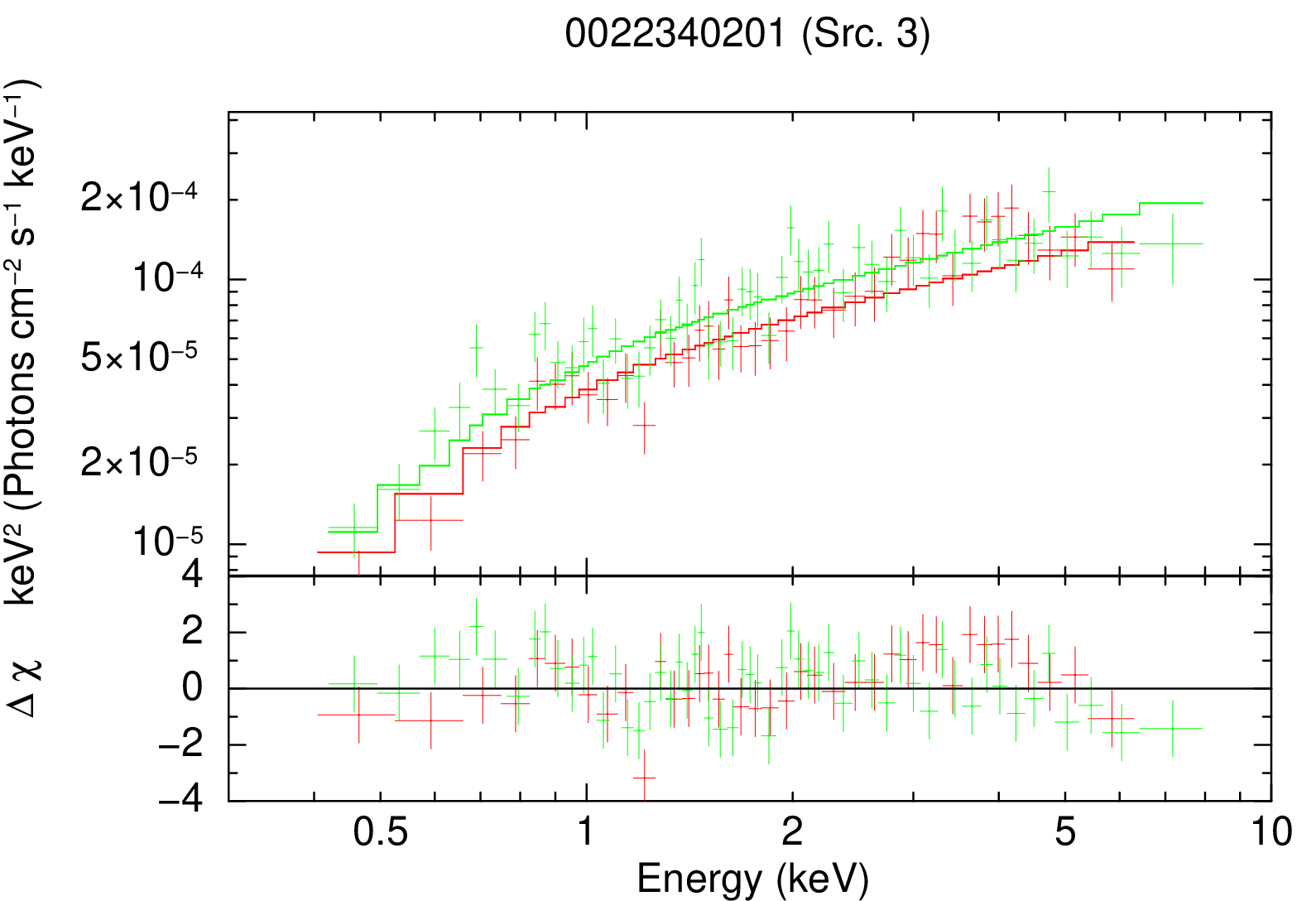}
\includegraphics[height=8.cm, angle=-90]{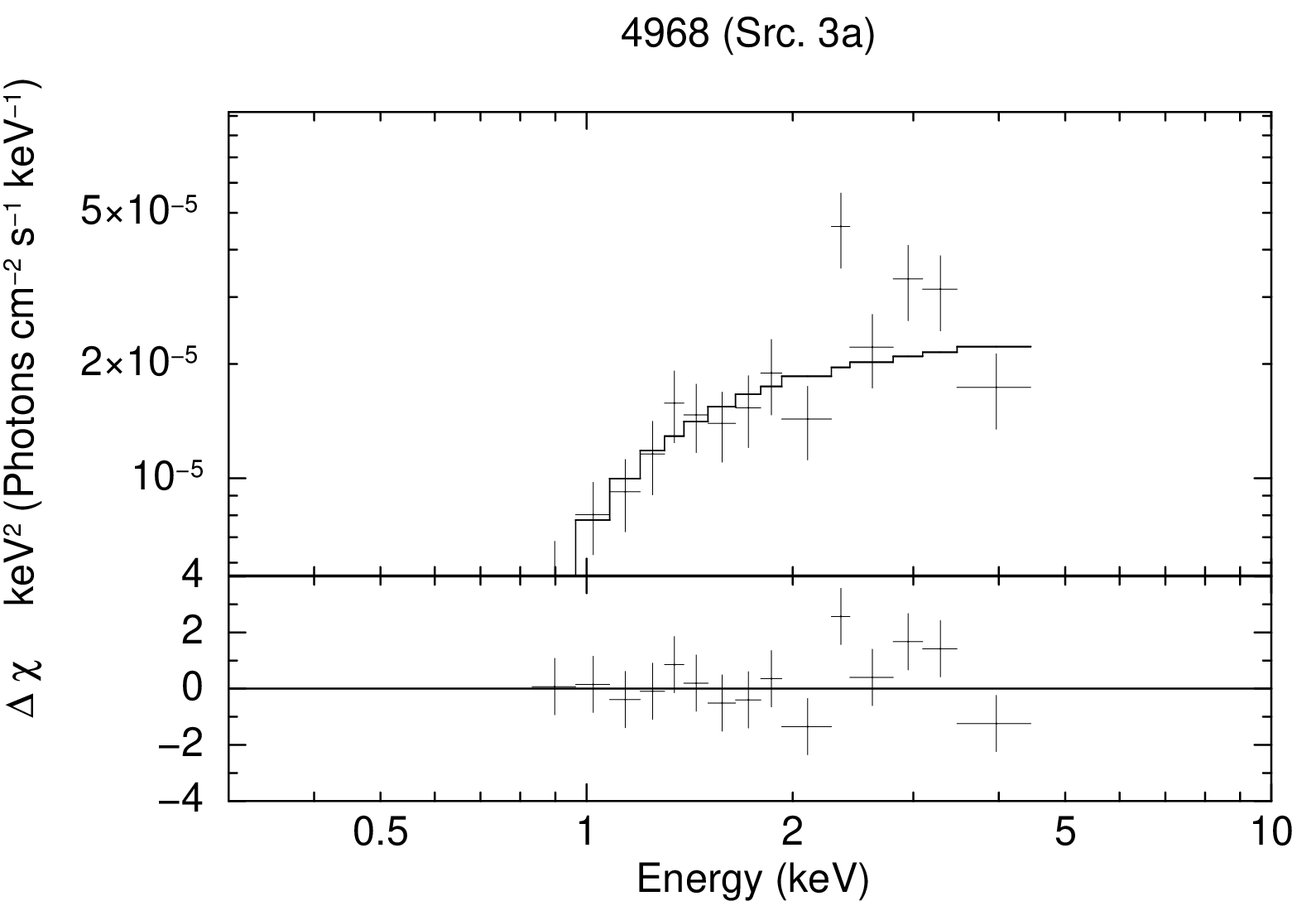}
\includegraphics[height=8.cm, angle=-90]{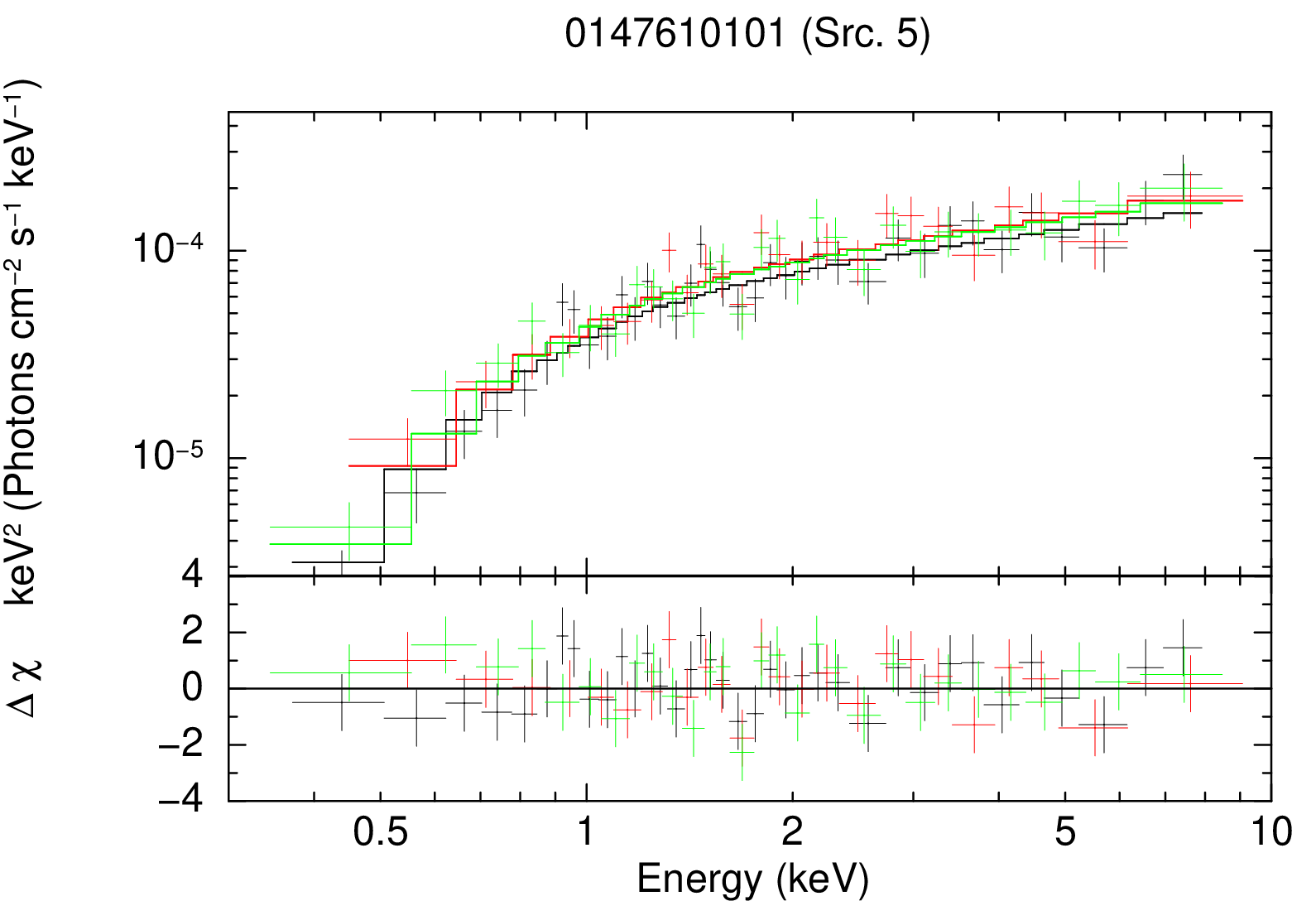}
\includegraphics[height=8.cm, angle=-90]{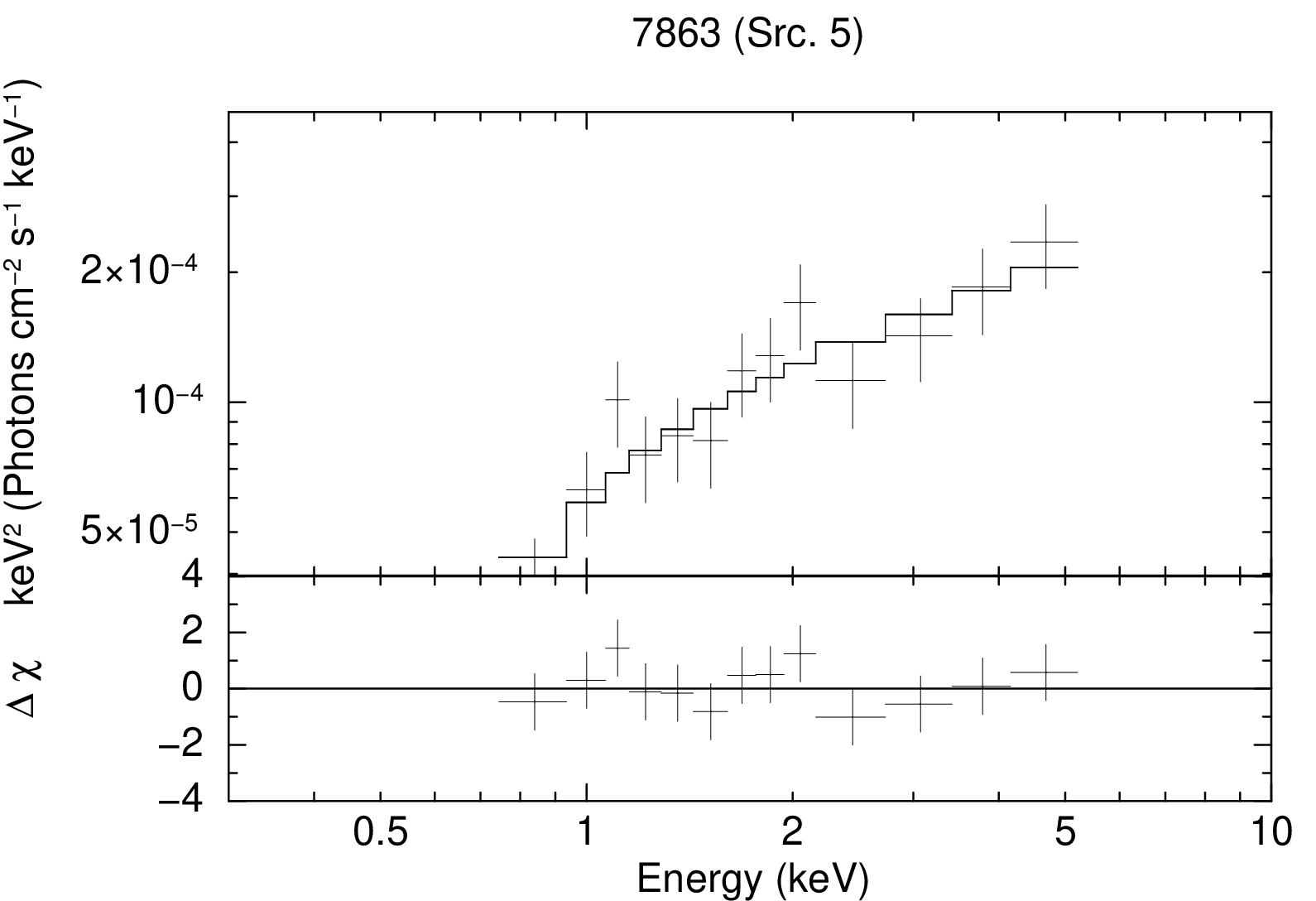}
\includegraphics[height=8.cm, angle=-90]{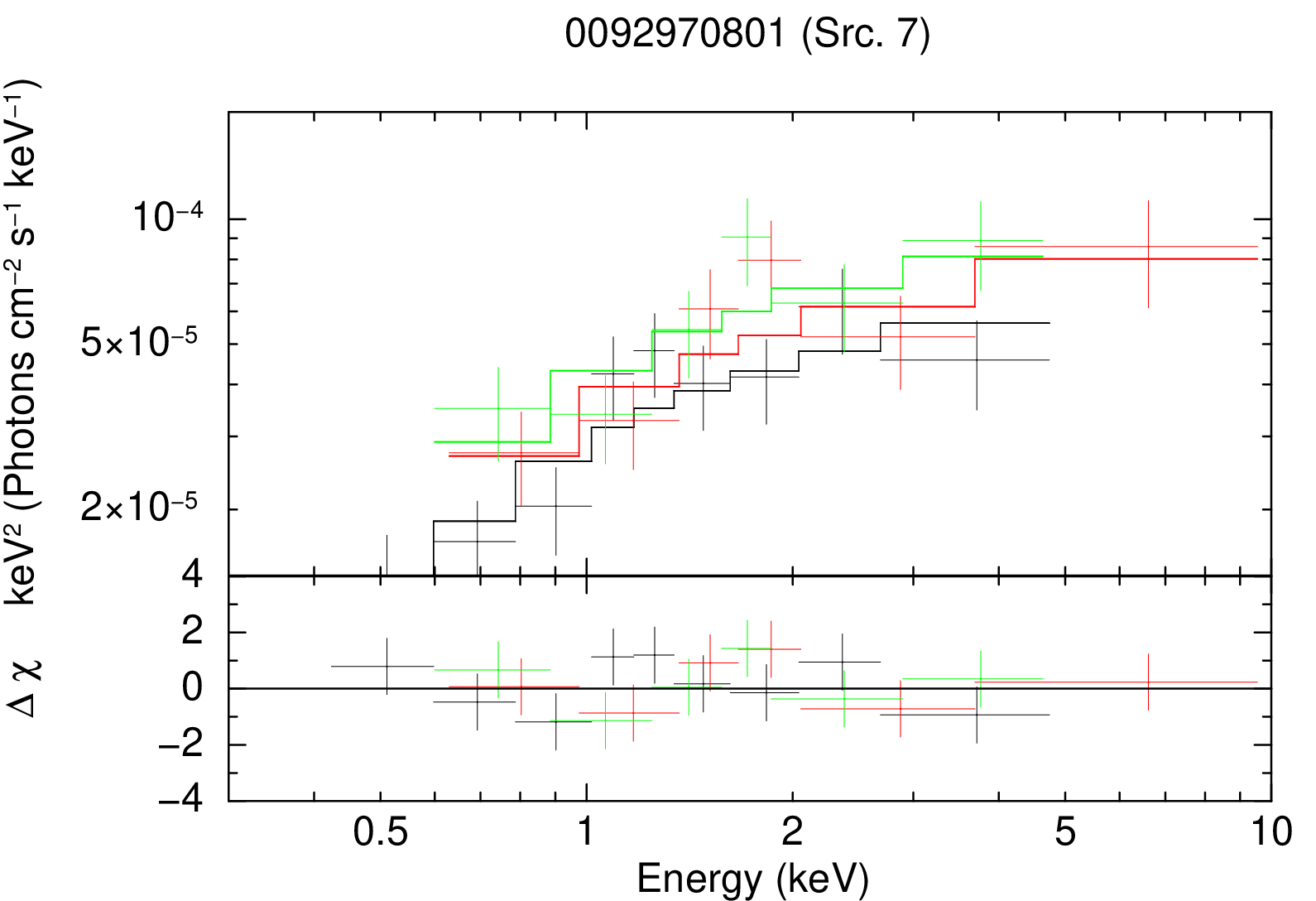}
\includegraphics[height=8.cm, angle=-90]{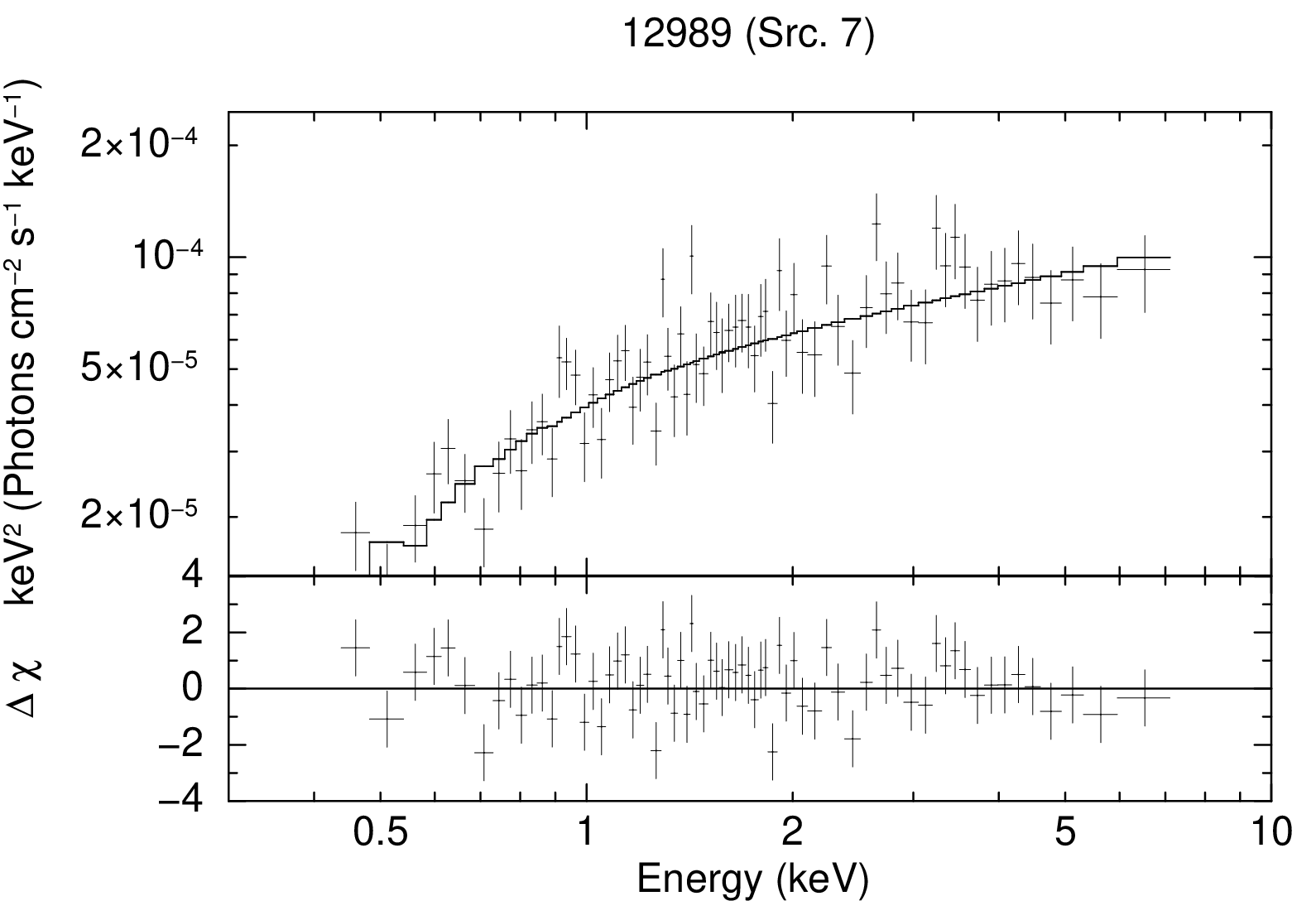}
\includegraphics[height=8.cm, angle=-90]{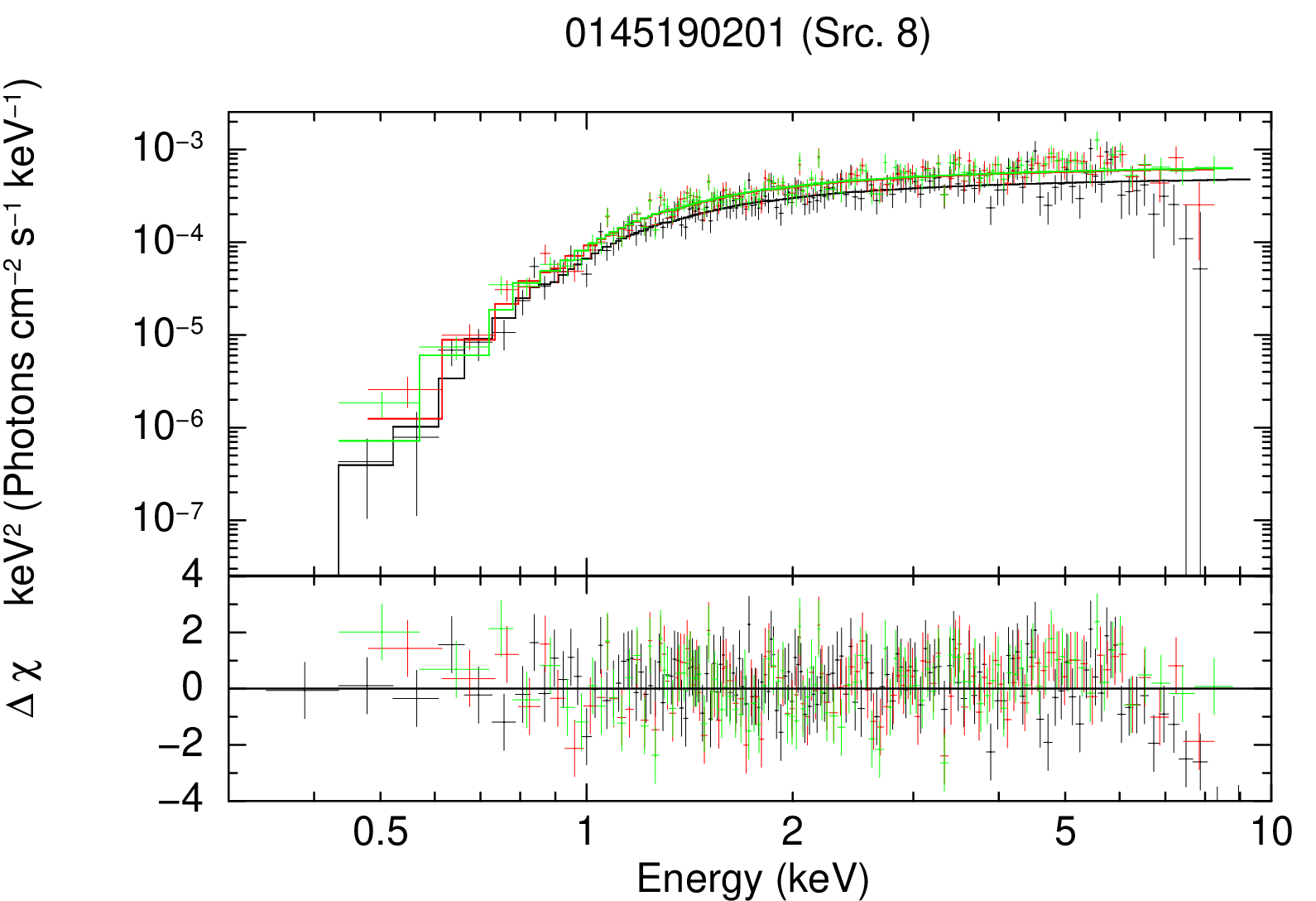}
\includegraphics[height=8.cm, angle=-90]{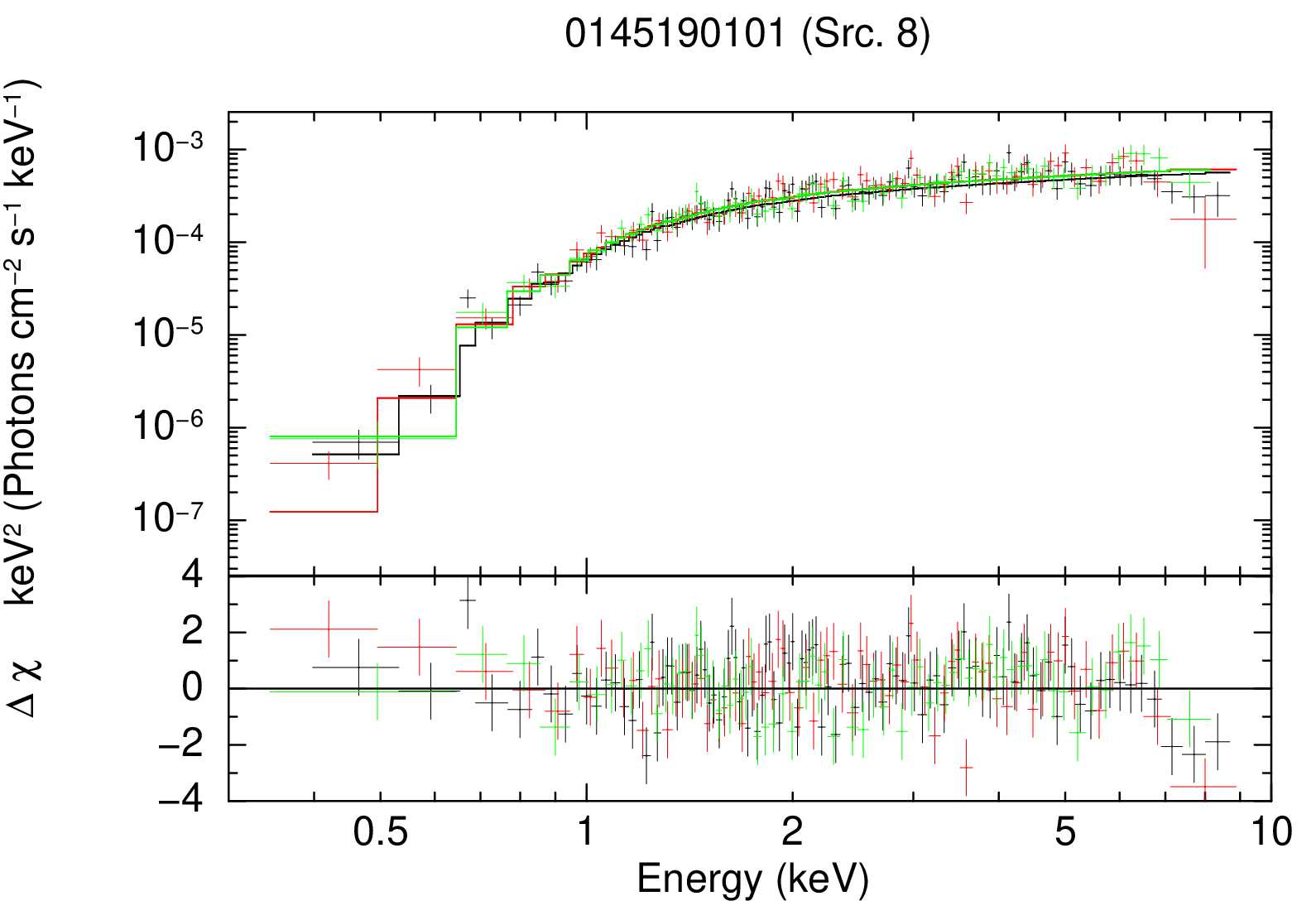}
\begin{flushleft}
\bf{Figure~\ref{spectra}.} (continued)
\end{flushleft}
\end{figure*}

Where sufficient counts were available to enable X-ray spectral analysis, the spectra were fitted with basic spectral models using {\sc xspec} v.12.6.0.  Observations with $\ge 250$ counts were grouped using the {\sc ftool} {\sc grppha} to a minimum of twenty counts per energy bin, to produce bins with Gaussian errors and hence permit the use of $\chi^2$ statistics in fitting the data.  Where only $100-250$ counts were available, models were fitted to ungrouped X-ray spectra using Cash statistics \citep{cash_1979}.  The data were not background subtracted for the Cash analyses; instead in each of these cases the background spectrum was first constrained from the background aperture, and then added to the fit as a constant component, normalised for the relative sizes of the extraction regions.  In all cases the fitting was limited to the 0.3-10 keV band.

The spectra were fitted with both an absorbed power-law continuum and an absorbed multi-colour disc blackbody ({\sc diskbb} in {\sc xspec}, hereafter MCD) models, with the latter representative of the emission from a standard optically-thick, geometrically thin accretion disc \citep{mitsuda_etal_1984}.  
Absorption was modelled by the {\sc tbabs} model, using abundances from \citet*{wilms_etal_2000}.  
Two absorption components were included, the first of which was fixed to the Galactic line-of-sight column to the object (\citealt{dickey_and_lockman_1990}, cf. Table~\ref{srcs}), and the second left as a free parameter to model intrinsic absorption in the source and/or its host galaxy.  Where multiple \xmmn EPIC detectors were operating, an additional multiplicative constant was included in the spectral model. This was fixed to 1.0 for the PN detector (or MOS1 where no PN data were available) and varied freely for the others, in order to account for the calibration uncertainties between the EPIC detectors.  In all cases with equally sized extraction apertures the values of this parameter remained within a few percent of unity, and the values scaled appropriately where smaller extraction regions were necessary on individual detectors.

Table~\ref{powerlaw}  shows the parameters resulting from fitting the two simple models to the source data.  Plots of the absorbed power-law spectral fits, unfolded from the detector response, are displayed in Fig.~\ref{spectra} for all observations with greater than 250 source counts.  An absorbed power-law continuum spectrum provides a statistically acceptable fit (\ie null hypothesis probability $> 5$ per cent, so rejection likelihood $< 95$ per cent) to data from all but two observations.  These exceptions are the least luminous of the resolved source triplet, \srcthree c, which was not well characterised by either model; and a marginal rejection of the latter, much less luminous observation of \srcten, which was the sole dataset acceptably fitted by the MCD model alone.  In contrast, eight 
spectra were acceptably fitted by the absorbed power-law model, but rejected a MCD fit at greater than the 95 per cent level, and 11 spectra were acceptably fitted by both models.  In all, the absorbed power-law fit was statistically preferred (\ie provided the closer fit to the data) in 15/21 
cases, compared to 5/21 
for the MCD, although many of these preferences were marginal in nature.

The large majority of the spectra that were acceptably characterised as absorbed power-law continua all had rather hard photon indexes, with $\Gamma = 1.4$--$2.2$.  The absorption column densities for these fits were mostly moderate (where well constrained), with values in the range $\sim 1 - 3 \times 10^{21}$ cm$^{-2}$, although the column for \srceight~was notably higher at $\sim 10^{22}$ cm$^{-2}$, which may be due to its location in an edge-on system. It should be noted however, that such an interpretation may bring into doubt any physical association with the particularly blue potential counterpart to this source.  We discuss this further in section 4.1. 
The spectra that were acceptably fitted by the absorbed MCD model had -- with one notable exception -- inner disc temperatures of $kT_{\rm in} \sim 1.0$--$1.9$ keV, and rather low modelled absorption columns at $< 1 \times 10^{21}$ cm$^{-2}$, where well-constrained.  The exception was the fainter observation of \srcten, where the MCD model was the only acceptable fit, and the spectrum was remarkably soft with an inner disc temperature $kT_{\rm in} \sim 0.4$ keV.

An inner disc temperature in the range $1$--$2$ keV is fairly typical for a sub-Eddington stellar mass black hole in the thermal dominant state, with the accretion disc extending to the innermost stable circular orbit \citep{mcclintock_and_remillard_2006}.  However, for objects at close to $L_X \sim 10^{41} \ergsec$ it would appear somewhat anomalous, and certainly not indicative of the presence of IMBHs, which should have much cooler disc-dominated spectra (\eg \citealt{kaaret_etal_2003}; \citealt{miller_etal_2003}; \citealt{miller_etal_2004}), as indeed does seem to be the case for the fainter dataset for \srcten.  We investigated whether there was any evidence for cool disc components in the other spectra by fitting a two-component, absorbed MCD plus power-law model to the best datasets (the 12 
spectra with $\ga 1000$ counts), where we had the best opportunity of deconvolving the spectra into multiple components.  In all but two cases there was no discernable improvement to the absorbed power-law spectral fit with the addition of a cool disc component.  Two cases offered marginal improvements: the fit to the original \xmmn detection of \srcthree~(before being resolved by {\it Chandra\/}) improved by $\Delta\chi^2 = 10.9$ for two additional degrees of freedom with a $\sim 0.1$ keV disc; and the first, most luminous detection of \srcone~improved by $\Delta\chi^2 = 13.1$ for two additional degrees of freedom with a $\sim 0.07$ keV disc.  However, in the latter case the MCD model alone still offered the most acceptable fit to the data.  This lack of strong cool disc components in the sample is perhaps unsurprising given the lack of strong soft residuals in the spectra presented in Fig.~\ref{spectra}.

However, one object stands out as possessing interesting residuals in Fig.~\ref{spectra}.  Both observations of \srceight, the highest quality data in the sample with $> 5000$ counts per dataset, appear to show a high energy spectral turnover beyond $\sim 5$ keV.  Interestingly, this feature is the key diagnostic of the putative ultraluminous state, that is seen in many less-luminous ULX spectra \citep{gladstone_etal_2009}.  We note that the other main diagnostic of this putative state, a soft excess, is not seen; although this may be due to the object being buried in an edge-on galaxy, and hence its soft X-rays being relatively highly obscured, as is suggested by its spectral fits (cf. Table~\ref{powerlaw}).  Interestingly, this is very comparable to the object NGC 4517 ULX1 (\citeauthor{walton_etal_2011a} 2011a), 
which is also located in an edge-on galaxy and has a highly absorbed soft spectrum, with a break at high energy.  Confirmation was sought of the significance of the spectral break using the method of \citet{stobbart_etal_2006}, \ie by fitting unabsorbed power-law and broken power-law models to the 2--10 keV X-ray spectra of both observations of \srceight. A broken power-law model provided a better fit to both observations, with very significant {\it F}-test probabilities of $1 \times 10^{-12}$ (0145190201) and $4 \times 10^{-8}$ (0145190101), hence the spectrum is confirmed to be consistent with the ultraluminous state.

Both Fig.~\ref{spectra} and Table~\ref{powerlaw} demonstrate that the individual objects also displayed both flux and spectral shape variations between different observing epochs.  In some cases the spectral variability was quite subtle, for example a hardening of the absorbed power-law spectrum with flux in \srctwo.  In other objects the contrasts were much starker, particularly in \srcone~and \srcten.  In the former case the object transits from an absorbed disc-like spectrum at high flux to an absorbed power-law spectrum in the two subsequent spectra where the flux is a factor $> 4$ lower, and in the latter case the spectrum changes from hard and power-law-like at high flux to much softer and disc-like after a drop in flux of nearly a factor 20.  We discuss these changes further in Section~\ref{discussion}.

\subsection{X-ray timing}

\subsubsection{Short term variability}

\begin{table}
\caption{Luminosity and variability of the extreme luminosity ULXs}
\begin{center}
\begin{tabular}{ccccc}
\hline
Obs ID & $L_{\rm X}~^a$ & $F_{\rm var} \rm{(200~s)} ^b$ & $F_{\rm var} \rm{(2~ks)} ^b$
\\
\hline
{\bf \srcone} & & &
\\
0200780101 & $10.2 ^{+0.5} _{-1}$ & $<0.2 ^c$ & - 
\\
10562 & $2.5 ^{+0.1} _{-0.3}$ & - & $0.11 \pm 0.06$
\\
0601670101 & $1.90 ^{+0.07} _{-0.06}$ & - & $<0.1 ^c$
\\
{\bf \srctwo} & & &
\\
0093630101 & $2.8 \pm 0.2$ & - & - 
\\
0306230101 & $3.6 ^{+0.1} _{-0.2}$ & - & $0.05 \pm 0.04$
\\
0553300401 & $4.6^{+0.1}_{-0.2}$ & $0.09 \pm 0.04$ & $0.10 \pm 0.02$
\\
12988 & $3.4^{+0.1}_{-0.2}$ & $<0.1 ^c$ & $0.13 \pm 0.02$
\\
{\bf \srcthree} & & &
\\
0022340201 & $6.1 \pm 0.3$ & - & $0.08 \pm 0.04$
\\
4968 - a & $1.01^{+0.08}_{-0.2}$ & - & - 
\\
4968 - b & $0.46 ^{+0.06}_{-0.1}$ & - & - 
\\
4968 - c & $0.52 ^{+0.08}_{-0.4}$ & - & - 
\\
{\bf \srcfive} & & &
\\
0147610101 & $5.8 ^{+0.2} _{-0.3}$ & $0.13 \pm 0.06$ & -
\\
7863 & $9^{+1}_{-3}$ & $<0.3 ^c$ & - 
\\
{\bf \srcseven} & & &
\\
0092970801 & $30 ^{+2} _{-4}$ & $<0.3 ^c$ & - 
\\
12989 & $35 ^{+2} _{-3}$ & - & $<0.1^c$
\\
{\bf \srceight} & & &
\\
0145190201 & $4.19 ^{+0.09} _{-0.1}$ & $0.07 \pm 0.02$ & - 
\\
0145190101 & $3.9 \pm 0.1$ & $0.11 \pm 0.02$ & - 
\\
{\bf \srcnine} & & &
\\
553 & $2.9 \pm 0.9$ & - & - 
\\
1454 & $3.2 \pm 0.6$ & - &  - 
\\
1666 & $3.8 ^{+0.7} _{-1}$ & - & - 
\\
0112980101 & $6.8 ^{+0.7} _{-1}$ & - & - 
\\
0112980401 & $5.9 \pm 0.8$ & - & - 
\\
0112980501 & $5 \pm 1$ & - & - 
\\
7698 & $1.9 \pm 0.7$ & - & - 
\\
{\bf \srcten} & & &
\\
0025541001 & $6.1 \pm 0.3$ & $0.10 \pm 0.06$ & - 
\\
11230 & $0.36^{+0.05}_{-0.2}$ & - & - 
\\
10120 & $0.40 \pm 0.05$ & - & - 
\\
\hline
\end{tabular}
\end{center}
\begin{minipage}{\linewidth}
Notes:
$^a$0.3--10 keV observed luminosity, in units of $10^{40} \ergsec$, calculated as per the text.  The quoted errors are the 1$\sigma$ confidence ranges on the luminosity.
$^b$Fractional variability amplitude. Values and errors are shown where variability is
detected at $>$ 1 $\sigma$ significance (the quoted errors are the 1$\sigma$ confidence ranges). Upper limits shown are $3 \sigma$ limits, and are calculated where the variability was less than white noise.  The {\sc ftool lcstats} was used to provide this upper limit.
\end{minipage}
\label{luminosity}
\end{table}

\begin{figure*}
\centering
\includegraphics[width=6.5cm, angle=0]{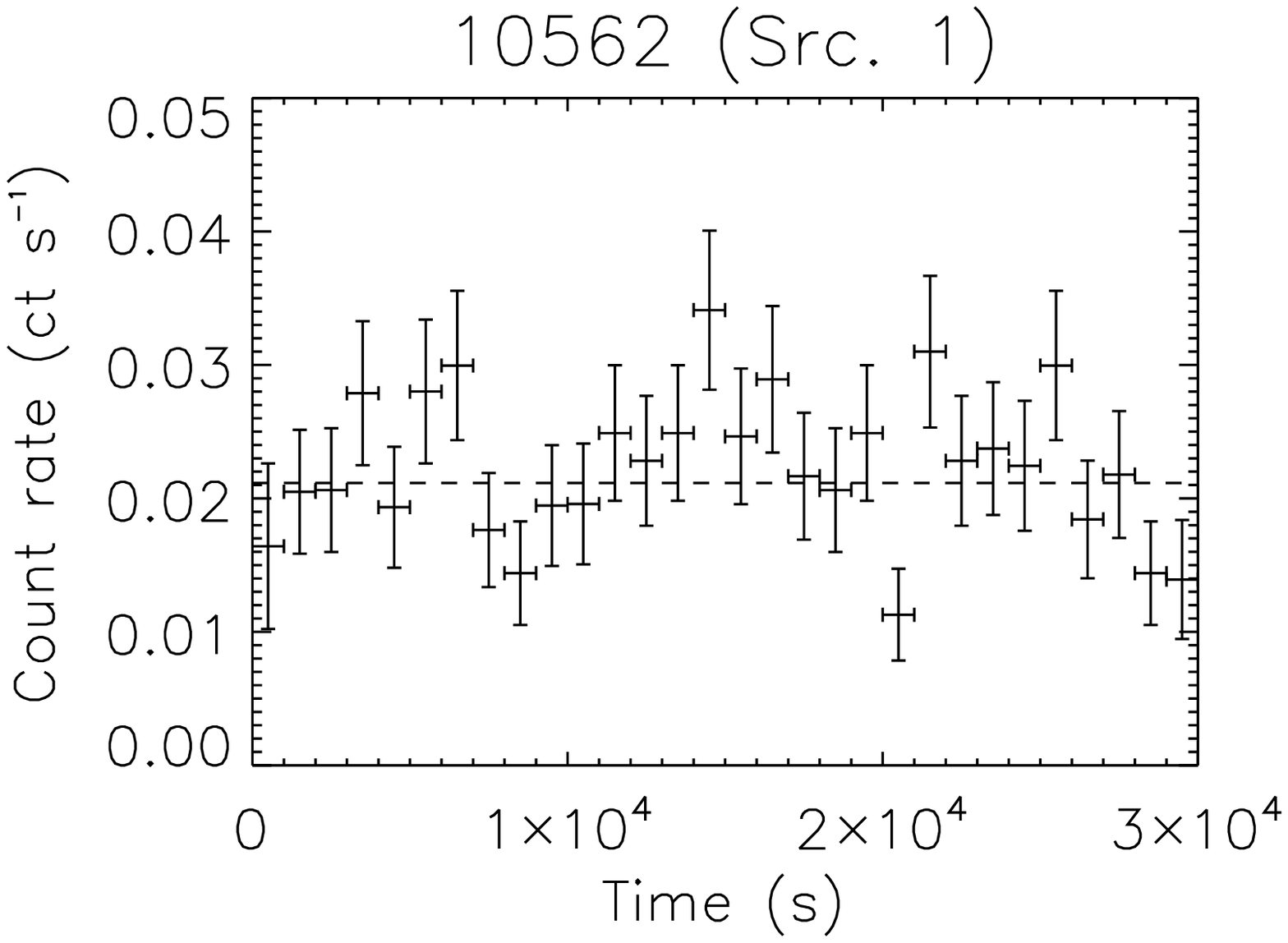}
\includegraphics[width=6.5cm, angle=0]{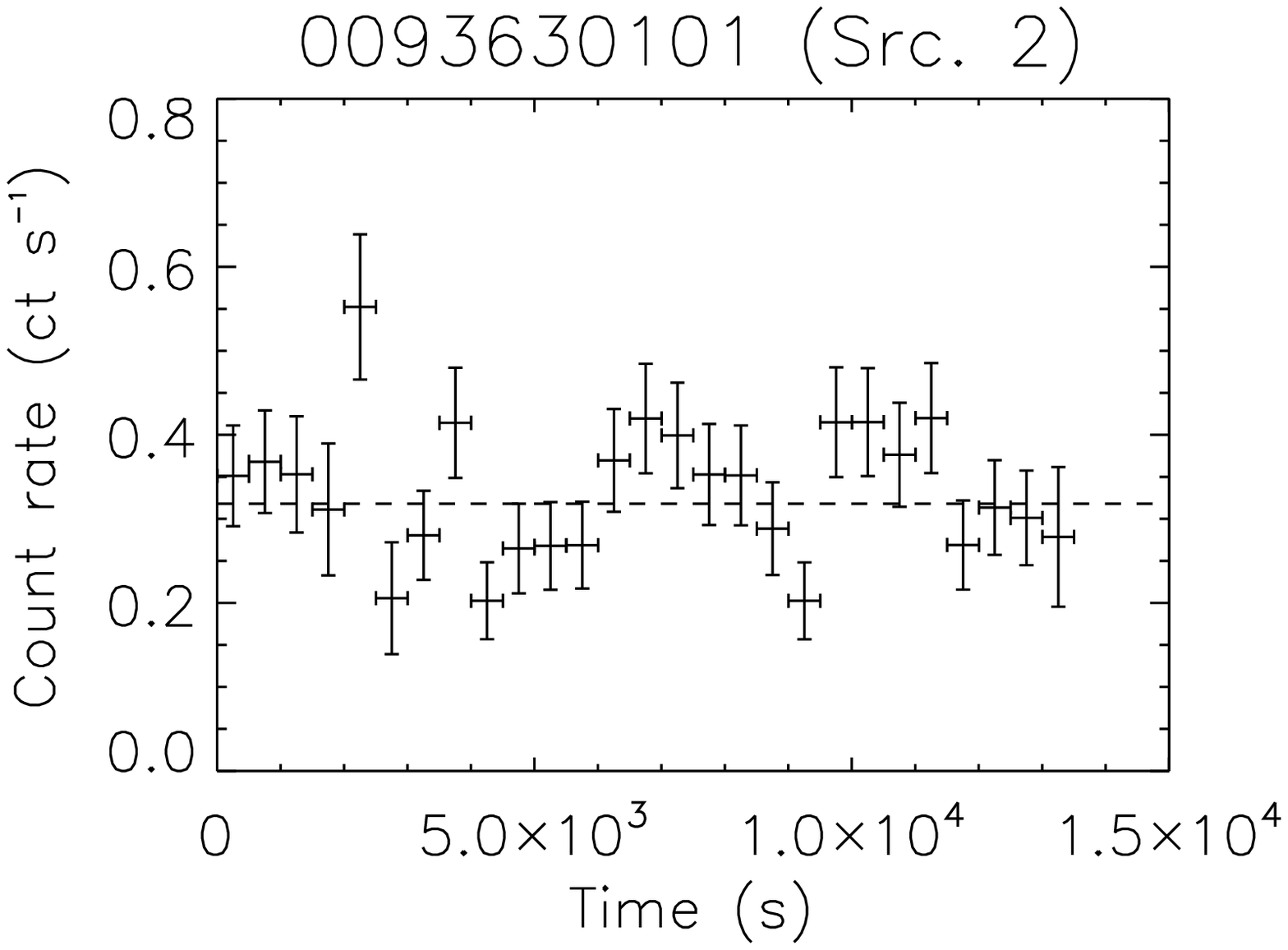}
\includegraphics[width=6.5cm, angle=0]{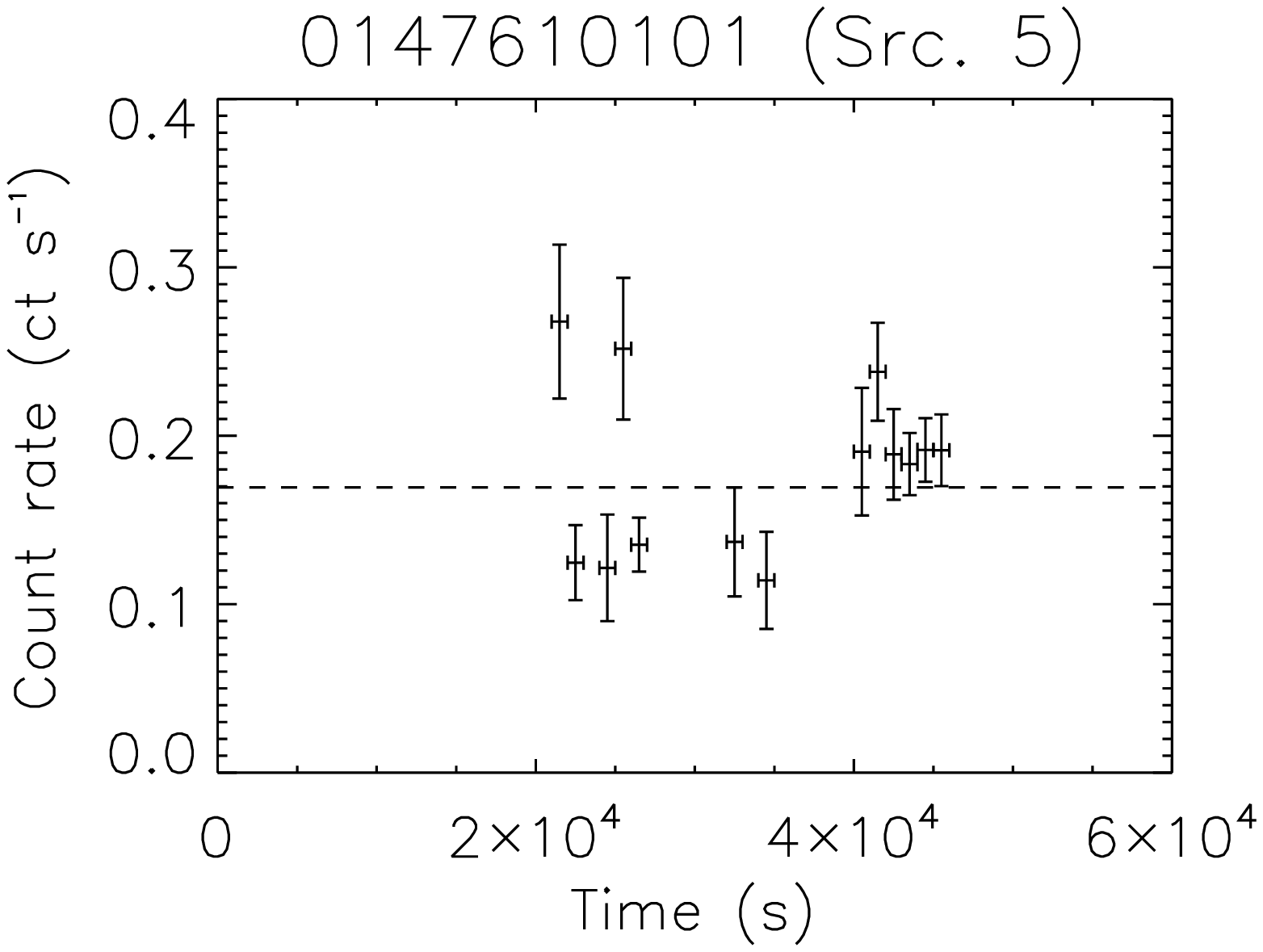}
\includegraphics[width=6.5cm, angle=0]{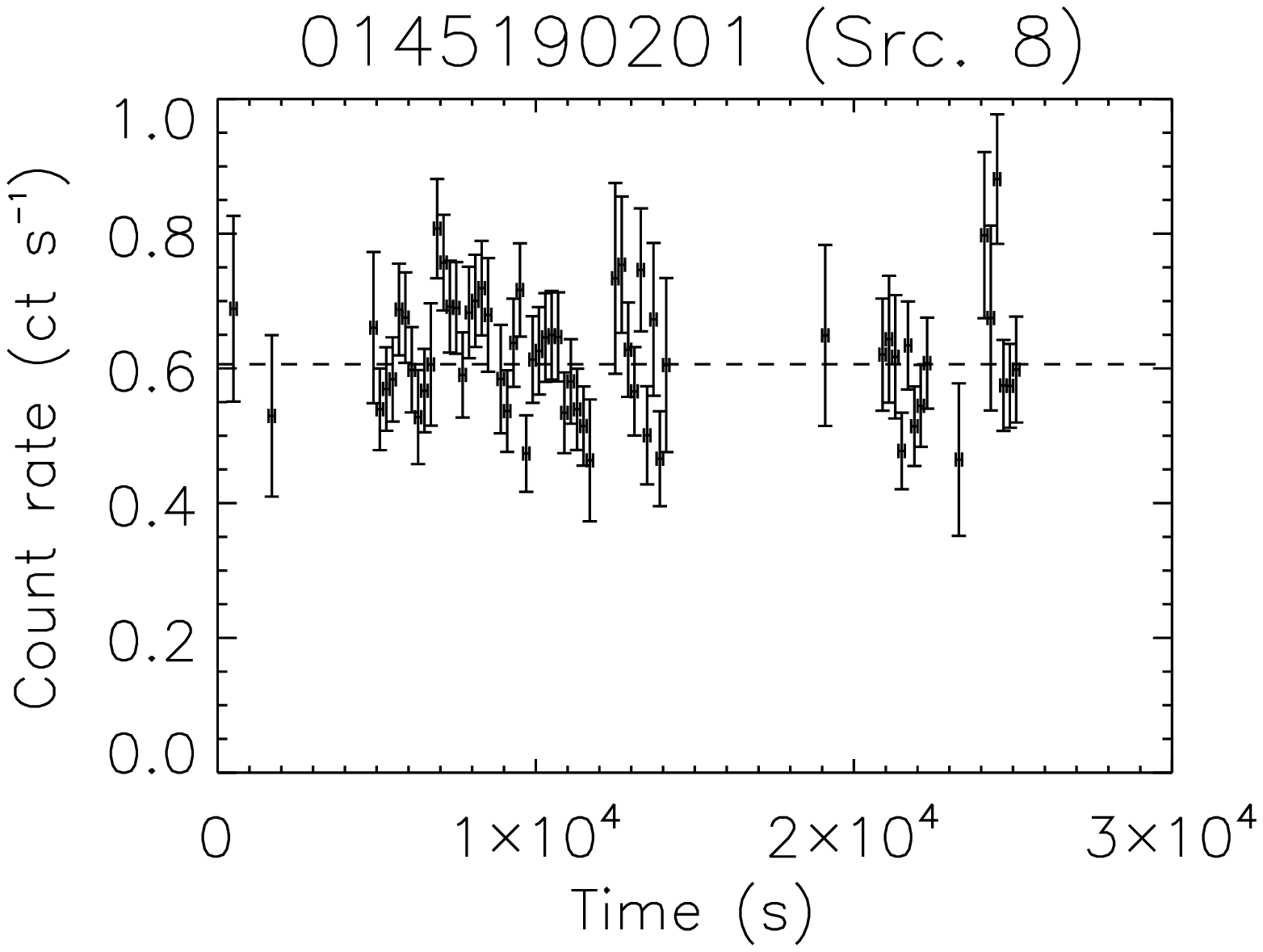}
\includegraphics[width=6.5cm, angle=0]{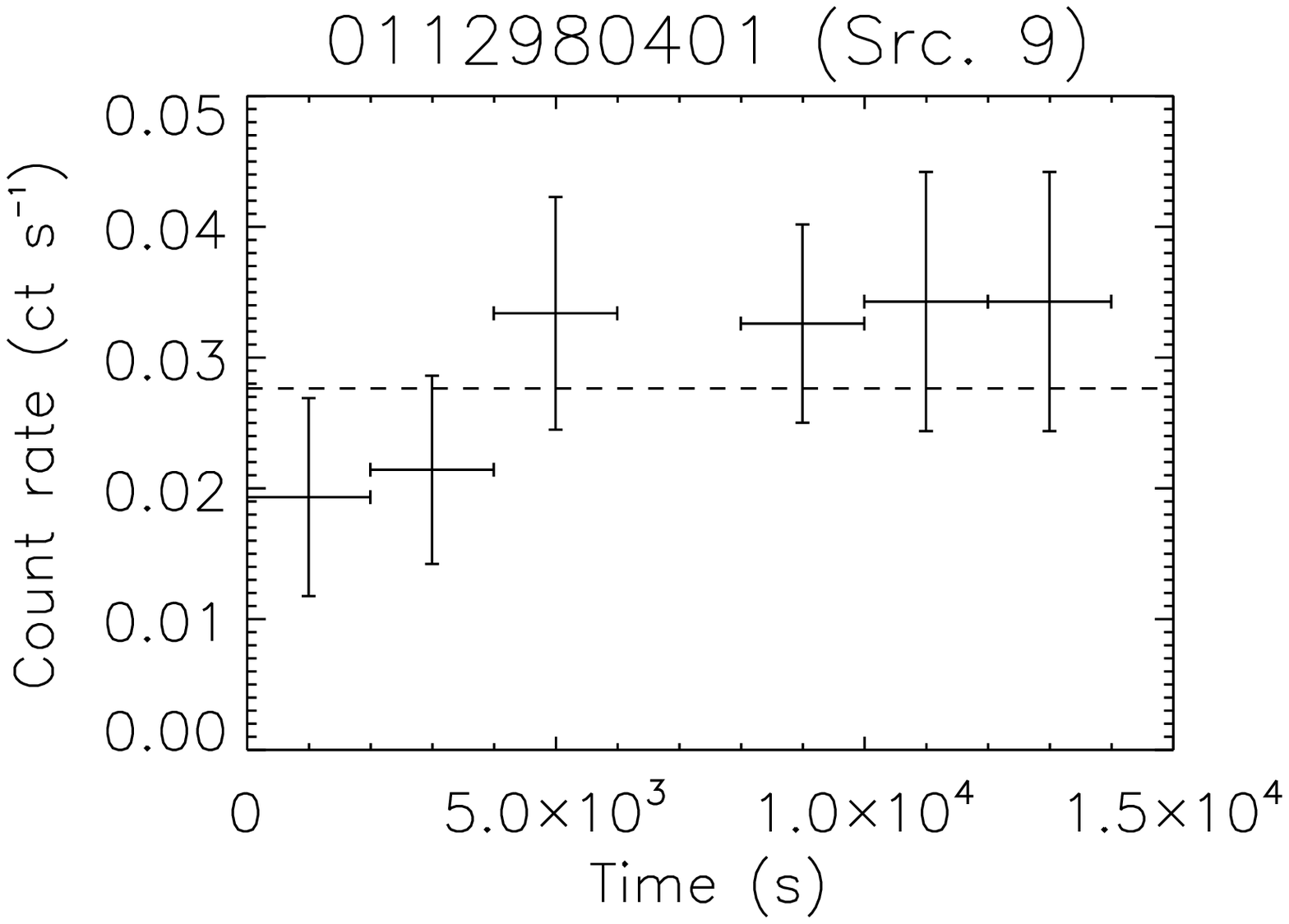}
\includegraphics[width=6.5cm, angle=0]{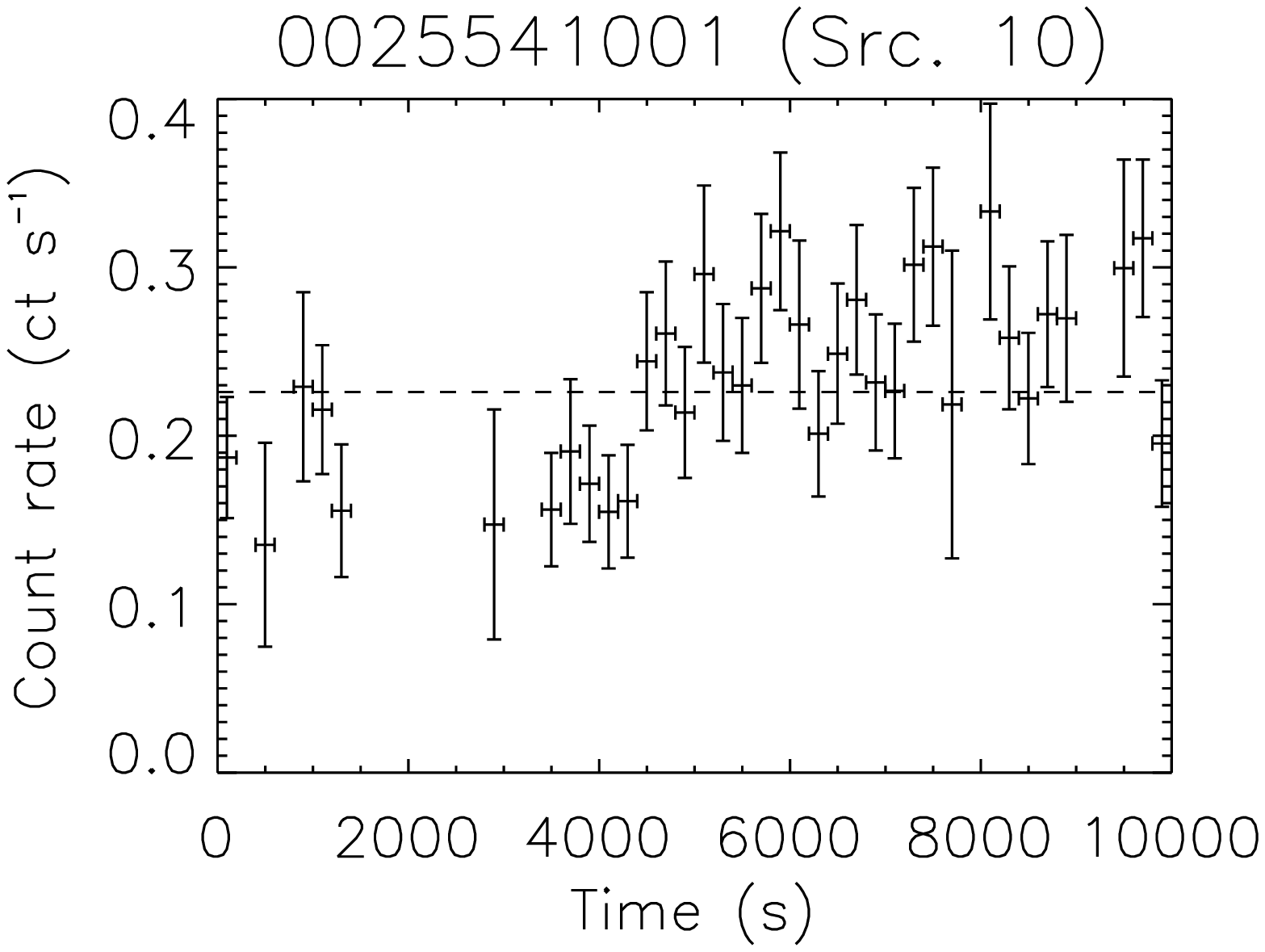}

\caption{A selection of background-subtracted light curves of the ULX 
candidates, chosen to show a variety of behaviours and data quality: continuous, moderate count rate data ({\it top left \& top right}); sparse data ({\it middle left \& middle right}); low count rate data ({\it bottom left}); variability about a mean count rate ({\it top left \& top right}) and an apparent maintained shift in the mean count rate ({\it bottom left \& bottom right}).   
\xmmn light curves are summed over all EPIC detectors, except for 
0093630101, where only PN and MOS2 detections were used. The 
dashed line shows the minimised $\chi^2$ best fitting constant count rate for
each observation.}
\label{lightcurves}
\end{figure*}

In addition to extracting X-ray spectra, we also extracted background-subtracted light curves for each observation.  A representative selection of these light curves is shown in Fig.~\ref{lightcurves}, covering the differing data qualities, exposure lengths and amounts of variability in the data (see below).  The light curves are binned such that there are $\sim 25$ counts
per element of temporal resolution.

The fractional variability amplitude, $F_{\rm var}$, was used to determine whether statistically significant short term variability was present in the incident count rate of the sources.  Here we define $F_{\rm var}$ as
\begin{equation}
F_{\rm var} = \sqrt{\frac{ \frac{1}{N-1} \sum_{i=1}^{N} (x_i - \bar{x})^2  
- \frac{1}{N} \sum_{i=1}^{N} \sigma^2_{{\rm err,}i}}
{\bar{x}^2}}
\label{fvar}
\end{equation}
as per \citet{vaughan_etal_2003}, where $N$ is the number of temporal bins used, $x_i$ are the individual count rates per bin, $\bar{x}$ is the mean count rate and $\sigma^2_{err,i}$ are the variances on the individual count rates.  Variability was tested for on two separate fixed timescales, 200 s and 2 ks, and the results are shown in Table \ref{luminosity}.  Unfortunately, in most of the 24 observations the low count rates resulted in insufficient counts per temporal bin ($\la 20$), or short exposure times limited the available number of temporal bins (again, $\la 20$), which made such an analysis impractical.  These datasets are shown as the blank spaces in Table~\ref{luminosity}.  In fact, the data was only adequate for these analyses in nine observations at 200 s, and seven at 2 ks.  In determining $F_{\rm var}$ temporally incomplete light curve bins (\eg due to overlapping with the end points of GTIs, or the on/off times of instruments) were excluded so that the errors on the data points were normally distributed.  
Short term variability was detected at greater than $2 \sigma$ significance in observations of four (and at greater than $3 \sigma$ in two) of the eight sources in the sample, most notably from both observations of \srceight~on 200 s timescales.  
The other marginally significant detections and upper limits were of sufficient quality to constrain values that were consistent with similar levels of variability, mainly at $\sim 10$ per cent of the total count rate.

\subsubsection{Long term variability}

In order to study the long term variability of the sources, fluxes
were obtained for each observation, that were converted to luminosities using the distances in Table~\ref{srcs}. Where an 
observation contained sufficient counts to allow spectral fitting, fluxes were
determined using the absorbed power-law continuum spectral fit. For observations
with insufficient counts for spectral fitting, light curves
were $\chi^2$ fitted with a constant count rate, and the resultant best fitting value was converted to a flux using {\sc pimms}\footnote{\tt http://heasarc.nasa.gov/Tools/w3pimms.html}.  A characteristic  spectral index of 
2.0 and an intrinsic absorption column of $1.0\times10^{21}$ ${\rm cm}^{-2}$ 
were assumed, except for observation 10120 of \srcten, where the parameters of the much 
softer disc-like spectrum seen in observation 11230 were used instead.
The observed luminosities are shown in Table \ref{luminosity}, alongside the $F_{\rm var}$ values.  

Most objects display long term luminosity variations at some level between the observations.  In three objects the initial high luminosity that led to them being included in the sample is not maintained in subsequent observations, and appears to drop by factors $> 4$.  Indeed, as a result of spectral fitting, two objects appear to be slightly under-luminous compared to the selection criteria of the sample (\srctwo~\& \srceight), although in both cases their peak luminosity is still considerable at $\sim 4 \times 10^{40} \ergsec$.  In contrast, the most luminous object, \srcseven, remains persistently extremely luminous at $\sim 3 \times 10^{41} \ergsec$ in both observations obtained to date.



\section{Comparing the characteristics of the sample sources to other ULXs}

In the previous sections we have examined the characteristics of a small sample of the most luminous ULXs.  Here, we will attempt to place these results in the wider context of ULX and HLX studies.

\subsection{Spatial properties}

Similarly to other ULXs, all but one of the sample remain unresolved and point-like at the X-ray spatial resolution of the \chan observatory, which is the highest currently available.  In the one case where the original \xmmn source detection was resolved to reveal underlying structure, \srcthree, it resolved into three luminous ULXs, each of which is point-like.  However, the sum of luminosities in the three resolved objects is substantially below the luminosity of the original \xmmn detection. Moreover, by virtue of the short-term variability in the original detection we can conclude that it is likely that just one of the three objects was substantially brighter in the first epoch, such that it was the dominant source of flux in the \xmmn detection.  As it is unclear which of the three subsequently resolved objects was brightest in the original data, we retain all three in this work.  However, for the purposes of constraining the relative changes within this object (e.g. its luminosity with time), we base the second epoch measurement on the brightest of the three resolved objects, \srcthree a, in the remainder of this paper, which effectively provides a lower limit on the long term change in flux.


The locations of the candidate ULXs within their host galaxies may also be used as a comparator.  The objects that lie coincident with face-on spiral galaxies all appear associated with the outer regions of their hosts, close to spiral arm structure.  In three galaxies that have had their H{\sc ii} regions mapped, the ULXs all lie close to known H{\sc ii} regions.  This evidence would seem indicative of these objects being associated with recent star formation; and this appears in common with less luminous ULXs.  Indeed, an association with star formation has long been qualitatively established by the discovery of populations of ULXs in star forming galaxies in the early years of the \chan mission (\eg \citealt{fabbiano_etal_2001}; \citealt{roberts_etal_2002}; \citealt{gao_etal_2003}), 
with subsequent studies quantifying the relationship with the amount of ongoing star formation (\eg \citealt{swartz_etal_2004}; \citealt*{gilfanov_etal_2004}; \citealt{swartz_etal_2009}).  
Furthermore, the quoted separation of the extreme ULXs from known H{\sc ii} regions is of the order $\sim 200$ pc in NGC 470 and NGC 7479, consistent with the separation of ULXs from stellar clusters in the Antennae (\citealt{zezas_etal_2002}) and other nearby starforming galaxies \citep{kaaret_etal_2004}. In the two face-on galaxies with \hst imaging, NGC 2276 and NGC 7479 (both $\sim 33$ Mpc distant, where the $\sim 1$ arcsecond diameter error region equates to $\approx 160$ pc), the objects appear near to likely star forming complexes in the spiral arms.  
All four of these ULXs possess candidate counterparts, consistent with stellar clusters in the three NGC 2276 ULXs (\srcthree a--c), and a possible supergiant stellar counterpart to the NGC 7479 object (\srcten).  Thus the objects remain at the very least associated with a star forming environment, even in \hst imaging.  
This is also a characteristic in common with two of the three best studied HLXs; M82 X-1 is located in the centre of an archetypal starburst galaxy (\eg \citealt{matsumoto_etal_2001}; \citealt{kaaret_etal_2001}), 
and object N10 is in the famous starforming ring of the Cartwheel galaxy \citep{gao_etal_2003}.

\srceight~presents an interesting case; it lies coincident with a dust lane in the edge-on galaxy NGC 5907, and is relatively highly absorbed in the X-rays, consistent with being embedded within the dust lane; yet has a very blue counterpart, that is also relatively luminous with $M_{F450W} = -9.4$.  It is difficult to interpret this as a young stellar cluster or a single giant star as there is no evidence for further star formation in its vicinity; but this object cannot be a background object as the high extinction through the NGC 5907 dust lanes would not permit a background object with blue colours to be visible.  Given the X-ray characteristics of this object discussed later, it would seem to be a plausible ULX; therefore its blue colour and high absolute magnitude suggest its optical emission is dominated by reprocessed X-ray emission from the outer regions of the accretion disc, as suggested for some fainter ULX optical counterparts (e.g. \citealt{Tao_etal_2011}; \citealt{grise_etal_2012}) and ESO 243-49 HLX-1 (\citealt{farrell_etal_2012}; \citealt{soria_etal_2011}), although the high X-ray absorption column remains a problem for this interpretation.


After the exclusion of two of the sample near elliptical galaxies as background QSOs, we are left with \srcseven~alone in this type of habitat.  It is somewhat surprising that we find any HLX in this type of environment, given their lack of star formation.  Indeed, previous results show that the X-ray luminosity functions of elliptical galaxies are steep (\eg complete sample of \citeauthor{walton_etal_2011b} 2011b; also \citealt{kim_and_fabbiano_2010}), 
and \citet*{irwin_etal_2004}
demonstrated that the numbers of ULX candidates with luminosities above $2 \times 10^{39} \ergsec$ found in and around elliptical galaxies were consistent with background expectations, with a couple of notable exceptions.  Hence this detection is quite remarkable, particularly at a luminosity of $\sim 3 \times 10^{41} \ergsec$ which makes it the second most luminous ULX candidate known after ESO 243-49 HLX-1 \citep{farrell_etal_2009}.  We caution that there is a non-negligible chance of it being an unidentified background object.  However, we note that its location -- close to a peculiar S0 galaxy that possesses a dust lane, somewhat like Cen A but with no indication of nuclear activity -- is also interesting, particularly as ESO 243-49 HLX-1 shares a very similar environment \citep{farrell_etal_2012}.  This object therefore deserves further study to elucidate its nature.

\subsection{X-ray characteristics}

\begin{figure*}
\centering
\includegraphics[width=15cm, angle=0]{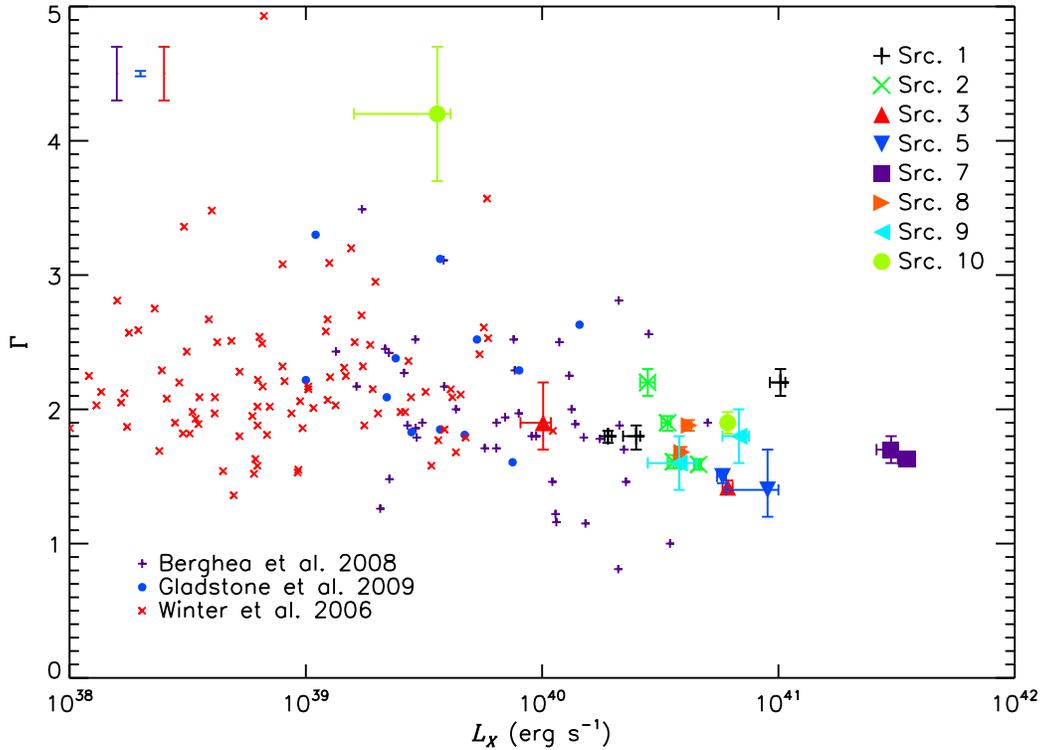}
\caption{The relationship between power-law photon index $\Gamma$ and observed luminosity for ULXs.  Data points from our sample are plotted only where sufficient data was available for spectral fitting, and are plotted against the observed 0.3--10 keV luminosity, calculated using the power-law spectral fit and assuming that sources are at the distance of the identified host galaxy.  All error bars are $1\sigma$ uncertainty regions.  Different sources are identified by the symbols shown in the key (top right).  Three samples of lower luminosity ULXs are included for comparison (smaller data points).  Data are taken from: {\it purple plus sign\/} -- \citet{berghea_etal_2008}, where luminosities were calculated from spectral parameters using a dummy response in {\sc xspec}; 
{\it blue filled circles\/} -- \citet{gladstone_etal_2009}, luminosities were obtained from a coupled disc-corona model (DKBBFTH, \citealt{done_and_kubota_2006}); {\it red crosses\/} -- \citet{winter_etal_2006}, unabsorbed luminosities were converted to observed luminosities using {\sc pimms} and assuming the power-law spectral fit.   Typical error bars for each of the three comparison samples are shown in the same order from left to right in the top left corner.}
\label{L_gamma}
\end{figure*}

In Section~\ref{spec_analysis} we investigated the X-ray spectra of the objects in this sample, finding a general preference for a power-law-like spectrum, which was intrinsically hard.  We place these results in the context of other work on ULXs in Fig.~\ref{L_gamma}, where we compare the photon indexes of the power-law fits to the objects in our sample to three more samples of ULXs fitted with power-law continua, all as a function of luminosity.  The three comparator samples are from: \citet{gladstone_etal_2009}, containing 12 of the best quality ULX datasets in the \xmmn archive, and covering the $10^{39} - 10^{40} \ergsec$ luminosity regime; the larger sample of \chan datasets analysed by \citet{berghea_etal_2008}, ranging from $10^{39}$ to $5 \times 10^{40} \ergsec$ in luminosity; and the large sample of ULXs with \xmmn data presented by \citet{winter_etal_2006}; although after removing their correction for absorption, we find the observed luminosities of many objects in the Winter et al. sample to lie significantly below $10^{39} \ergsec$.  

Fig.~\ref{L_gamma} suggests a hardening of the average photon index with increased observed luminosity; we investigate this further in Table~\ref{wmeans}, where we calculate the weighted mean value for the photon index for various samples shown in Fig.~\ref{L_gamma}.  This clearly shows that the average spectrum of ULXs is harder at higher luminosities, as originally suggested by \citet{berghea_etal_2008}.  Indeed, the effect appears quite pronounced; the two samples composed of sources below $10^{40} \ergsec$ (\citealt{gladstone_etal_2009}, and the less luminous \citealt{berghea_etal_2008} objects) have weighted mean photon indexes $\langle\Gamma\rangle > 2$, whereas the more luminous samples are significantly harder with $\langle\Gamma\rangle \sim 1.6$.  This is clear evidence of a difference in the properties of the most luminous ULXs from the more ordinary ULX population.  Interestingly, similarly hard spectral indexes have been seen in two well-studied HLXs: M82 X-1 with \eg $\Gamma = 1.55 \pm 0.05$ \citep{kaaret_etal_2009}, and Cartwheel N10 with $\Gamma = 1.75 \pm 0.25$ \citep*{wolter_etal_2006}.  However the most luminous ULX, ESO 243-49 HLX-1, appears somewhat different; in its brightest phases it is distinctly soft, with \eg $\Gamma = 3.4 \pm 0.2$ \citep{godet_etal_2009}\footnote{The best quality X-ray spectral data currently available for ESO 243-49 HLX-1 at its peak luminosities show its spectrum to be dominated by a cool ($kT_{\rm in} \leq 0.26$ keV) disc component, see \cite{servillat_etal_2011}.}, although it hardens significantly to $\Gamma \sim 2$ when at fainter fluxes (\citealt{farrell_etal_2011}; \citealt{servillat_etal_2011}), such that it appears similar to the sources presented here (see section \ref{imbh}).

\begin{table}
\caption{Mean photon index for various ULX samples}
\begin{center}
\begin{tabular}{lc}
\hline
Sample	& $\langle\Gamma\rangle$ \\
\hline
\citeauthor{gladstone_etal_2009}	& $2.19 \pm 0.01$ \\
\citeauthor{berghea_etal_2008}, $< 10^{40} \ergsec$	& $2.10 \pm 0.07$ \\
\citeauthor{berghea_etal_2008}, $> 10^{40} \ergsec$	& $1.54 \pm 0.06$ \\
This sample (all)	& $1.72 \pm 0.07$ \\
This sample, $> 5 \times 10^{40} \ergsec$	& $1.61 \pm 0.07$\\
\hline
\end{tabular}
\label{wmeans}
\end{center}
\end{table}

\begin{figure*}
\centering
\includegraphics[width=15cm, angle=0]{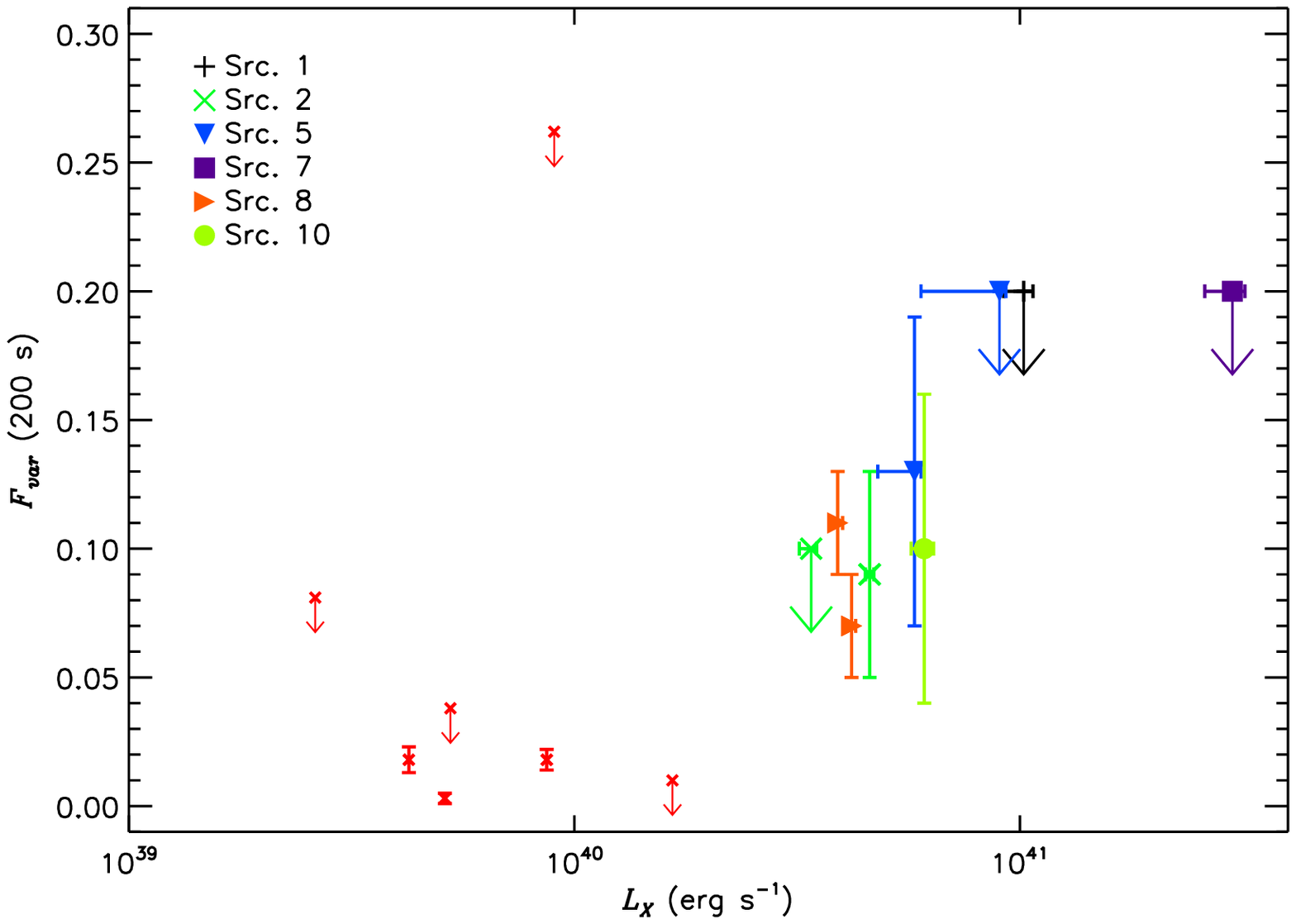}
\includegraphics[width=15cm, angle=0]{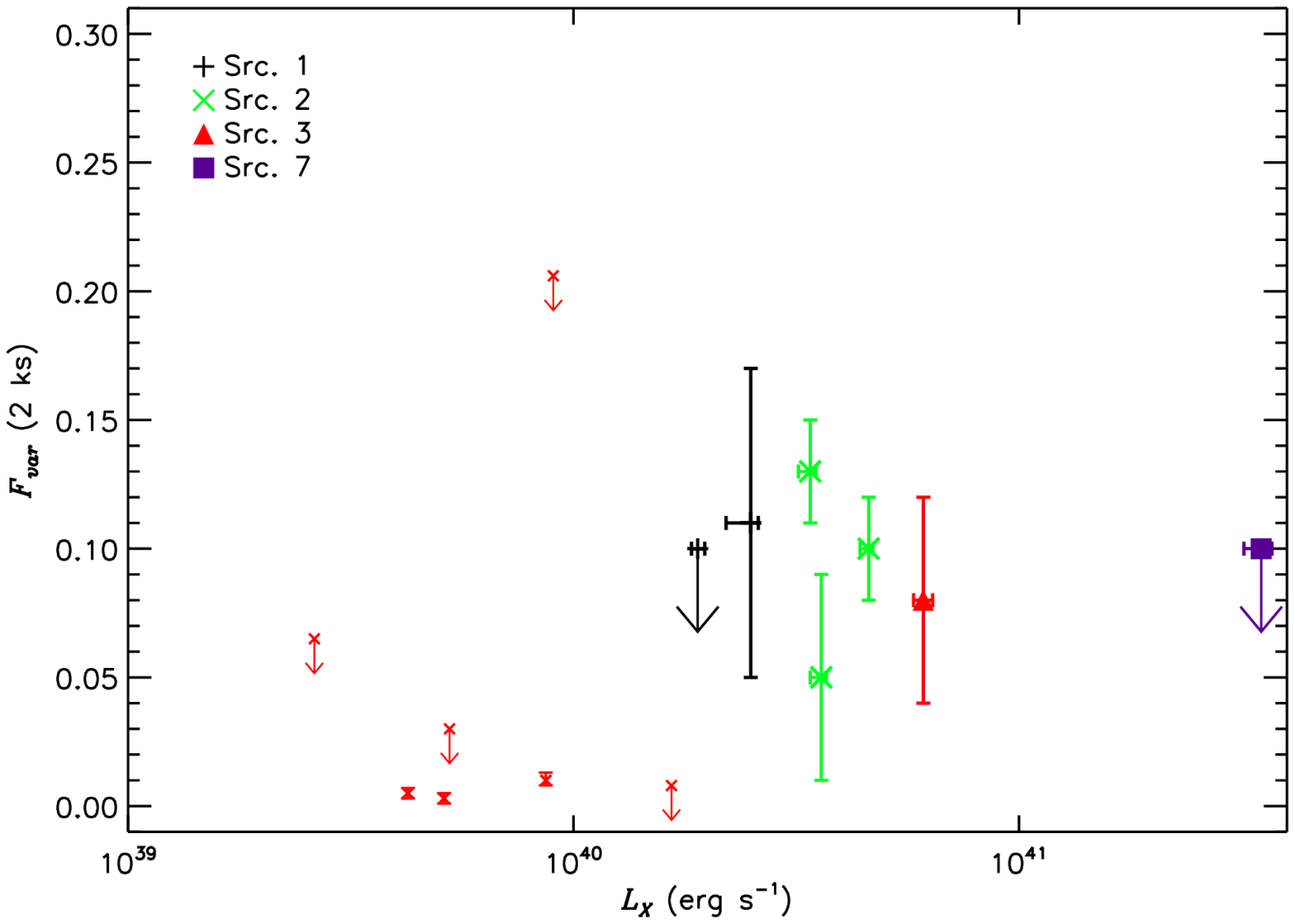}
\caption{Fractional variability amplitude ($F_{\rm var}$) shown as a function of 0.3--10 keV luminosity.  This is plotted for time resolutions of 200 s ({\it top}) and 2 ks ({\it bottom}). The upper limits shown are $2\sigma$ significance levels in all cases, and error bars are $1\sigma$ confidence regions.  Each object from the luminous ULX sample is labelled by a different symbol; a key is provided in the top left of each panel.  Lower luminosity ULXs (small crosses) are shown for comparison, these data points were obtained by the integration of fits to the power spectra for a selection of sources from \citet{heil_etal_2009}.  Count rates from \citet{gladstone_etal_2009} for the same observations were used to calculate $F_{\rm var}$. The comparison sample includes (from left to right) - NGC 2403 X-1, NGC 1313 X-1, NGC 1313 X-2, NGC 5204 X-1, Ho IX X-1, NGC 4559 X-1 and Ho II X-1.}
\label{LFvar}
\end{figure*}

We also previously examined evidence for intra-observation variability in our objects.  Here, we compare the results to well-constrained variability measurements from less-luminous ULXs.  Specifically, in Fig.~\ref{LFvar} we compare the fractional variability over the 0.3--10 keV range for our sample of luminous ULX candidates to the fractional variability of a selection of lower luminosity ULXs, on time scales of both 200 s and 2 ks.  The comparison sample were selected for having both a well constrained power spectrum in \citet{heil_etal_2009}, and a well-constrained flux by \citet{gladstone_etal_2009}.  Their fractional variabilities were calculated by integrating the power-law and broken power-law power spectra from \citet{heil_etal_2009}, and normalising by flux using the values from \citet{gladstone_etal_2009}.  As Fig.~\ref{LFvar} demonstrates, the variability levels of the less luminous ULXs are low (typically 2 per cent fractional variability or less where well constrained), consistent with many reports of low short-term variability levels in ULXs (\eg \citealt{swartz_etal_2004}; \citealt{feng_and_kaaret_2006}; \citealt{heil_etal_2009}).  
In contrast, the measurements and constraints on the higher luminosity objects are consistent with higher levels of variability, $\sim 10$ per cent or more fractional variability in most cases, although most error bars are relatively large due to the poorer data quality in these objects.  This, then, tentatively points to a second observational difference between `normal' ULXs and their higher luminosity cousins; more short-term variability in the more luminous objects.

\begin{figure*}
\centering
\includegraphics[width=15cm, angle=0]{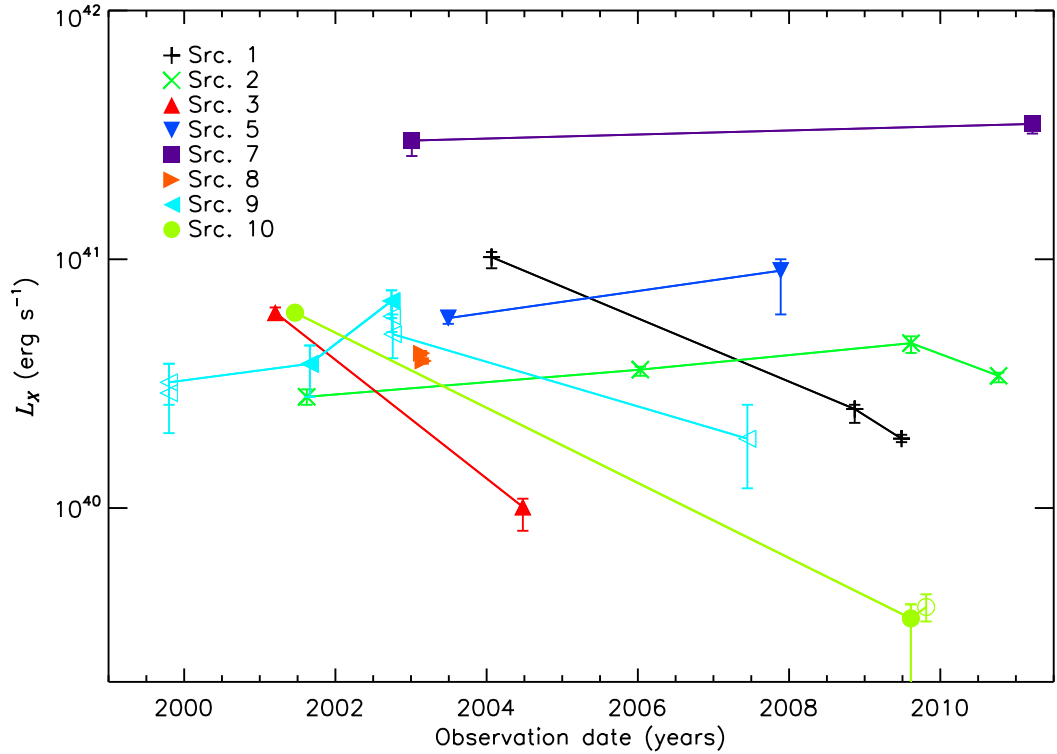}
\caption{Long term light curves of the objects in our sample.  We show the observed 0.3--10 keV  luminosity, plotted against observation date. Filled symbols correspond to observations with
greater than 100 counts, where the fitted absorbed power-law models were used to calculate
luminosity.  Unfilled symbols correspond to observations with insufficient counts for spectral fitting, where instead spectral parameters were assumed as per Section 3.4. The low luminosity data point for \srcthree~corresponds to the most luminous of the resolved sources (\srcthree a) in the \chan observation.}
\label{date_L}
\end{figure*}

In Fig.~\ref{date_L} we show the changes in the luminosity of the objects with time.  It is immediately obvious the data is extremely sparse, with no more than 2 -- 6 measurements per object over a baseline of up to a decade; hence any conclusions made on their long-term behaviour must be very tentative as we simply cannot tell how the objects behaved when they were not being observed.  With this caveat in mind, we can tentatively sort the objects into two groups: firstly three objects that appear to fade substantially from an initial peak in luminosity, down to more `usual' ULX luminosities ($\la 2 \times 10^{40} \ergsec$).  This includes the only HLX candidate with a spiral host, \srcone~in NGC 470, the \xmmn detection subsequently resolved by {\it Chandra\/}, \srcthree, and \srcten~(which we discuss further below).  The remaining five objects all sustain their extreme luminosities, within a factor $\sim 2$, most notably \srcseven~which is detected at a luminosity $\sim 3 \times 10^{41} \ergsec$ in both epochs.  Given the sparse data available from our sample, and the lack of extensive monitoring of less luminous ULXs, we are also able to do little to compare the long-term variability properties of our objects to other ULXs.  Recently {\it Swift\/} monitoring programmes have been undertaken for a small handful of objects, that appear persistently bright with typical flux variations of factors up to $\sim 5$ over the course of the monitoring campaigns (\eg \citealt*{kaaret_and_feng_2009}; \citealt{grise_etal_2010}).  
The HLX candidate M82 X-1 also appears to vary similarly in {\it RXTE\/} monitoring, although it also displays a periodic signal on 62 day timescales \citep{kaaret_and_feng_2007}.  
The brightest HLX, ESO 243-49 HLX-1, again appears different -- its variability resembles an outburst cycle in an ordinary X-ray binary, with a fast rise to peak luminosities followed by a slow decay in flux, and a $375 \pm 13$ day gap between two outburst peaks (\citealt{lasota_etal_2011}; \citealt{servillat_etal_2011}).




Further inspection of Fig.~\ref{L_gamma} shows that several of the objects display significant X-ray spectral variability, in this case represented by a change in photon index at different flux levels (or, indeed, at different observation epochs with similar flux levels, as is the case for \srceight).  However, these changes are not consistent across the small sample, with some objects hardening with increased flux (\eg \srctwo), and others softening (\eg \srcone).  This at least is consistent with the varied picture seen across the population of lower luminosity ULXs, where many different behaviours are observed with changing luminosity (\eg \citealt{feng_and_kaaret_2009}; \citealt{kajava_and_poutanen_2009}; \citealt{grise_etal_2010}; \citealt{vierdayanti_etal_2010}; \citealt{servillat_etal_2011}).  
Perhaps more revealing are the situations in which the preferred spectral model changes with luminosity, as they may indicate direct analogies with known spectral state transitions.  We discuss this in the following section.

\section{Discussion: on the nature of the most extreme ULXs}\label{discussion}



This work has highlighted that there are possible differences between the standard population of lower luminosity ULXs, and those most luminous objects that we study here.  Most notably, the brightest objects are spectrally harder, well described by power-law continua with photon indexes $\Gamma \sim 1.7$, and there is tentative evidence they display higher amplitude flux variability, with typical $F_{\rm var}$ of $\sim 10$ per cent on timescales of 200 - 2000 seconds.  But what does this mean physically for the objects in our sample?

\subsection{Intermediate-mass black holes?}\label{imbh}

Perhaps the most interesting question we can ask is: do these objects finally offer evidence for a population of IMBHs underlying some ULXs?  As discussed earlier, most super-Eddington models for stellar-mass black holes struggle to explain the high luminosities of this sample (\eg \citealt{zampieri_and_roberts_2009}), and there are suggestions that a different class of objects is required to explain sources above the X-ray luminosity function break \citep{swartz_etal_2011}.  It is therefore quite a straightforward step to infer that we may be looking at IMBHs.

Crucially, this inference does appear to be supported by the observational characteristics of the sample.  In particular, the X-ray spectra and timing properties are consistent with the hard (formerly low/hard) state as described by \citet{mcclintock_and_remillard_2006}, sharing both a power-law spectrum (with $\Gamma \sim 1.7$) and a fractional variability level consistent with $\sim 10$ per cent.  Given that the hard state is typical only found at Eddington fractions of 10 per cent or less, this immediately constrains the masses of our black holes to be in the IMBH regime.  We calculate individual lower limits for each object, based on the highest observed luminosity with a power-law-like spectrum and the 10 per cent $L_{\rm Edd}$ limit, in Table~\ref{mass_limits} as mass $M_1$.  This implies black hole mass lower limits of $1.9 - 27 \times 10^3~\Msun$ across our sample.

One issue with the timing result is that \citet{mcclintock_and_remillard_2006} note that 10 -- 30 per cent fractional variability is expected in the 0.1 - 10 Hz range in the hard state, whereas we only probe frequencies as high as $\sim 0.1$ Hz in the best of our data.  However, if we are dealing with large black holes we would expect their variability timescales to increase proportionally to their mass for the same accretion rate \citep{mchardy_etal_2006}; 
hence a $1000~\Msun$ black hole would vary on 100 times longer timescales than a $10~\Msun$ black hole.  So, the 0.1 -- 10 Hz section of the PSD for a stellar mass black hole scales to the $10^{-3} - 0.1$ Hz regime, that we probe here, for a $\sim 1000~\Msun$ IMBH.

An important test of whether these objects really are in the hard state would be to observe them transit into other states that are also seen at sub-Eddington luminosities in Galactic binaries.  Such transitions have been suggested for ESO 243-49 HLX-1 \citep{servillat_etal_2011} and M82 X-1 \citep{feng_and_kaaret_2010}, with in both cases the high luminosities at which these transitions occur providing a strong argument for IMBHs.  Interestingly, ESO 243-49 HLX-1 appears to possess a hard power-law X-ray spectrum at similar luminosities to the objects in this sample.  We potentially see a similar transition in one case: \srcone~is far better described by a MCD, representative of the thermal dominant state, than a power-law at its highest luminosity ($\sim 10^{41} \ergsec$), and vice versa in subsequent lower luminosity datasets ($\le 2.5 \times 10^{40} \ergsec$).  Unfortunately, the timing data is of insufficient quality to confirm these states (with the little or no variability expected in the thermal dominant state potentially providing a distinction).  However, the temperature of the disc appears highly anomalous for an IMBH, at $\sim 1$ keV, making the identification of a classical transition for an IMBH very dubious in this case.

One prediction that we can make is that if these are IMBHs in the hard state, then they should be emitting a steady radio jet.  If so, the peak observed X-ray luminosity can be used in conjunction with the radio luminosity to provide limits on the black hole mass via the so-called
X-ray-radio fundamental plane \citep*{merloni_etal_2003}.  We have attempted to do this using the currently available data.  To this end, the {{\it VLA~\/}} FIRST \citep*{becker_etal_1995} and NVSS \citep{condon_etal_1998} 1.4 GHz radio surveys were searched at the source positions for any radio counterparts.  However, none were detected, so the radio catalogue  completeness limits were used as an upper-limit to the radio flux instead. A flat radio spectrum was assumed to extrapolate the radio flux from the observed band to 5 GHz, and 2-10 keV X-ray fluxes were extracted from the highest 0.3-10 keV flux of each source using {\sc pimms} and spectral parameters from the appropriate power-law fit.  Before presenting the limits we highlight two important caveats: firstly that the fundamental plane is not calibrated for IMBHs/ULXs, which may lead to some uncertainty; and secondly we are severely 
limited in what conclusions we can draw by the lack of simultaneous X-ray and radio observations.

Nevertheless, we present the upper limits on the IMBH masses in Table~\ref{mass_limits} as $M_2$.  For a number of the sources the fundamental plane estimated limits on black hole mass are several orders of magnitude greater than the hard state mass limits, and in the SMBH mass regime. For these sources large improvements in radio detection limits would be necessary to use the X-ray-radio fundamental plane to test for IMBHs. However, for the closest source (\srceight) the upper limit is already in the IMBH regime, and only $\sim$ an order of magnitude above the lower limit from the hard state identification.  For two more objects (\srctwo~and \srcfive) a modest improvement of less than an order of magnitude would begin to constrain the mass within the IMBH regime.

However, there are problems with interpreting these objects as IMBHs.  One such issue is the location of the objects, the majority of which are found in the same star forming environments as lower-luminosity ULXs.  While the necessity to form only one such object in any galaxy neatly sidesteps the strong arguments for the bulk of the populations of ULXs in these regions having to be predominantly stellar-mass objects (\citealt{king_2004}; \citealt{mapelli_etal_2008}), one still has to both form these very massive objects, and then capture a high mass star, in order to appear as a ULX.  Even so, processes such as the assembly and subsequent collapse of a very massive star at the centre of a young, massive star cluster have been suggested as viable IMBH formation channels in star forming regions (\eg \citealt{portegies_zwart_etal_2004}), and may provide a natural environment in which to capture a star to feed the IMBH \citep*{hopman_etal_2004}.  However, in this case the lack of particularly bright and massive stellar clusters coincident with the ULXs may be an acute problem, 
even more so than the lack of such coincidences noted for less luminous ULXs (\eg \citealt{zezas_etal_2002}).  Alternatively, if the IMBHs are not newly formed, the small number of such objects we observe may be commensurate with the predicted difficulties of capturing a young star into a close orbit by an IMBH \citep{madhusudhan_etal_2006}, presuming such objects could find their way into a star forming region.  

For the one object not readily related to star formation, \srcseven~in the S0 galaxy IC 4320, assuming the identification as a hard state binary is correct then an older IMBH must be required.  In this case the possibilities include a primordial Population III star remnant (\eg \citealt{madau_and_rees_2001}), or the stripped nucleus of an accreted dwarf galaxy (\eg \citealt{king_and_dehen_2005}; \citealt{Bellovary_etal_2010}), although the requirement to accrete from a massive stellar reservoir likely remains.  Again, we note the only established HLX with a luminosity in excess of \srcseven, ESO 243-49 HLX-1, is also associated with an S0 host \citep{farrell_etal_2012}.

\begin{table}
\caption{Limits on the black hole mass for an IMBH from hard state Eddington arguments and the 
X-ray - radio fundamental plane.}
\begin{center}
\begin{tabular}{lcc}
\hline
Source ID & $M_{1}/$\Msun$^a$ & $M_2/$\Msun$^b$
\\
\hline
\srcone & $\ga 1900$ & $\la 2 \times 10^6$
\\
\srctwo & $\ga 3500$ & $\la 2 \times 10^5$ 
\\
\srcthree & $\ga 4700$ & $\la 2 \times 10^6$
\\
\srcfive & $\ga 6900$ & $\la 5 \times 10^5$
\\
\srcseven & $\ga 2.7 \times 10^4$ & $\la 2 \times 10^8$
\\
\srceight & $\ga 3200$ & $\la 3 \times 10^4$
\\
\srcnine & $\ga 5200$ & $\la 3 \times 10^7$
\\
\srcten & $\ga 4700$ & $\la 4 \times 10^6$
\\
\hline
\end{tabular}
\end{center}
\begin{minipage}{\linewidth}
Notes:
$^a$Black hole mass limit from Eddington arguments, assuming
that the source is in the hard state and accreting at $\le 10$ per cent of the
Eddington rate when at maximum luminosity with a power-law spectrum.  For seven of eight objects this is the highest luminosity seen; for \srcone~its highest luminosity spectrum is far better described by a MCD, so we use the second highest luminosity dataset for this object.  
$^b$Black hole mass derived from the X-ray-radio fundamental plane.
\end{minipage}
\label{mass_limits}
\end{table}


\subsection{The most extreme super-Eddington objects?}

While there are possible means of explaining the most luminous ULXs in star forming regions as IMBHs, as discussed above, it is incontrovertible that this is the same environment as that in which the majority of lower luminosity ULXs are found.  We should therefore explore whether it is possible that these most luminous objects are driven by a more extreme version of the super-Eddington processes now thought to power the majority of the lower luminosity ULXs.

There are possible scenarios in which we might explain most of these luminous objects by super-Eddington emission.  Firstly, we note that \citet{zampieri_and_roberts_2009} argue that the black holes underlying some ULXs in low metallicity regions may be formed from the collapse of a single low-metallicity star which, by virtue of reduced wind-driven mass loss during its life time, can leave a black hole remnant as large as $\sim 90~\Msun$ (see also \eg \citealt{belczynski_etal_2010} 
for mass transfer binaries).  A combination of this high mass and similar maximal Eddington rates as are inferred in some stellar mass ULX models could raise the maximum luminosity for stellar remnant ULX models to $\sim 10^{41} \ergsec$.  In support, we note that the apparent locations of many of the ULXs studied here, on the outskirts of their host galaxy (cf. Fig.~\ref{DSSimgs}), may indeed be consistent with the lower metallicity regions required to form such objects.  Secondly, \citet{king_2009} suggests that the degree of beaming due to geometric collimation by the inflated inner disc and outflowing wind at high accretion rates, that constrains the escape path of the X-rays from the central region, scales directly proportionally to the accretion rate.  Hence, extreme luminosities may be explained by a high beaming factor (which may also help explain the rarity of such objects, as significant numbers could be present, but would not be visible outside their narrow beam).

However, both these scenarios would still have to explain why the X-ray characteristics of these ULXs appear so similar to the very sub-Eddington hard state, and not akin to the ultraluminous state spectra described by \citet{gladstone_etal_2009}.  One possibility is the low quality of the data in most cases masks both the soft excess and the subtler high energy break of the ultraluminous state, and the underlying spectra are similar to those observed in Ho IX X-1 or NGC 1313 X-1 by \citeauthor{gladstone_etal_2009}, where the best fitting power-law continua have indexes $\sim 1.6 - 1.85$.  However, these observations both have particularly low fractional variability (cf. Fig.~\ref{LFvar}), unlike the objects in the higher luminosity sample, which appear variable on short timescales (albeit with large errors, or only upper limits in some cases).  One way around this would be if the higher luminosity objects are more prone to displaying extrinsic variability.  This is suggested by \citet{middleton_etal_2011a} to explain the variability of NGC 5408 X-1, where it could originate in blobs of outflowing material driven off the surface of the accretion disc by the extreme radiation pressures from its central regions, that cross the line-of-sight to those regions.  Better data is required to resolve these issues; but it is notable that in the one dataset within this sample with reasonably good data quality (\srceight~in NGC 5907) an ultraluminous state spectrum is observed.  However, we note that the distance to NGC 5907 is uncertain, and recent estimates (\eg $d = 13.4$ Mpc by \citealt{tully_etal_2009}) place the host galaxy near enough that the luminosity of \srceight~is reduced by a further 20 per cent, such that it is consistent with the top end of the `normal' ULX population.  It is therefore questionable whether it should be included within this sample.  Even so, its hard power-law-like spectrum and detected fractional variability provide important testimony that an ultraluminous state object can possess similar observational characteristics to the sub-Eddington hard state in moderate quality data.

One further mass limit could be placed assuming that the sample contains high accretion rate objects.  Both \citet{gladstone_and_roberts_2009} and \citet{yoshida_etal_2010} note that the low luminosity ULXs in NGC 4490 and (in the latter paper) M51 appear to transit between a power-law spectral state at $\sim 10^{39} \ergsec$ to a MCD-like model at slightly higher luminosities.  \citeauthor{gladstone_and_roberts_2009} suggest this could be a transition from a sub-Eddington steep power-law state to a $\sim$ Eddington rate state; if so the highest luminosity in which a power-law spectrum is observed provides an estimate of the black hole mass from the requirement $L_{\rm X} / L_{\rm Edd} < 1$.  However, without seeing the transition luminosity to a MCD spectrum itself this measurement does not provide strong constraints (as with further observations we might see the power-law at higher luminosities); for the sources in the sample this estimated upper limit is in the range $\sim 200$--$2700~\Msun$, with the only possible constrained limit (from \srcone) as $\le 800~\Msun$ from the observed luminosity of the observation where a MCD fit is much preferred.


\subsection{Contaminants?}

A final point to consider is whether any more of our objects could be background (or, plausibly, foreground) contaminants.  A calculation based on that performed by \citeauthor{walton_etal_2011b} (2011b) was used to establish the expected contamination of this sample by background objects; this found that three were expected out of the ten selected sources with luminosities above $5 \times 10^{40} \ergsec$ within 100 Mpc.  In the analysis we were able to identify two background QSOs near elliptical galaxies, including one new $z = 3.25$ QSO near NGC 4874.  The calculation predicts one more background object, likely to be associated with a spiral galaxy, but no such object has been identified at present.  We note that we would be unlikely to identify such an object from its X-ray properties alone.  In Table A.1 of Appendix A we present X-ray spectral fits to the QSO datasets, using the same models as for the ULXs (cf. Table~\ref{powerlaw}).  Although \srcfour~does appear somewhat softer than the luminous ULXs, errors on its spectral parameters are large.  Large errors also dominate the measurements for \srcsix, and we were not able to derive short-term variability characteristics for either source due to the paucity of data.  We would not, therefore, have been able to distinguish these objects from ULXs without the optical follow-up data.  We therefore caution that without further optical follow-up (which, as the case of \srcsix~shows, must go several magnitudes deeper than the DSS) we cannot rule out a QSO nature for the several ULXs currently lacking follow-up data.

Further possible contaminants include foreground objects and other rare types of objects within the host galaxies.  We can at least rule out foreground stars as none are evident in the DSS images in Fig~\ref{DSSimgs}.  However, one possibility that should be mentioned is that of recent supernovae in the host galaxies, which are not accounted for in the contamination calculation of \citeauthor{walton_etal_2011b} (2011b).  Such supernovae have been seen with typical X-ray luminosities 
in the range of $10^{37}$--$10^{41}~{\rm erg~s^{-1}}$ and fade over a variety of timescales, \eg SN 1988Z decreased by $\sim 5$ over 2 years, whilst SN 1980K decreased by a similar factor between days 35 and 82 after outburst \citep{immler_and_lewin_2003}.  Interestingly, the X-ray spectra of some objects have been seen to soften as the flux decreases, for example in SN 1999em \citep{pooley_etal_2002}, where the early X-rays are thought to originate in the circumstellar gas shocked by the initial blast wave, whereas the later soft emission comes from the lower temperature reverse shock region (see \citealt*{fransson_etal_1996} for details). We do see several objects in the sample whose X-ray emission appears to fade in the years following their initial detection, including one object (\srcten) that also softens dramatically.  However, each of these objects shows some degree of fractional variability within at least one observation, consistent with a compact object and rejecting the recent supernova hypothesis.  This, and the lack of any identified optical supernovae associated with the ULXs in this sample, argues that we do not have any such objects amongst our sources.

A second rare class of objects that may be associated with the host galaxies are recoiling SMBHs, as has been suggested for the HLX candidate CXO J122518.6+144545 \citep{jonker_etal_2010}. Such objects are predicted to form via the binary coalescence of SMBHs during galaxy mergers, where anisotropic gravitational wave emission imparts a kick of up to several thousand km s$^{-1}$ on the new, single SMBH (see \citealt{Komossa_2012} and references therein). This kick is sufficient to displace the SMBH from the centre of its host galaxy, and in some cases even expel it completely.  Such objects might appear both X-ray and optically luminous, as they are predicted to retain a reservoir of gas and stars to accrete from; a key indicator of their SMBH nature would then be broad emission lines in their optical spectra.  However, we again regard the chances of detecting such an object in our sample to be low; none of our spiral galaxies are undergoing (or appear to have recently undergone) the major merger necessary to create such an object.  Interestingly, the only early-type galaxy in our sample, IC 4320, possesses a dust lane, a characteristic that has been linked to minor galaxy mergers in such systems \citep{shabala_etal_2011}.  It also hosts the most luminous ULX in our sample.  The detection and characterisation of the optical counterpart to this object is therefore of great interest.

\section{Conclusions}

Even in the absence of definitive black hole mass measurements for ULXs, various arguments have resulted in the wide acknowledgement that the majority of these extraordinary objects are most probably powered by super-Eddington accretion onto stellar remnant black holes. However, the luminosities that can be produced in this manner are limited by both a maximum Eddington ratio and maximum stellar remnant black hole mass, which complicates the unification of the brightest ULXs with the less luminous population. These particularly luminous, extreme ULXs may instead remain as potential IMBHs, with their rarity consistent with arguments as to the difficulties in producing stable high accretion rates in many such objects.

Here we have presented a study of 10 candidate extreme ULXs.  We were able to identify 2 as background QSOs, resulting in a sample of 8 good candidates for further study.  By examining the X-ray emission characteristics of this sample we note that these extreme sources do indeed appear to be distinctly different from the lower luminosity ULXs.  Not only are they typically spectrally harder, they are also consistent with being more variable (albeit at low statistical significance in many objects). This combination of spectral and timing characteristics is highly suggestive of the sub-Eddington hard state, which given the luminosities involved would strongly argue for large ($10^3$--$10^4$~\Msun) IMBHs.  This is however not entirely conclusive, as accretion at the highest plausible super-Eddington ratios onto the most massive stellar remnant black holes could potentially produce ULXs up to $\sim 10^{41} \ergsec$.  Further high quality X-ray observations are therefore required to test this, with the capability to detect ultraluminous state-like signatures, especially for the sources that lie in the transition zone between the brightest local ULXs ($\sim 3 \times 10^{40} \ergsec$) and the HLX luminosity range.

The sample of 8 candidate ULXs includes 2 candidate HLXs.  One is a transient HLX, that subsequently fades to high ULX luminosities; but the other appears persistent between two snapshots separated by 8 years, at a luminosity that is second only to ESO 243-49 HLX-1 amongst the known HLXs.  These join the very small number of HLXs identified in the literature. As they are particularly difficult to explain as stellar remnants, they must be regarded as some of the best known IMBH candidates.  Given the cosmological and astrophysical importance of IMBHs, HLXs are highly compelling targets for further observations with current and future missions, to both better constrain the presence of a massive black hole and better understand the astrophysics of these remarkable objects.




\section*{Acknowledgements}

The authors thank the anonymous referee for their careful reading of the paper, and useful suggestions for improvements to the text.
ADS, DJW and AES gratefully acknowledge funding from the Science and 
Technology Facilities Council in the form of PhD studentships, and TPR 
in the form of a standard grant. JCG thanks the Avadh Bhatia Fellowship and Alberta Ingenuity.
We thank Sean Farrell and Russell Smith for useful discussions and contributions whilst 
producing this paper.  This work is based on observations obtained with {\it XMM-Newton\/}, an ESA science 
mission with instruments and contributions directly funded by ESA Member 
States and NASA; and the \chan X-ray observatory.
It has also included observations made with the NASA/ESA Hubble Space 
Telescope, and obtained from the Hubble Legacy Archive, which is a 
collaboration between the Space Telescope Science Institute 
(STScI/NASA), the Space Telescope European Coordinating Facility 
(ST-ECF/ESA) and the Canadian Astronomy Data Centre (CADC/NRC/CSA); the 
Digitized Sky Surveys, which were produced at the Space Telescope 
Science Institute under U.S. Government grant NAG W-2166; and 
observations obtained at the Gemini Observatory, which is operated by 
the Association of Universities for Research in Astronomy, Inc., under a 
cooperative agreement with the NSF on behalf of the Gemini partnership.

\bibliography{bibtex_refs_2}
\bibliographystyle{mn2e}

\appendix

\section[]{The two identified background contaminants}

\begin{figure*}
\centering
\includegraphics[width=15cm, angle=0]{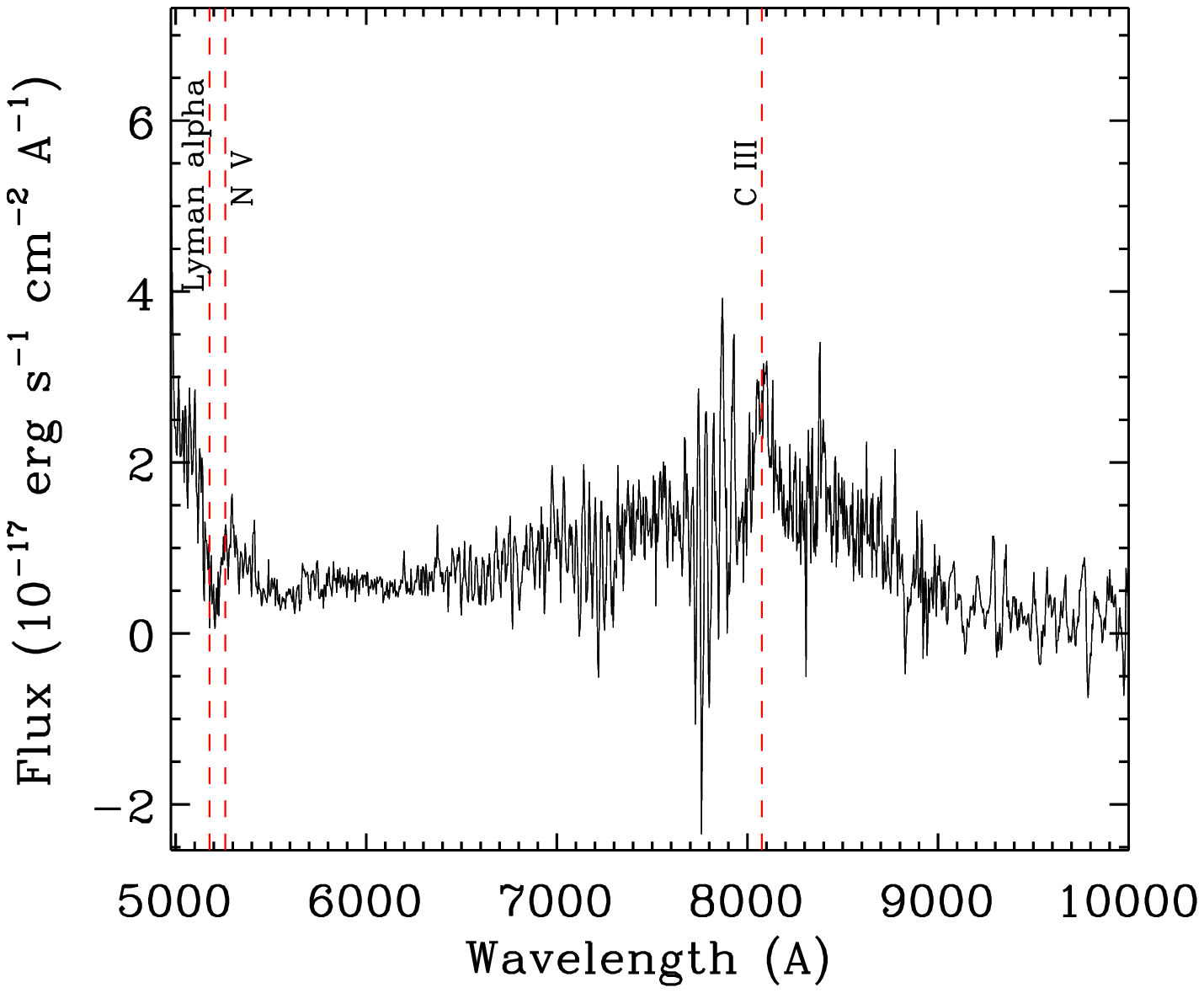}
\caption{Gemini GMOS-N longslit spectrum of the optical counterpart to Src. 6, taken using the R150 grating and the 0.75 arcsecond slit.  It was suggested that this source may be a bright ULX in a UCD in the Coma cluster, however lines identified in this spectrum (shown by dashed red lines; 
see text for details) instead identify it as a QSO at redshift $\approx$ 3.25.}
\label{GMOSspec}
\end{figure*}

\begin{table*}
\caption{X-ray spectral fitting results for the background contaminants}
\begin{center}
\begin{tabular}{cccccccccc}
\hline
& & \multicolumn{3}{c}{Absorbed power-law} & \multicolumn{3}{c}{Absorbed 
multi-colour-disc}\\
Source ID & Obs ID & ${N_H} ^a$ & $\Gamma ^b$ & $\chi ^2 / \rm{dof}$ & 
${N_H} ^a$ & $kT_{\rm{in}} ^c$ & $\chi ^2 / \rm{dof}$ & ${
L_{\rm ULX}}^d$ & ${L_{\rm QSO}}^e$ \\
\hline
\srcfour & 0112270601 & - & - & - & - & - & - & $7.6 \pm 0.9 ^f$ & $1.2 
\pm 0.2$\\
         & 0112271001 & $<0.2$ & $2.4^{+0.6}_{-0.5}$ & 41\%$^g$ & 
$<0.04$ & $0.6^{+0.2}_{-0.1}$ & 88\%$^g$ & $12 \pm 3$ & $1.9 \pm 0.5$\\
         & 0112271101 & $<0.1$ & $2.5 ^{+0.5} _{-0.3}$ & 62\%$^g$ & 
$<0.02$ & $0.43 ^{+0.09} _{-0.07}$ & 88\%$^g$ & $12^{+4}_{-2}$ & 
$1.9^{+0.6}_{-0.3}$\\
         & 12990 & - & - & - & - & - & - & $7 \pm 1 ^f$ & $1.1 \pm 0.2$\\
\srcsix  & 556 & - & - & - & - & - & - & $4.5 \pm 0.8 ^h$ & $37 \pm 7$\\
         & 0124711401 & $<0.3$ & $2.3^{+3}_{-0.6}$ & 23.9/32 & $<1$ & 
$<0.4$ & 27.4/32 & $<9$ & $<70$\\
         & 0153750101 & $<0.1$ & $1.4^{+0.4}_{-0.3}$ & 41.3/42 & $<0.1$ 
& $1.3^{+0.8}_{-0.4}$ & 39.9/42 & $11^{+2}_{-4}$ & $90^{+20}_{-30}$\\
         & 0300530701 & $<0.4$ & $1.7^{+1}_{-0.5}$ & 42.0/45 & $<0.2$ & 
$1.1^{+0.8}_{-0.5}$ & 41.3/45 & $<10$ & $<80$\\
         & 0300530601 & $<0.2$ & $1.6^{+0.6}_{-0.4}$ & 36.6/31 & $<0.1$ 
& $1.1^{+0.8}_{-0.4}$ & 36.6/31 & $<20$ & $< 200$\\
         & 0300530501 & $<0.3$ & $2.3^{+2}_{-0.8}$ & 45.7/44 &  $<0.2$ & 
$0.5^{+0.3}_{-0.2}$ & 45.6/44 & $7^{+2}_{-3}$ & $60 \pm 20$\\
         & 0300530301 & $<4$ & $2.0^{+4}_{-0.7}$ & 45\%$^g$ & $<1$ & 
$1.0^{+0.6}_{-0.5}$ & 35\%$^g$ & $<120$ & $< 1000$\\
         & 0300530101 & $<0.2$ & $1.5^{+0.8}_{-0.4}$ & 33.7/41 & $<0.1$ 
& $1.3^{+2}_{-0.8}$ & 36.4/41 & $10^{+2}_{-5}$ & $80^{+20}_{-40}$\\
         & 9714 & - & - & - & - & - & - & $1.9 \pm 0.4 ^h$ & $15 \pm 3$\\
         & 10672 & $<0.2$ & $2.2 ^{+0.9} _{-0.5}$ & 46\%$^g$ & $<0.06$ & 
$0.6 ^{+0.3} _{-0.2}$ & 75\%$^g$ & $2.9^{+0.6}_{-0.8}$ & $24^{+5}_{-7}$\\
\hline

\end{tabular}
\end{center}
\begin{minipage}{\linewidth}
Notes:
$^a$Absorption column density external to our own Galaxy ($\times 10^{22}~\rm{cm}^{-2}$).
$^b$Power-law photon index.
$^c$Inner-disc temperature (keV).
$^d$0.3--10 keV luminosity ($\times 10^{40} \ergsec$), estimated from a 
power-law spectral fit, and assuming the object is at the distance of the foreground galaxy.
$^e$0.3--10 keV luminosity ($\times 10^{44} \ergsec$), estimated from a 
power-law spectral fit, and assuming the object is at the distance derived from its optical spectral redshift measurement.
$^f$This observation had insufficient counts for spectral fitting, so a 
power-law spectrum with the same spectral index and absorption column as 
observation 0112271001 was assumed when estimating the flux.
$^g$Goodness of the minimised cash-statistic fit.
$^h$Again, insufficient counts were available for spectral fitting, so a 
power-law spectrum was assumed with a typical spectral index of 2 and an 
intrinsic absorption column of $1 \times 10 ^{21}~{\rm cm^{-2}}$.
\end{minipage}
\label{appendix1}
\end{table*}

In the course of our work we were able to identify two extreme ULX candidates as background contaminants.  These were \srcfour, initially associated with NGC 4065 by Walton et al. (2011) and identified as a background QSO on the basis of SDSS spectroscopy; and \srcsix, initially associated with NGC 4874 but later identified as a background QSO using Gemini GMOS spectroscopy.  In Table~\ref{appendix1} we present the results of the simple X-ray spectral fits to the data for these objects, for comparison with the presumed {\it bona fide\/~}extreme ULXs analysed in the body of the paper (see Section 5.3 for discussion related to this comparison).  Here we also present the Gemini GMOS spectrum of the optical counterpart to \srcsix~in Figure~\ref{GMOSspec}.  The broad emission lines present at $\sim 5177,~5265~\&~8076$ \AA~ are identifiable with the Lyman $\alpha$ 1216 \AA, N{\small V} 1240 \AA~ \& C{\small III} 1909 \AA~ lines (respectively) at a redshift of $z \approx 3.25$, demonstrating this is very likely to be a high redshift QSO.

\bsp

\label{lastpage}

\end{document}